# Unitarity Corrections in High Energy QCD *

**Jochen Bartels**

*II. Institut für Theoretische Physik, Universität Hamburg,*
*Luruper Chaussee 149, D-22761 Hamburg, Germany*
*E-mail:* bartels@x4u.desy.de

**Carlo Ewerz**

*Cavendish Laboratory, Cambridge University,*
*Madingley Road, Cambridge CB3 0HE, U.K.*
*and*
*DAMTP, Cambridge University,*
*Silver Street, Cambridge CB3 9EW, U.K.*
*E-mail:* carlo@hep.phy.cam.ac.uk

Abstract: The high energy limit of QCD is investigated in the generalized leading logarithmic approximation. We study unitarity corrections to the BFKL Pomeron containing $t$-channel states with up to six gluons. Special attention is given to the field theory structure of the corresponding multi–gluon amplitudes. We discuss the transition from two to six gluons in the $t$-channel.

Keywords: QCD, Deep Inelastic Scattering.

*Work supported in part by the EU Fourth Framework Programme 'Training and Mobility of Researchers', Network 'Quantum Chromodynamics and the Deep Structure of Elementary Particles', contract FMRX-CT98-0194 (DG 12 - MIHT). CE was also supported in part by Deutsche Forschungsgemeinschaft and by the German Bundesministerium für Bildung, Wissenschaft, Forschung und Technologie.

# Contents

---

# 1. Introduction

The scattering of hadrons at high energy and small momentum transfer is very interesting. On the one hand, hadronic scattering in this Regge limit has been very successfully described [1] by Regge theory [2], based on analyticity and unitarity of the $S$-matrix. The behaviour of hadronic scattering amplitudes is encoded in the positions of so–called Regge poles and cuts in the plane of complex angular momentum. With Gribov's reggeon calculus [3] a consistent field theory of interacting reggeons was found and extensively studied [4]. To the present day, Regge theory remains one of the deep truths of particle physics. The positions of the Regge singularities, however, cannot be calculated in the framework of Regge theory.

On the other hand, QCD has been firmly established as the quantum field theory describing the physics of strong interactions. It should therefore – at least in principle – be possible to derive Regge theory from QCD and to understand it in terms of quark and gluon degrees of freedom. This important and fundamental problem has not yet been solved. The main difficulty lies in the fact that the Regge limit is characterized by high parton densities. In addition, many hadronic scattering processes in the Regge kinematics are dominated by small momentum scales. Therefore non–perturbative effects are expected to become important.

Fortunately, there are a few scattering processes that can be treated perturbatively even in the Regge limit. These processes involve the scattering of small color dipoles. An example is the scattering of highly virtual photons, in which the virtuality of the photons provides a hard scale. This process is of phenomenological interest [5, 6], as it can be measured at LEP [7] and, even better, at a future Linear Collider. For our present investigation we will look at it from a more theoretical perspective. The basic idea is to approach the difficult dynamics of QCD in the Regge limit from a perturbative starting point, namely highly virtual $\gamma^*\gamma^*$ scattering. Our hope is that the results obtained in the study of this special process are of more general relevance to the general dynamics of QCD in the Regge limit. In a later step, one should then of course try to understand the effects of non–perturbative dynamics on the emerging picture. One such non–perturbative effect would be the diffusion of momenta into the infra–red [8, 9], which can be controlled to a certain extent in $\gamma^*\gamma^*$ scattering by demanding sufficiently large photon virtualities.

The first step towards a QCD-based description of hadronic scattering in the high energy limit was done when the BFKL (or perturbative) Pomeron was derived [10, 11]. It describes a $t$-channel exchange carrying vacuum quantum numbers and resums the leading logarithms of the squared energy $s$ which compensate the smallness of the strong coupling constant. The BFKL Pomeron can be understood as the $t$-channel exchange of two (reggeized) gluons. It is expected to apply to processes which are governed by a single hard momentum scale.

The BFKL Pomeron results in a power–like growth $\sigma \sim s^{0.5}$ of the total cross section at high energy. This eventually leads to a violation of unitarity, since according to the Froissart bound [12, 13] the growth of total cross–sections can at most be logarithmic with the energy. Even though the actual violation of that bound would occur only well above the Planck scale it makes the leading logarithmic approximation inconsistent as far as our main goal is concerned. Namely, we cannot expect to derive a consistent picture of the Regge limit from an approximation that does not respect unitarity.

Unitarity can be restored by including exchanges with more than two gluons in the $t$-channel. A set of so–called unitarity corrections can be identified that results in a unitary amplitude. This leads to the generalized leading logarithmic approximation [14, 15, 16]. In order to satisfy unitarity also in all sub–channels, the number of gluons in the $t$-channel should not be fixed, and arbitrary numbers of gluons should be taken into account. The natural objects to consider are therefore amplitudes describing the production of $n$ gluons in the $t$-channel. These $n$-gluon amplitudes obey a tower of coupled integral equations, the first of which ($n = 2$) coincides with the BFKL equation. In the Regge limit the dynamics of the interaction is effectively reduced to the two transverse dimension. Consequently, the integral equations are equations in two–dimensional tranverse momentum space. The complex angular momentum acquires the meaning of an energy–like variable, whereas its conjugate variable, i. e. rapidity, can be interpreted as a time–like variable. This is in agreement with the fact that real gluon emissions in the (generalized) leading logarithmic approximation are strongly ordered in rapidity.

So far, the amplitudes have been investigated for up to $n = 4$ gluons. Although a complete analytic solution for the four–gluon amplitude is still missing, a series of remarkable properties of the amplitudes has been found. The three–gluon amplitude and a part of the four–gluon amplitude can be written as superpositions of two–gluon amplitudes. This is called the reggeization of that parts, generalizing the well–known reggeization of the gluon in non–abelian gauge theories [17]. A consequence of this phenomenon is that the amplitudes exhibit a very interesting field theory structure. They consist of only very few building blocks: states of two and four interacting $t$-channel gluons and a vertex coupling the two– to the four–gluon state [18, 19]. The latter is a number–changing element and turns the quantum–mechanical problem of the $n$-gluon states [15, 20] into a field theory of unitarity corrections. A further striking feature is observed after a Fourier transformation to two–dimensional impact parameter space. The two–gluon state [21], the four–gluon states as well as the two–to–four vertex are invariant under conformal transformations of the gluon coordinates [22]. These properties indicate that the whole set of unitarity corrections in the generalized leading logarithmic approximation can be described by an effective conformal field theory in two–dimensional impact parameter space with rapidity as an additional parameter. It would be a great step forward to identify this conformal field theory. That would allow one to apply the extremely powerful methods of conformally invariant field theory in two dimensions [23] (for a review see [24]).

These interesting properties of the unitarity corrections have until now only been observed in the amplitudes with up to four gluons, and only the most basic elements of the potential field theory have been calculated explicitly. More generally, the existence of such a field theory structure has not yet been proved beyond four gluons. To find further evidence for the conjectured field theory structure and to extract its general properties requires the investigation of higher $n$-gluon amplitudes. A series of questions arise naturally in this program. An important goal is of course to find out about the existence of new elements of the effective field theory, like a possible two–to–six vertex, and to determine their properties. A further natural question is whether the two–to–four vertex can be generalized to the case in which the two incoming gluons are not in a color singlet state, and how this generalization looks like. Closely related is the question of how a repeated two–to–four transition takes place. The study of the six–gluon amplitude will provide an answer to the question whether a Pomeron can split into two Odderons, the Odderon being the $C = -1$ partner of the Pomeron. An important issue will also be to test the conformal invariance of the higher $n$-gluon amplitudes in impact parameter space.

In the present paper we study the amplitudes with five and six gluons in the $t$-channel. We show that the five–gluon amplitude reggeizes completely, i. e. is a superposition of two– and four–gluon amplitudes. The corresponding mechanism appears to apply to each $n$-gluon amplitude with odd $n$. We extract a reggeizing part from the six–gluon amplitude and derive an integral equation for the remaining part. This equation contains a term which gives rise to at least one new element in the effective field theory. The other terms in the equation can be expressed through the known two–to–four vertex, indicating that further reggeization takes place in the six–gluon amplitude. We discuss our findings in the light of the potential effective field theory. The six–gluon amplitude allows us for the first time to determine the perturbative Pomeron–Odderon–Odderon vertex. Its precise form and properties, as well as those of the generalization of the two–to–four vertex to the color non–singlet will be discussed in a separate publication. Also the conformal invariance of the five– and six–gluon amplitudes will be investigated in a separate publication.

The problem of unitarity in high energy QCD has been addressed in many different ways, among them the following. The first attempt was made in ref. [25], but in that approach a set of diagrams has been left out which is included in our approach. These diagrams turn out to be essential for the emergence of higher $n$-gluon states. Considerable progress has been made in investigating these $n$-gluon states in the $t$-channel. Much interest was attracted by the symmetric three–gluon state (the Odderon) [26, 27]. More generally, the large $N_c$ limit of the $n$-gluon states has been shown to be equivalent to a completely integrable model [28], namely the XXX Heisenberg model with non–compact $\mathrm{SL}(2,\mathbb{C})$ spin zero [29]. This remarkable result was subsequently used to obtain further interesting results [30], including the Odderon as a special case. These results are relevant for our present approach since the $n$-gluon states – although with finite $N_c$ – are elements of the potential effective field theory. The two–to–four transition vertex and its relation to the triple–Pomeron vertex has been studied in [31, 32]. The latter can be obtained from the two–to–four vertex after projection on BFKL Pomerons. It has been shown to have the structure of a conformal three–point function in [33]. That reference also discusses the analytic structure of the four–gluon state for finite $N_c$. An alternative approach to QCD at high energy is Mueller's dipole picture [34], which has been proved to be equivalent to the BFKL formalism [35, 36]. The dipole picture has been used to study unitarity effects as well [37, 38, 39]. The value of the triple–Pomeron coupling obtained in the dipole picture [40] agrees with the one calculated from the two–to–four vertex arising in our present approach [41]. Other related approaches are the formulation of an effective action for the Regge limit [42, 43], a similar approach also aiming at a simplified effective theory for high energy scattering [44], and the method of operator expansion [45]. Renormalization group methods have been used to study unitarization in [46]. In [47] a supercritical phase of reggeon field theory is studied in which the Pomeron consists of a single gluon and a 'wee parton'. Approaches of more non–perturbative nature are for example the eikonal approximation in a soft gluon background [48], the model of the stochastic vacuum [49], and the semiclassical approach [50]. Also the recently conjectured AdS/CFT correspondence [51] has been applied to the Regge limit [52, 53]. This short overview is of course not complete. All the different approaches should be regarded as complementary. As far as this has been tested, different approaches often lead to equivalent or similar results. The most difficult problem will eventually be the combination of the perturbative and non–perturbative approaches.

The paper is organized as follows. In section 2 we recall important facts about the BFKL equation and briefly describe the violation of unitarity and the reggeization of the gluon. In section 3 we motivate the use of $n$-gluon amplitudes and outline their formal definition. The integral equations for the amplitudes are described in detail, including color algebra and the inhomogeneous terms consisting of quark loops with $n$ gluons attached. We introduce reggeon momentum diagrams and use them to classify the occurring momentum space integrals in terms of a small set of standard integrals. Section 4 contains a review of the known results about the three– and

four–gluon amplitudes as well as a discussion of the field theory structure found in these amplitudes. In section 5 we solve the equation for the five–gluon amplitude and discuss the general mechanism that is expected to cause complete reggeization in all amplitudes with an odd number of gluons. Section 6 is devoted to the study of the six–gluon amplitude. We extract a reggeizing part and derive the equation for the remaining part which is then discussed in detail. From that equation we also derive the Pomeron–Odderon–Odderon vertex. We conclude with a summary and an outlook. The two appendices contain algorithms for computing contractions of color tensors (appendix A) and for bringing the momentum space integrals to their standard forms (appendix B).

## 2. The BFKL equation and violation of unitarity

In this section we give a very brief account of the basic properties of the BFKL equation. For more detailed accounts of the BFKL theory we refer the reader to the reviews [43, 54, 55].

### 2.1 The BFKL equation

The total hadronic cross section can be related to the elastic forward scattering amplitude via the optical theorem,

$$\sigma_{\text{tot}} = \frac{1}{s} \operatorname{Im} A_{\text{el}}(s, t=0) \,, \tag{2.1}$$

where $t$ is the momentum transfer. It is convenient to use partial wave amplitudes, which in the high energy limit amounts to performing a Mellin transformation

$$A(s,t) = is \int_{\delta - i\infty}^{\delta + i\infty} \frac{d\omega}{2\pi i} \left( \frac{s}{M^2} \right)^{\omega} A(\omega, t) \,, \tag{2.2}$$

thereby changing from squared energy $s$ to complex angular momentum $\omega$. The high energy behavior of the total cross section is then determined by the singularities of $A(\omega, t)$ in the $\omega$-plane, i. e. Regge poles and Regge cuts. The rightmost singularity gives the leading contribution and is identified with the Pomeron. As it describes an elastic scattering process (see (2.1)) it carries vacuum quantum numbers.

If there is a hard momentum scale in the process the use of perturbation theory is justified. The smallness of the strong coupling constant $\alpha_s$ at large momentum scales can at high energy be compensated by large logarithms of the energy. The resummation of Feynman diagrams contributing to the leading logarithmic approximation (LLA) — $\alpha_s \ll 1$ and $\alpha_s \log(s) \sim 1$ — results in the BFKL equation [10, 11]. The longitudinal degrees of freedom decouple in high energy scattering, and the dynamics takes place in transverse space only. The full amplitude can be written in factorized form

$$A(\omega, t) = \int \frac{d^2\mathbf{k}}{(2\pi)^3} \frac{d^2\mathbf{k}'}{(2\pi)^3} \phi_\omega(\mathbf{k}, \mathbf{k}'; \mathbf{q}) \phi_1(\mathbf{k}, \mathbf{q}) \phi_2(\mathbf{k}', \mathbf{q}) \,, \tag{2.3}$$

$\phi_1$, $\phi_2$ being the impact factors of the scattered colorless states. The color neutrality implies

$$\phi_{1,2}(\mathbf{k}=0,\mathbf{q}) = \phi_{1,2}(\mathbf{k}=\mathbf{q},\mathbf{q}) = 0\,, \tag{2.4}$$

which is important for the infrared finiteness of the amplitude. The function $\phi_\omega$ can be interpreted as the partial wave amplitude for the scattering of virtual gluons with virtualities $-\mathbf{k}^2$, $-(\mathbf{q}-\mathbf{k})^2$, $-\mathbf{k}'^2$, and $-(\mathbf{q}-\mathbf{k}')^2$ respectively. It is described by the BFKL equation, which is an integral equation in the two–dimensional space of transverse momenta and of Bethe–Salpeter type. In detail it has the form

$$\omega\phi_\omega(\mathbf{k},\mathbf{k}';\mathbf{q}) = \phi^0(\mathbf{k},\mathbf{k}';\mathbf{q}) + \int \frac{d^2\mathbf{l}}{(2\pi)^3}\,\frac{1}{\mathbf{l}^2(\mathbf{q}-\mathbf{l})^2} K_{\mathrm{BFKL}}(\mathbf{l},\mathbf{q}-\mathbf{l};\mathbf{k},\mathbf{q}-\mathbf{k})\,\phi_\omega(\mathbf{l},\mathbf{k}';\mathbf{q})\,. \tag{2.5}$$

$\phi^0$ is an inhomogeneous term, depending on the process under consideration. The integral kernel, the so-called BFKL or Lipatov kernel, is given by

$$\begin{aligned} K_{\mathrm{BFKL}}(\mathbf{l},\mathbf{q}-\mathbf{l};\mathbf{k},\mathbf{q}-\mathbf{k}) \;=\; & -N_c g^2 \left[\mathbf{q}^2 - \frac{\mathbf{k}^2(\mathbf{q}-\mathbf{l})^2}{(\mathbf{k}-\mathbf{l})^2} - \frac{(\mathbf{q}-\mathbf{k})^2\mathbf{l}^2}{(\mathbf{k}-\mathbf{l})^2}\right] \\ & +(2\pi)^3\mathbf{k}^2(\mathbf{q}-\mathbf{k})^2\left[\beta(\mathbf{k})+\beta(\mathbf{q}-\mathbf{k})\right]\delta^{(2)}(\mathbf{k}-\mathbf{l}). \end{aligned} \tag{2.6}$$

The strong coupling constant is normalized to $\alpha_s = \frac{g^2}{4\pi}$ and is kept fixed in LLA. The function $\beta$ in the kernel is defined as

$$\beta(\mathbf{k}^2) = \frac{N_c}{2}g^2\int\frac{d^2\mathbf{l}}{(2\pi)^3}\frac{\mathbf{k}^2}{\mathbf{l}^2(\mathbf{l}-\mathbf{k})^2}\,. \tag{2.7}$$

The function $\alpha(\mathbf{k}^2) = 1+\beta(\mathbf{k}^2)$ is known as the gluon trajectory function. It passes through the physical spin 1 of the gluon at vanishing argument $\mathbf{k}^2 = 0$. Since it is the function $\beta$ that will frequently occur throughout this paper we call it (in obvious abuse of language) the trajectory function of the gluon as well.

The factor $(-N_c)$ in the BFKL kernel is a color factor. If the two gluons entering the amplitude $\phi_\omega$ are not in a color singlet state the color factor $C_I$ will be different. (In that case the amplitude is not infrared finite and has to be regularized.) In general, $C_I$ depends on the irreducible representation $I$ of the two gluons. If $N_c = 3$, the factor $C_I$ equals $-3$, $-\frac{3}{2}$, $-\frac{3}{2}$, 0, 1 for the irreducible representations $\mathbf{1}$, $\mathbf{8_A}$, $\mathbf{8_S}$, $\mathbf{10}+\overline{\mathbf{10}}$, $\mathbf{27}$, respectively. We will use the symbol $K_{\mathrm{BFKL}}$ for the BFKL kernel only if the two gluons are in a color singlet state.

The complex angular momentum $\omega$ acts as an energy variable in the BFKL equation. It can be shown to be conjugate to rapidity which thus acquires the meaning of a time variable in the BFKL equation.

The general form of the solution of the BFKL equation can be derived from the integral equation by iteration. Accordingly, the elastic scattering amplitude at high energies has in LLA the structure of a gluon ladder in the $t$-channel,

$s \to \infty$ 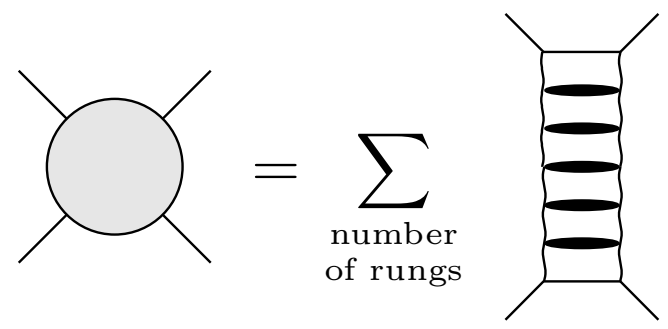 , (2.8)

and the ladder rungs represent BFKL kernels. The BFKL equation can be solved analytically, but we will not make use of the explicit form of the solution. In the present paper we will also not need the conformal invariance [21] of the BFKL equation in impact parameter space.

Finally, we mention the $t$-channel reggeon unitarity relation for the BFKL amplitude. Let $C(\omega; \mathbf{k}, \mathbf{k}'; \mathbf{q})$ be the amputated BFKL amplitude, i. e. the amplitude $\phi_\omega$ without the reggeon propagator $(\omega - \beta(\mathbf{k}^2) - \beta((\mathbf{q}-\mathbf{k})^2))^{-1}$,

$$C(\omega; \mathbf{k}, \mathbf{k}'; \mathbf{q}) = \Big(\omega - \beta(\mathbf{k}^2) - \beta((\mathbf{q}-\mathbf{k})^2)\Big)\, \phi_\omega(\mathbf{k}, \mathbf{k}'; \mathbf{q}) \qquad (2.9)$$

with $t = -\mathbf{q}^2$. After a continuation to the physical region of the $t$-channel it is possible to show [10] with the help of the BFKL equation that

$$\mathrm{disc}_\omega C = \int d^2\mathbf{l}\, (2\pi)^3 \frac{1}{\mathbf{l}^2(\mathbf{q}-\mathbf{l})^2} \delta\left(\omega - \beta(\mathbf{l}^2) - \beta((\mathbf{q}-\mathbf{l})^2)\right) C(\omega; \mathbf{k}, \mathbf{l}; \mathbf{q}) C^*(\omega; \mathbf{l}, \mathbf{k}'; \mathbf{q})\,. \qquad (2.10)$$

The right hand side can be understood as a unitarity integral for the two–gluon amplitude $C$.

## 2.2 Violation of unitarity

The leading singularity in the $\omega$-plane can be determined analytically from the BFKL equation, and leads a power–like growth of the amplitude $A \sim s^{(1+\omega_{\text{BFKL}})}$ with the exponent $\omega_{\text{BFKL}} = \alpha_s N_c\, 4 \ln 2/\pi \simeq 0.5$. Consequently, the total cross section in the leading logarithmic approximation grows like $\sigma_{\text{tot}} \sim s^{\omega_{\text{BFKL}}}$. This result is in conflict with the Froissart–Martin theorem [12, 13] which derives a bound on the total hadronic cross section from unitarity. In detail, the Froissart–Martin bound is $\sigma_{\text{tot}} \leq \text{const.} \log^2(s)$. A power–like growth will eventually violate this bound — and thus unitarity — at asymptotically large energies. This observation is the starting point for the considerations in this paper.

## 2.3 Reggeization of the gluon

The phenomenon of reggeization in non–abelian gauge theories [17] is the following. The $t$-channel exchange in the BFKL equation carrying the quantum numbers of a gluon, i. e. a color octet[1] exchange, gives rise to a special solution. For antisymmetric color octet exchange the color factor in the BFKL kernel is $N_c/2$ instead of $N_c$. (In this color representation the amplitude is not infrared finite and a regularization has to be applied.) Let us further assume that the inhomogeneous term $\phi_0$ is a function of $(\mathbf{k}_1 + \mathbf{k}_2)$. Then the equation exhibits the solution

$$\phi^{\mathbf{8_A}}(\mathbf{k}_1 + \mathbf{k}_2) = \frac{\phi_0^{\mathbf{8_A}}(\mathbf{k}_1 + \mathbf{k}_2)}{\omega - \beta(\mathbf{k}_1 + \mathbf{k}_2)} \,. \tag{2.11}$$

[1]We speak of 'octet' to mean the adjoint representation also for general $N_c$.

This solution has a pole and can thus be interpreted as describing the propagation of a single particle with momentum $(\mathbf{k}_1 + \mathbf{k}_2)$ and the quantum numbers of a gluon. In a sense the gluon turns out to be a bound state of two gluons here. The fact that the gluon is a composite state of gluons is often termed 'bootstrap'. It indicates that the correct degrees of freedom in high energy QCD are not elementary gluons but so–called reggeized gluons. The reggeized gluon can be understood as a collective excitation of the gauge field.

When we interchange the two gluons in the color octet amplitude above we find that its sign changes. This fact gives rise to the notion of signature. It characterizes the behavior under the exchange of two gluons, that is the simultaneous interchange of color and momentum labels. The reggeized gluons obviously carries negative signature.

# 3. Integral equations for $n$-gluon amplitudes

## 3.1 The $n$-gluon amplitudes

The method suited to restore unitarity in the perturbative approach is known as the generalized leading logarithmic approximation (GLLA). It constitutes an approximation scheme in which a minimal set of non–leading corrections is identified that leads to a unitary amplitude. The minimal set of contributions required here comprises subleading corrections with a larger number $n$ of reggeized gluons in the $t$-channel. These are what we call unitarity corrections to the BFKL Pomeron. It is necessary to include all possible $n$ to eventually fulfill the requirement of unitarity. Quark exchanges in the $t$-channel are always suppressed by powers of the energy $s$ with respect to the corresponding gluon exchanges and are not taken into account in GLLA.

The most complete approach to a systematic treatment of unitarity corrections in a perturbative framework was formulated in [14, 15, 16]. Its aim is to arrive at an effective description of QCD in the Regge limit in the spirit of a reggeon field theory [3, 4], the requirement of unitarity being built in from the very beginning. Of course, the BFKL amplitude with its two $t$-channel gluons should be incorporated into the whole approach as the lowest order contribution. It appears natural to define partial wave amplitudes similar to the BFKL amplitude but now with $n$ reggeized gluons in the $t$-channel. In eq. (2.10) we have seen the $t$-channel reggeon unitarity equation for the BFKL amplitude. It can be summarized symbolically as

$$\text{disc}_\omega A(\omega, t) \sim C_2 C_2^* \,, \tag{3.1}$$

where $C_2$ is an amputated amplitude. We include a reggeon propagator to arrive at $D_2 = C_2[\omega - \beta(\mathbf{k}_1) - \beta(\mathbf{k}_2)]^{-1}$. We have chosen a new symbol $D_2$ for the BFKL amplitude here since we now want to consider the special physical process of $\gamma^*\gamma^*$-scattering. The amplitude $D_2$ obeys the BFKL equation (2.5) with a special choice of the inhomogeneous term, namely the coupling of the two gluons to the photons via a quark loop. The unitarity relation (3.1) can be generalized to include $n$-gluon intermediate states. Symbolically, the generalization has the form

$$\text{disc}_\omega A(\omega, t) \sim \sum_{n=2}^{\infty} C_n C_n^* \,. \tag{3.2}$$

We include a reggeon propagator to find $D_n = C_n(\omega - \sum_{i=1}^n \beta(\mathbf{k}_i))^{-1}$. Here the $D_n$ describe the production of $n$ on–shell gluons in the $t$-channel. They are non–amputated amplitudes, i. e. have propagators on the external gluon lines. The correct treatment of $t$-channel unitarity relations including multi–particle amplitudes is highly non–trivial. To our knowledge the most complete survey of this extensive technology is [56], and the reader is referred to that reference for the details. We will content ourselves here with having motivated the physical meaning of the $n$-gluon amplitudes $D_n$ we are going to study. Further below we will briefly outline the formal definition of the amplitudes $D_n$.

Once we take into account subleading corrections with more reggeized gluons in the $t$-channel and consider multi–particle amplitudes like the $n$-gluon amplitudes $D_n$ there exist of course subchannels of the scattering amplitude and we have to care about unitary also in the subchannels. The approach initiated in [14, 15, 16] is designed to ensure unitarity not only in the direct channel but also in all subchannels. This implies that the number of gluons in the $t$-channel gluons is not conserved. Due to that the set of integral equations for the $n$-gluon amplitudes is turned into a tower of coupled equations including number–changing integral kernels. (A detailed description of the integral equations will follow in section 3.2.)

The non–conservation of the number of reggeized gluons in the $t$-channel evolution contrasts sharply with the situation in the Bartels–Kwieciński–Praszałowicz (BKP) equations [15, 20]. The latter describe the $t$-channel evolution of a compound state of a fixed number of reggeized gluons in the Regge limit. Their large-$N_c$ limit turned out to be equivalent to a completely integrable model [28], namely the XXX–Heisenberg model with conformal $\text{SL}(2, \mathbb{C})$ spin $s = 0$ [29]. Although the BKP equations do not apply directly to our $n$-gluon amplitudes $D_n$ they will play an important role in the effective field theory of unitarity corrections that we are heading for. As we will explain in more detail in section 4.3 there will be different elements in the effective field theory. First we will have $n$-particle Green functions, which are number–conserving elements. Their behavior will be governed by the BKP equations. In addition, there will be number–changing elements which we will call vertices. They arise as a unique feature of the approach pursued here, and turn the quantum mechanical problem described by the BKP equations into a quantum field theory.

We will now outline the formal definition of the $n$-gluon amplitudes $D_n$. The way the amplitudes $D_n$ are defined is inspired by Regge theory. A condensed but still rather extensive description of the methods that have to be used here can be found in [47], more complete reviews are contained in [57] and [58]. The procedure starts from a physical $2+n$ multi–particle scattering process. One identifies certain kinematical variables with the use of so–called Toller diagrams and hexagraphs, and after defining partial waves one can eventually get the desired amplitude by taking an appropriate mixed Regge and helicity–pole limit. Although it is in principle possible, we will not carry out this program for the amplitudes under consideration in this paper. The procedure becomes technically complicated very quickly with the increasing number of gluons in the amplitude. Moreover, the physical processes we would have to start with for larger $n$ are rather artifical. However, the method outlined here appears to be very natural for a special phenomenological application of the four–gluon amplitude. In [19] a part of the four–gluon amplitude $D_4$ in which the four gluons form two pairs of color singlets was used for the description of the process of high-mass diffractive dissociation in deep inelastic electron–proton scattering. The rapidity gap between the proton and the diffractively produced system is caused by a colorless exchange between the proton and the photon. The latter is modelled by a two–gluon exchange. In the amplitude for this process the initial state is therefore indeed a three–particle state, and the cross–section takes the form of a three–to–three scattering process. In [19] the appropriate limit for this process was identified as the well–known triple Regge limit in which $s \gg M^2 \gg Q^2 \gg \Lambda^2_{\mathrm{QCD}}$, where $M^2$ is the invariant mass of the diffractively produced particles.

The $n$-gluon amplitudes are defined to have as external lines two photons and $n$ gluons. The photons are coupled to the gluons via their splitting into a quark–antiquark pair to which the gluons are attached. Being $n$-gluon amplitudes the $D_n$ carry as arguments $n$ color labels $a_i$ in addition to the transverse momenta $\mathbf{k}_i$ of the gluons. The color labels correspond to generators $t^{a_i}$ of the gauge group $\mathrm{SU}(N_c)$ in the adjoint representation. As partial waves the $D_n$ have also as an argument the complex angular momentum $\omega$. Since all $D_n$ will carry the same argument $\omega$ we will suppress it in our notation. In our notation we will suppress the photon momenta as well. The $n$-gluon amplitudes are thus characterized as

$$D_n^{a_1 \dots a_n}(\mathbf{k}_1, \dots, \mathbf{k}_n)\,. \tag{3.3}$$

The transverse gluon momenta $\mathbf{k}_i$ in the amplitude are all chosen to be outgoing.

The $D_n$ are non–amputated amplitudes, i. e. they have propagators for the outgoing reggeized gluons. Further they are multiply–cut amplitudes. We take discontinuities in the $n-1$ energy variables defined from one photon and the $i$ first gluons ($1 \leq i \leq n-1$),

$$s_i = \left( p_{\gamma^*;1} + \sum_{j=1}^{i} p_j \right)^2 . \tag{3.4}$$

Here $p_{\gamma^*;1}$ and $p_j$ are the four–momenta of the left photon and the gluons, respectively. The amplitudes $D_n$ can be defined for the non–forward direction

$$\sum_{i=1}^{n} \mathbf{k}_i \neq 0 \tag{3.5}$$

as well. All results in this paper will hold for the non–forward direction, and we will not mention this in each case separately.

The simplest of the $n$-gluon amplitudes is $D_2$. It is identical with the well–known BFKL amplitude discussed in section 2.1. There the inhomogeneous term in the BFKL equation was not specified. In $D_2$ it is given by the lowest order coupling of the two $t$-channel gluons to the virtual photons through a quark loop. The two outgoing gluons in the BFKL amplitude $D_2^{a_1a_2}$ are in a color singlet such that we can factorize the color structure and define the momentum part $D_2$ by

$$D_2^{a_1a_2}(\mathbf{k}_1, \mathbf{k}_2) = \delta_{a_1a_2} D_2(\mathbf{k}_1, \mathbf{k}_2) \,. \tag{3.6}$$

We recall two simple but very important properties of the momentum part $D_2$ of the BFKL amplitude. The first is that it vanishes when one of its momentum argument vanishes,

$$D_2(\mathbf{k}_1, \mathbf{k}_2)|_{\mathbf{k}_1=0} = D_2(\mathbf{k}_1, \mathbf{k}_2)|_{\mathbf{k}_2=0} = 0 \,. \tag{3.7}$$

The second is the symmetry in its two momentum arguments,

$$D_2(\mathbf{k}_1, \mathbf{k}_2) = D_2(\mathbf{k}_2, \mathbf{k}_1) \,. \tag{3.8}$$

Concluding this section, we introduce a shorthand notation for the arguments of $D_2$. Later we will use it for the arguments of other functions as well. In the case that an argument of a function, say $D_2$, is a sum of two or more transverse momenta we will only give the indices of these momenta, and a string of indices stands for the sum of the corresponding momenta. So we have for example

$$D_2(12, 3) = D_2(\mathbf{k}_1 + \mathbf{k}_2, \mathbf{k}_3) \,. \tag{3.9}$$

### 3.2 Integral equations

The $n$-gluon amplitudes $D_n$ obey a tower of coupled integral equations. These have been derived in [16] by means of $s$-channel dispersion relations. Like the BFKL equation they are equations in two–dimensional transverse momentum space describing the $t$-channel evolution of the amplitudes under investigation. In this evolution, the complex angular momentum $\omega$ again plays the role of an energy variable. Its conjugate, i. e. rapidity, acquires the meaning of the time–like variable in the evolution.

The integral equation for the two–gluon amplitude $D_2^{a_1a_2}$ is of course identical to the BFKL equation,

$$\left(\omega - \sum_{i=1}^{2} \beta(\mathbf{k}_i)\right) D_2^{a_1a_2} = D_{(2;0)}^{a_1a_2} + K_{2\to 2}^{\{b\}\to\{a\}} \otimes D_2^{b_1b_2} \,. \tag{3.10}$$

We have moved the trajectory functions to the left hand side of the equation to make the generalization to larger $n$ more transparent. The inhomogeneous term $D_{(2;0)}$ denotes the lowest order coupling of the two gluons to the photons via the quark loop. The quark loop will be the subject of section 3.4. The integral kernel $K_{2\to 2}^{\{b\}\to\{a\}}$ is, roughly speaking, the BFKL kernel (2.6) without the gluon trajectory functions $\beta$. An exact definition of the integral kernels will be given in section 3.6. The superscript $\{b\} \to \{a\}$ corresponds to the color labels of the in– and outgoing gluons. The convolution symbol $\otimes$ stands for an integral $\int \frac{d^2\mathbf{l}}{(2\pi)^3}$ over the loop momentum and a propagator $\frac{1}{\mathbf{l}^2}$ for each of the two gluons entering the kernel from above.

The integral equation for the three–gluon amplitude $D_3^{a_1a_2a_3}$ has the form

$$\left(\omega - \sum_{i=1}^{3} \beta(\mathbf{k}_i)\right) D_3^{a_1a_2a_3} = D_{(3;0)}^{a_1a_2a_3} + K_{2\to 3}^{\{b\}\to\{a\}} \otimes D_2^{b_1b_2} + \sum K_{2\to 2}^{\{b\}\to\{a\}} \otimes D_3^{b_1b_2b_3} \,. \tag{3.11}$$

The inhomogeneous term $D_{(3;0)}$ is now the quark loop with three gluons attached to it. In (3.11) we find for the first time a new kernel in the equation. $K_{2\to 3}^{\{b\}\to\{a\}}$ is a transition kernel from two to three reggeized gluons. The second term on the right hand side of the equation therefore tells us that at some point in the $t$-channel evolution we can have a transition from two to three gluons. The last term describes the evolution of a system of three gluons, and the sum extends over all pairwise interactions of the three reggeized gluons via the kernel $K_{2\to 2}^{\{b\}\to\{a\}}$. Let us look at the term in which the first and second gluon interact via a kernel. In this term the momentum and the color label of the third gluon are not affected. The kernel should thus be understood to contain a factor $\delta_{a_3b_3}$. The symbol $\otimes$ denotes again the integration over the loop momentum in the first two gluons and propagator factors for each of them. The other terms are obtained in analogy.

The equations for higher $n$ are built in a very similar way. They contain as the respective inhomogeneous term the lowest order coupling of $n$ gluons to the quark loop. We denote this lowest order term as $D_{(n;0)}$. A detailed discussion of the quark loop and explicit formulae for $n \leq 6$ will follow in section 3.4. In addition, the higher equations contain also higher transition kernels $K_{2\to m}^{\{b\}\to\{a\}}$ from two to $m$ gluons. A general formula for arbitrary $m$ as well as the explicit formulae for $m \leq 6$ are contained in section 3.6.

Since we will make use of the integral equations for up to $n = 6$ in this paper, we now state them explicitly. The general rule should then be obvious. For $n = 4$ we have

$$\begin{aligned}\left(\omega - \sum_{i=1}^{4} \beta(\mathbf{k}_i)\right) D_4^{a_1a_2a_3a_4} &= D_{(4;0)}^{a_1a_2a_3a_4} + K_{2\to 4}^{\{b\}\to\{a\}} \otimes D_2^{b_1b_2} + \sum K_{2\to 3}^{\{b\}\to\{a\}} \otimes D_3^{b_1b_2b_3} \\ &\quad + \sum K_{2\to 2}^{\{b\}\to\{a\}} \otimes D_4^{b_1b_2b_3b_4} \,, \end{aligned} \tag{3.12}$$

for $n = 5$ the equation is

$$\begin{aligned}\left(\omega - \sum_{i=1}^{5} \beta(\mathbf{k}_i)\right) D_5^{a_1a_2a_3a_4a_5} &= D_{(5;0)}^{a_1a_2a_3a_4a_5} + K_{2\to 5}^{\{b\}\to\{a\}} \otimes D_2^{b_1b_2} \\ &\quad + \sum K_{2\to 4}^{\{b\}\to\{a\}} \otimes D_3^{b_1b_2b_3} + \sum K_{2\to 3}^{\{b\}\to\{a\}} \otimes D_4^{b_1b_2b_3b_4} \\ &\quad + \sum K_{2\to 2}^{\{b\}\to\{a\}} \otimes D_5^{b_1b_2b_3b_4b_5} \,, \end{aligned} \tag{3.13}$$

and finally for $n = 6$ the integral equation reads

$$\begin{aligned}\left(\omega - \sum_{i=1}^{6} \beta(\mathbf{k}_i)\right) D_6^{a_1a_2a_3a_4a_5a_6} &= D_{(6;0)}^{a_1a_2a_3a_4a_5a_6} + K_{2\to 6}^{\{b\}\to\{a\}} \otimes D_2^{b_1b_2} \\ &\quad + \sum K_{2\to 5}^{\{b\}\to\{a\}} \otimes D_3^{b_1b_2b_3} + \sum K_{2\to 4}^{\{b\}\to\{a\}} \otimes D_4^{b_1b_2b_3b_4} \\ &\quad + \sum K_{2\to 3}^{\{b\}\to\{a\}} \otimes D_5^{b_1b_2b_3b_4b_5} \\ &\quad + \sum K_{2\to 2}^{\{b\}\to\{a\}} \otimes D_6^{b_1b_2b_3b_4b_5b_6} \,. \end{aligned} \tag{3.14}$$

Here we again have to explain the meaning of the convolutions and the summation symbols. In short, the sums contain all combinations of the respective amplitudes and kernels in which the $t$-channel gluons do not cross. Before we give an example of the combinatorics we write the integral equations in pictorial language, which makes them easier to understand.

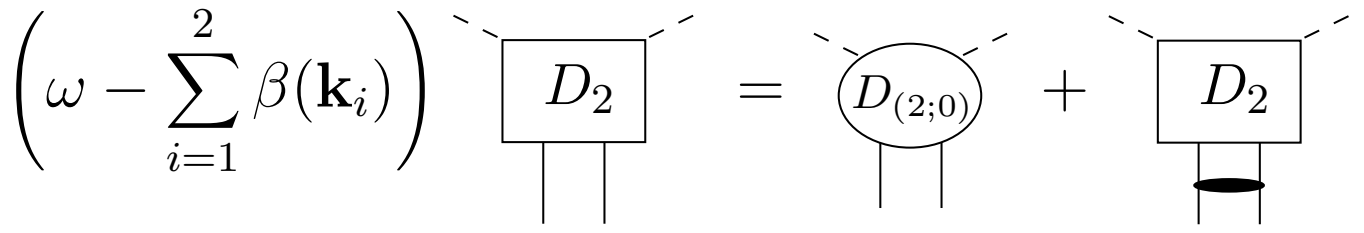

(3.15)

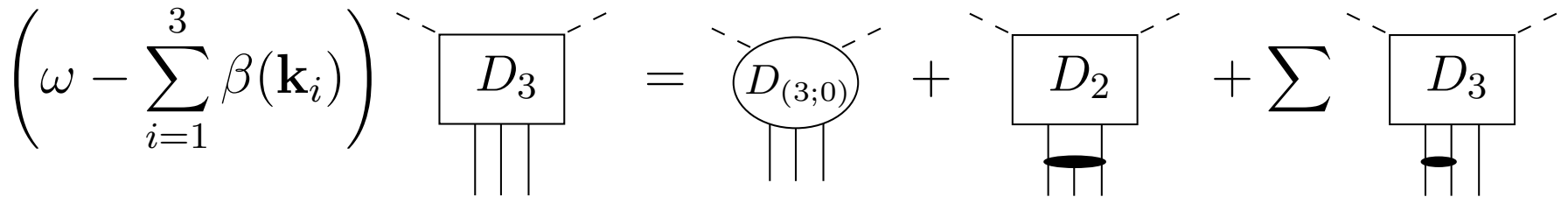

(3.16)

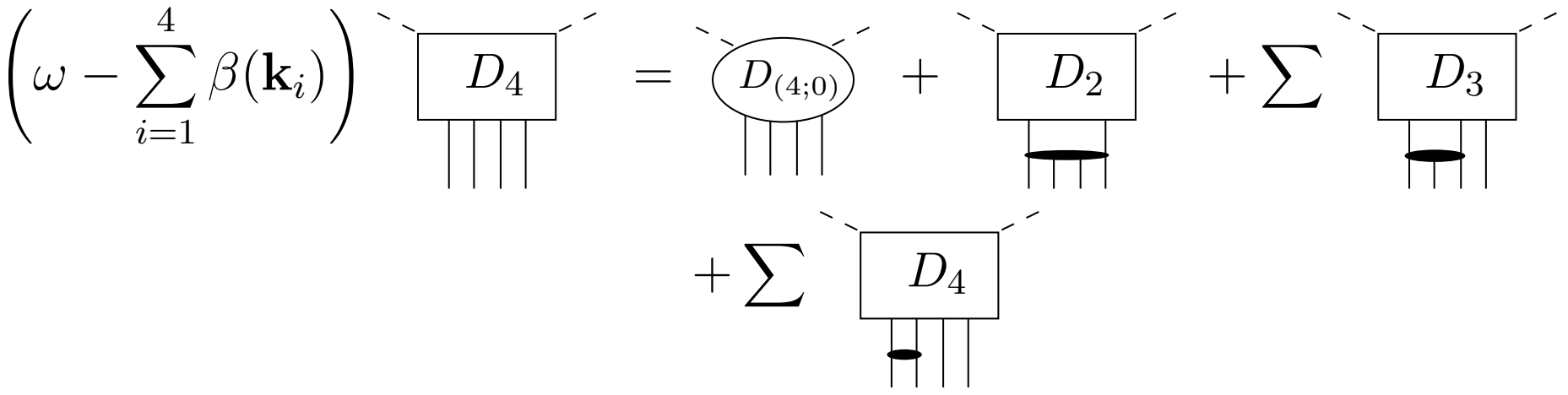

(3.17)

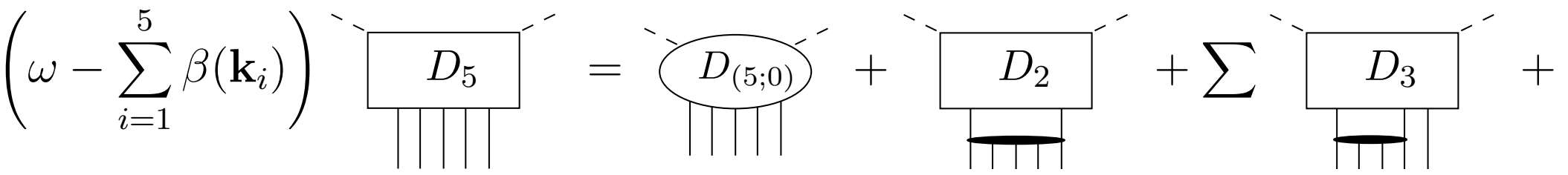

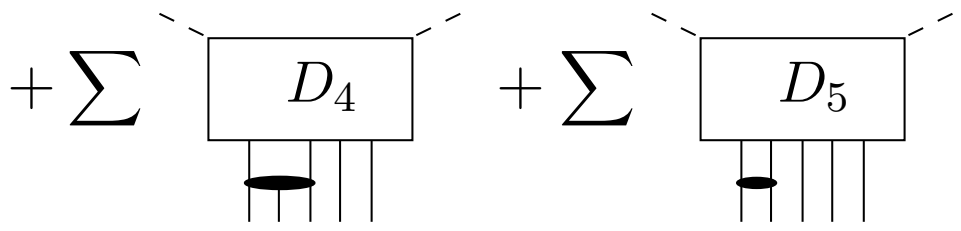

(3.18)

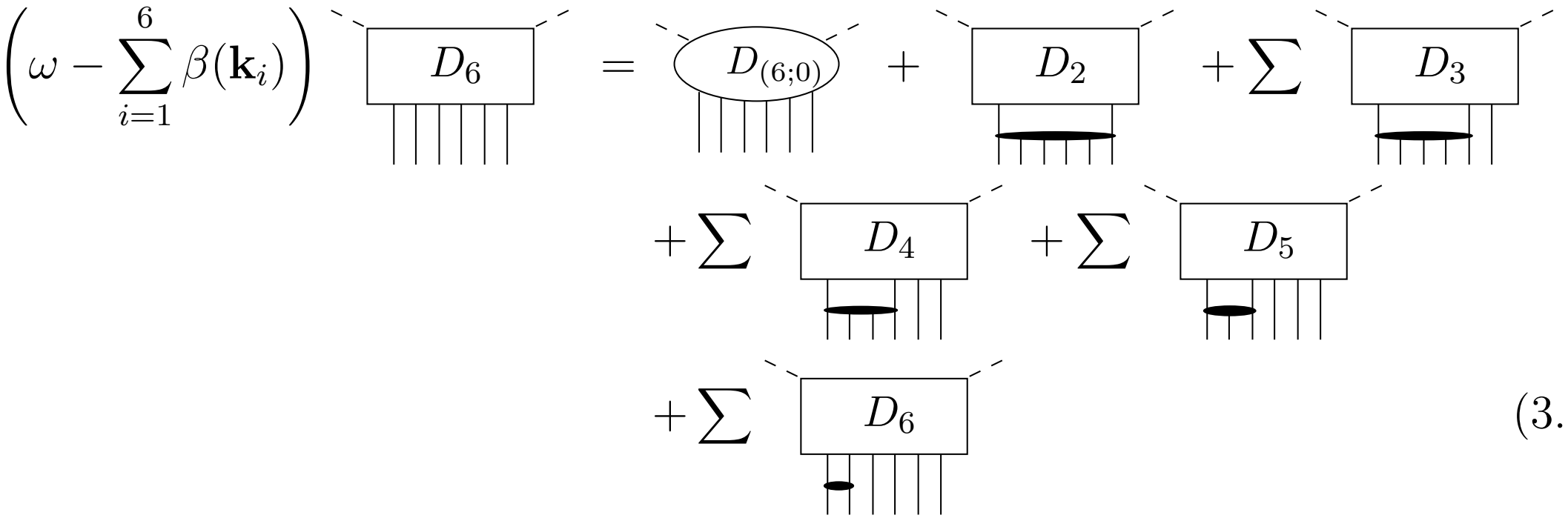

(3.19)

In each diagram only two gluon lines from the amplitudes enter a kernel. An integration $\int \frac{d^2\mathbf{l}}{(2\pi)^3}$ over the loop momentum and a propagator $\frac{1}{\mathbf{l}^2}$ for each of the two gluons entering the kernel from above are implied again. The momenta and color labels of the other gluons are not changed. With the help of the pictorial notation it is also very easy to understand which combinations of amplitudes and kernels have to be convoluted such that $t$-channel gluons do not cross. For example, the sum in the last but one term in the equation (3.17) for the four–gluon amplitude extends over the four convolutions

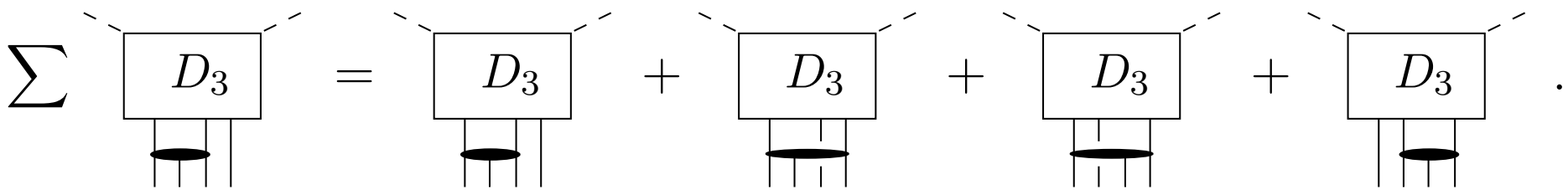

(3.20)

We will now in turn discuss the elements occurring in the integral equations: the inhomogeneous terms $D_{(n;0)}$ representing the coupling of $n$ gluons to a quark loop and the integral kernels $K_{2\to n}$. But before doing so, we first have to consider some color algebra.

### 3.3 Color structure

In this section we collect some essential facts about color algebra. While doing so we also introduce the so–called birdtrack notation[2] for structure constants. This diagrammatic notation is very useful for the problem of contracting indices of arbitrary color tensors. Such a powerful tool is needed here since tensor contractions constitute an essential part of the computations in the study of the integral equations. The diagrammatic method that serves this purpose is described in detail in appendix A.

We are interested in the structure of the gauge group $G = \mathrm{SU}(N_c)$ with generators $t^a$ ($a = 1, \ldots, N_c^2 - 1$) in the Lie algebra $\mathrm{su}(N_c)$. The algebra is

$$[t^a, t^b] = i f_{abc} t^c \,. \tag{3.21}$$

[2]A more complete account of this notation can be found in [59] where it is also applied to other Lie groups. Our normalization convention slightly deviates from [59].

For the case of su(3) the $t^a$ are given by the well–known Gell–Mann matrices $\lambda^a$, $t^a = \lambda^a/2$. The antisymmetric structure constants $f_{abc}$ can be expressed in terms of generators as

$$f_{abc} = -f_{acb} = -2i\left[\mathrm{tr}(t^a t^b t^c) - \mathrm{tr}(t^c t^b t^a)\right], \tag{3.22}$$

diagrammatically

$f_{abc} =$ 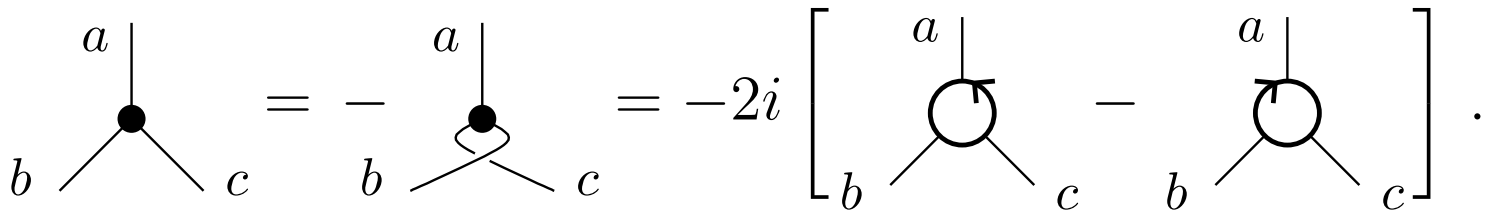 (3.23)

The thicker oriented lines stand for quark color representations, the unorientated lines correspond to gluon color lines. The $f_{abc}$ are obviously invariant under cyclic permutations of the indices. Normalization of generators is such that for the Killing form

$\mathrm{tr}(t^a t^b) = \frac{1}{2}\delta_{ab}$ or 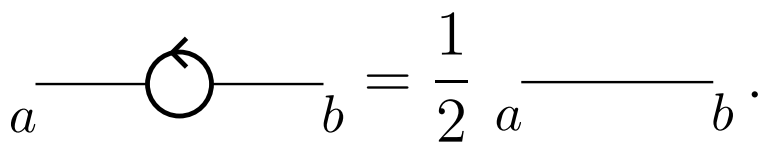 (3.24)

Using birdtrack notation the algebra (3.21) becomes

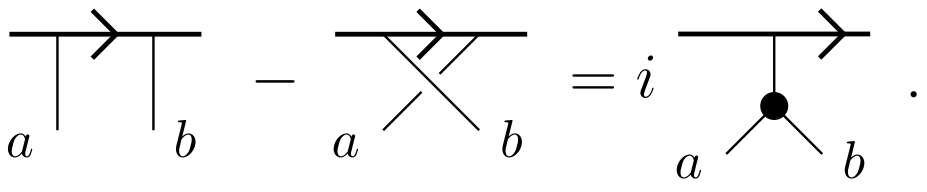 (3.25)

The anticommutator of two generators is

$$\{t^a, t^b\} = \frac{1}{N_c}\delta_{ab} + d_{abc}t^c\,, \tag{3.26}$$

and the symmetric structure constants $d_{abc}$ are expressed in terms of generators as

$$d_{abc} = d_{acb} = 2\left[\mathrm{tr}(t^a t^b t^c) + \mathrm{tr}(t^c t^b t^a)\right], \tag{3.27}$$

in diagrams

$d_{abc} =$ 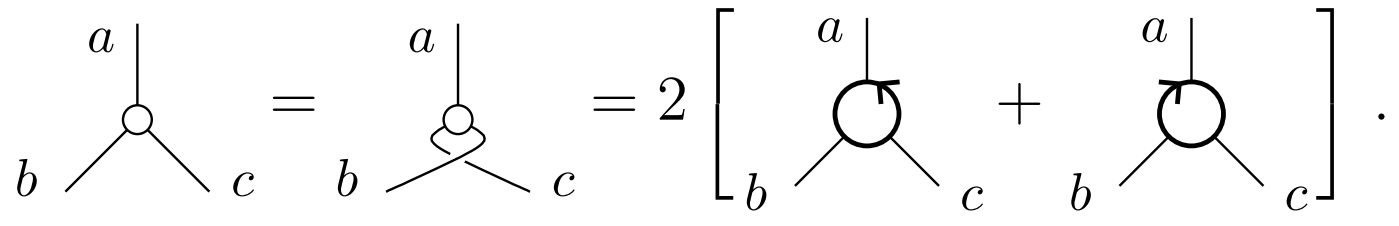 (3.28)

With this we have collected all basic elements of the birdtrack notation. We will slightly extend the birdtrack notation by the definition

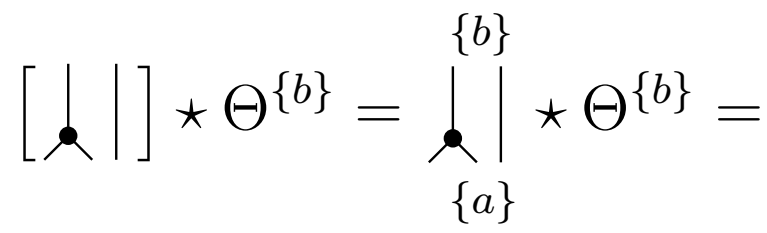 $= f_{a_1a_2b_1}\delta_{a_3b_2}\Theta^{b_1b_2}$ (3.29)

for the contraction of the set of color labels $\{b\}$ of an arbitrary color tensor $\Theta^{\{b\}}$. The symbol $\star$ stands for the contraction of the set $\{b\}$ in color space. The extension to more than two elements in the set $\{b\}$ is straightforward.

Generalizing (3.22) and (3.27) we define further tensors that are built from traces of generators in the following way[3]

$$d^{b_1 b_2 \dots b_n} = \mathrm{tr}(t^{b_1} t^{b_2} \dots t^{b_n}) + \mathrm{tr}(t^{b_n} \dots t^{b_2} t^{b_1}) \tag{3.30}$$

$$f^{b_1 b_2 \dots b_n} = \frac{1}{i} \left[ \mathrm{tr}(t^{b_1} t^{b_2} \dots t^{b_n}) - \mathrm{tr}(t^{b_n} \dots t^{b_2} t^{b_1}) \right] . \tag{3.31}$$

The definitions are valid for any $n \in \mathbb{N}$. For $n = 2, 3$, however, the tensors arising from (3.30), (3.31) are proportional[4] to $\delta_{b_1 b_2}$, $f_{b_1 b_2 b_3}$, and $d_{b_1 b_2 b_3}$ respectively. For the cases $n = 2, 3$ we will stick to the conventional definitions of structure constants given earlier in (3.22), (3.27).

The following three tensors are special cases of (3.30), (3.31). We will make extensive use of them throughout this paper. The tensor

$d^{abcd} =$ 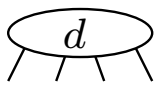 $= \mathrm{tr}(t^a t^b t^c t^d) + \mathrm{tr}(t^d t^c t^b t^a) =$ 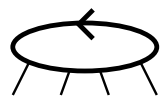 $+$ 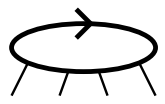 (3.33)

was used already in [19] in the investigation of the four–gluon amplitude. We now add to this the two tensors

$f^{abcde} =$ 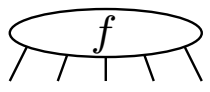 $= \frac{1}{i} \left[ \mathrm{tr}(t^a t^b t^c t^d t^e) - \mathrm{tr}(t^e t^d t^c t^b t^a) \right] = \frac{1}{i} \Big[$ 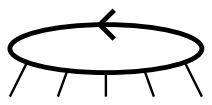 $-$ 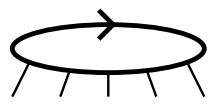 $\Big]$ (3.34)

and

$$d^{abcdef} = \mathrm{tr}(t^a t^b t^c t^d t^e t^f) + \mathrm{tr}(t^f t^e t^d t^c t^b t^a) . \tag{3.35}$$

All three are invariant under cyclic permutations of the color labels (as are obviously all tensors defined according to (3.30), (3.31)). It will turn out that these color tensors are very well suited for the whole problem of solving the integral equations. When interpreting the results in terms of reggeon color representations, the decomposition of these tensors into the lower order tensors $f_{abc}$, $d_{abc}$ and $\delta_{ab}$ is also useful:

$$d^{abcd} = \frac{1}{2N_c} \delta_{ab} \delta_{cd} + \frac{1}{4} (d_{abk} d_{kcd} - f_{abk} f_{kcd}) \tag{3.36}$$

$= \frac{1}{2N_c} \cap\cap + \frac{1}{4} \Big($ 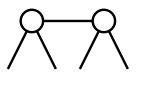 $-$ 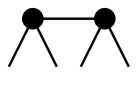 $\Big)$ , (3.37)

as is easily derived using (3.21) and (3.26). From this we get by cyclic permutation

$$d^{abcd} = \frac{1}{2N_c} \delta_{ad} \delta_{bc} + \frac{1}{4} (d_{adk} d_{kbc} + f_{adk} f_{kbc}) . \tag{3.38}$$

[3]In appendix A we will for brevity refer to tensors built from traces over generators in this way as 'standard tensors'. To the best of our knowledge this term does not carry a fixed meaning in the literature on Lie algebras.

[4]In detail we have according to (3.24), (3.27), and (3.22)

$$d^{ab} = \delta_{ab} \, ; \qquad d^{abc} = \frac{1}{2} d_{abc} \, ; \qquad f^{abc} = \frac{1}{2} f_{abc} \, . \tag{3.32}$$

We further mention the property $d^{bacd} = d^{abdc}$, which turns out to be useful for calculational purposes. For $f^{abcde}$ we have

$$f^{abcde} = \frac{1}{4N_c}(\delta_{ab}f_{cde} + f_{abc}\delta_{de}) + \frac{1}{8}(f_{abk}d_{kcl}d_{lde} + d_{abk}f_{kcl}d_{lde} + d_{abk}d_{kcl}f_{lde} - f_{abk}f_{kcl}f_{lde}) \qquad (3.39)$$

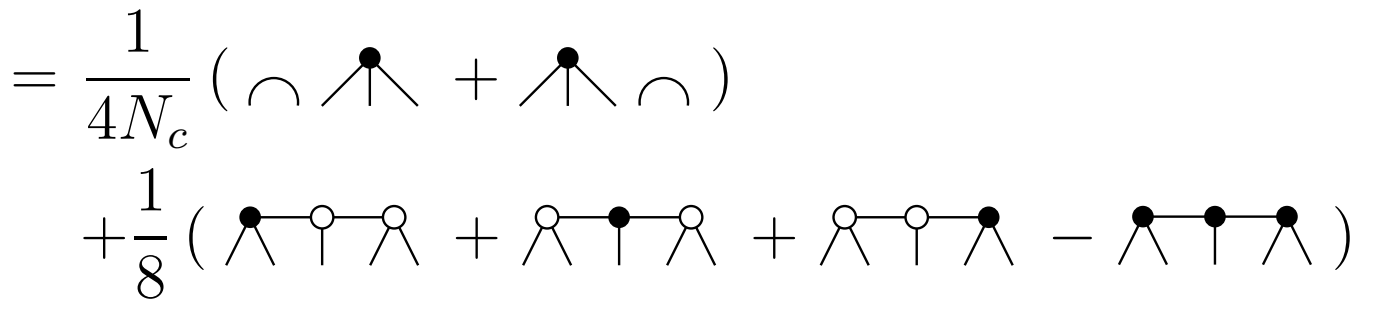

(3.40)

and further identities can be obtained from (3.39) by making use of the invariance under cyclic permutations. The tensor $d^{abcdef}$ can be decomposed in a similar way,

$$\begin{aligned} d^{abcdef} = & \frac{1}{4N_c^2}\,\delta_{ab}\delta_{cd}\delta_{ef} \\ & + \frac{1}{8N_c}\,(\delta_{ab}d_{cdk}d_{kef} - \delta_{ab}f_{cdk}f_{kef} + d_{abk}d_{kcd}\delta_{ef} - f_{abk}f_{kcd}\delta_{ef}) \\ & + \frac{1}{16}\,(d_{abk}d_{kcl}d_{ldm}d_{mef} - d_{abk}d_{kcl}f_{ldm}f_{mef} - d_{abk}f_{kcl}d_{ldm}f_{mef} \\ & - d_{abk}f_{kcl}f_{ldm}d_{mef} - f_{abk}d_{kcl}d_{ldm}f_{mef} - f_{abk}d_{kcl}f_{ldm}d_{mef} \\ & - f_{abk}f_{kcl}d_{ldm}d_{mef} + f_{abk}f_{kcl}f_{ldm}f_{mef})\,. \end{aligned} \qquad (3.41)$$

In birdtracks it becomes

$d^{abcdef} =$ 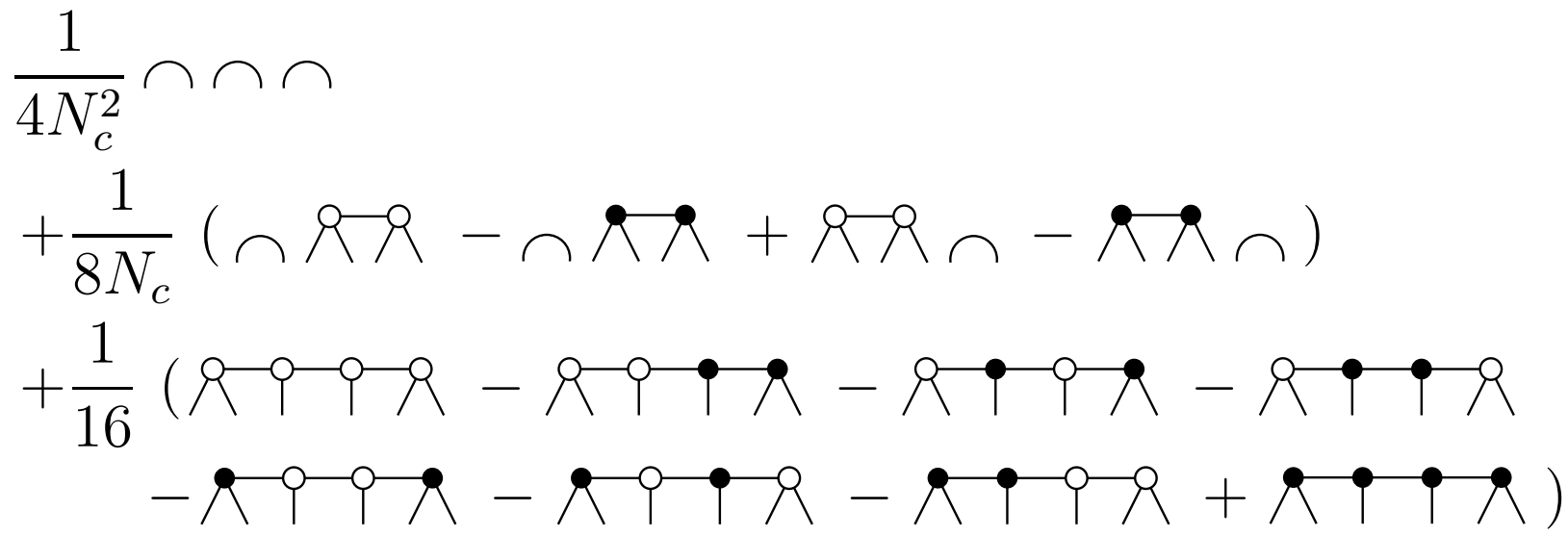

(3.42)

and again other possible decompositions of $d^{abcdef}$ are obtained from this by cyclic permutations.

To conclude this section about color tensors, we mention the well–known Jacobi identity

$$f_{abk}f_{kcd} + f_{ack}f_{bkd} + f_{adk}f_{kbc} = 0\,. \qquad (3.43)$$

### 3.4 The quark loop

Let us now consider the inhomogeneous terms in the integral equations (3.10)–(3.14). The terms $D^{a_1 \ldots a_n}_{(n;0)}$ describe the lowest order coupling of $n$ gluons to the quark loop. $D^{a_1 a_2}_{(2;0)}$ is the sum of four cut diagrams

$D^{a_1 a_2}_{(2;0)}(\mathbf{k}_1, \mathbf{k}_2) =$ 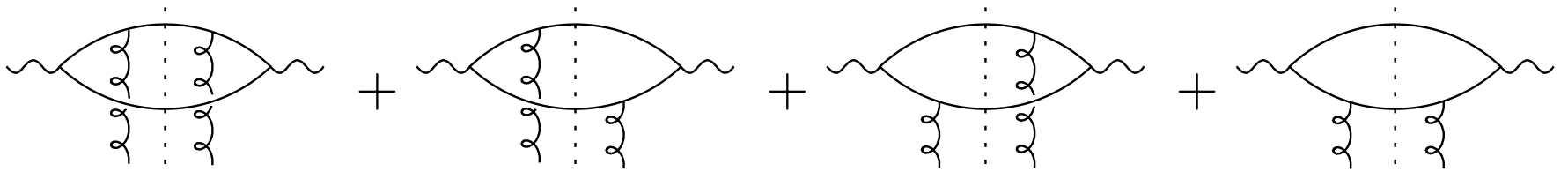 .

(3.44)

The two gluons are attached to the quark loop in all possible ways to preserve gauge–invariance. The $s$-channel cut (indicated in each diagram by the dotted vertical line) implies that the cut quark lines are set on–shell. $D_{(2;0)}^{a_1a_2}$ depends on the transverse momenta of the two gluons and on their color indices. The latter dependence is of course trivial and we can define the momentum part $D_{(2;0)}(\mathbf{k}_1, \mathbf{k}_2)$ of this amplitude by

$$D_{(2;0)}^{a_1a_2}(\mathbf{k}_1, \mathbf{k}_2) = \delta_{a_1a_2} D_{(2;0)}(\mathbf{k}_1, \mathbf{k}_2) \,. \tag{3.45}$$

An analytic expression for $D_{(2;0)}(\mathbf{k}_1, \mathbf{k}_2)$ was first found in [60]. The explicit formula for $D_{(2;0)}$ will not be needed in the following. We will only make use of the fact that $D_{(2;0)}$ is symmetric under the exchange of its transverse momentum arguments

$$D_{(2;0)}(\mathbf{k}_1, \mathbf{k}_2) = D_{(2;0)}(\mathbf{k}_2, \mathbf{k}_1) \tag{3.46}$$

and vanishes if one of its two arguments vanishes,

$$D_{(2;0)}(\mathbf{k}_1, \mathbf{k}_2)\Big|_{\mathbf{k}_1=0} = D_{(2;0)}(\mathbf{k}_1, \mathbf{k}_2)\Big|_{\mathbf{k}_2=0} = 0 \,. \tag{3.47}$$

In addition, it will be important for the consistency of the integral equations that

$$D_{(2;0)}(\mathbf{k}_1, \mathbf{k}_2) < \text{const.} \log \mathbf{k}_i^2 \tag{3.48}$$

in the ultraviolet region, i. e. the growth with the momenta is not stronger than logarithmic.

For the amplitudes $D_{(n;0)}$ with $n > 2$ we again have to attach the $n$ gluons to the quark loop in all possible ways to preserve gauge invariance. But the fact that we are dealing with multiply–cut amplitudes reduces the number of diagrams to consider. The cuts forbid the crossing of $t$-channel gluons as indicated in Fig. 1. The ordering of the gluons along the loop is therefore fixed up to the possibility of coupling the gluons to the quark or the antiquark. We thus have $2^n$ cut diagrams for the inhomogeneous term $D_{(n;0)}$.

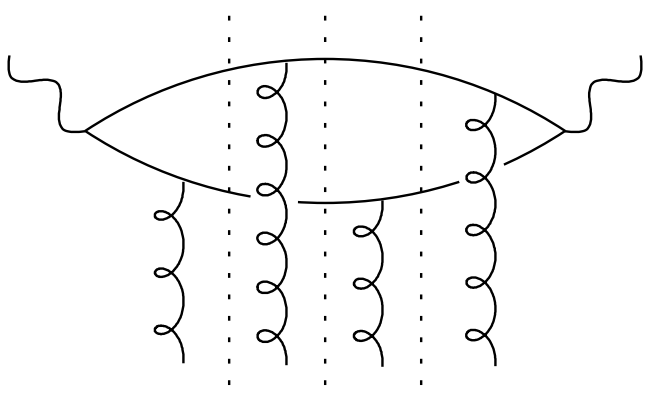

**Figure 1:** Cut amplitude contributing to the coupling of $n = 4$ gluons to a quark loop

It turns out that all the amplitudes $D_{(n;0)}^{a_1\dots a_n}$ can be expressed in terms of $D_{(2;0)}$, the momentum part of $D_{(2;0)}^{a_1a_2}$ as defined in (3.45). To see how this reduction mechanism works let us have a look at two neighbouring gluons along the quark loop. (They are not necessarily neighbouring as arguments of the amplitude, c. f. gluons 1 and 3 in Fig. 1.) In the high energy limit the quark–gluon vertices have to be contracted with a longitudinal momentum $p$. The Dirac trace over the quark loop then contains

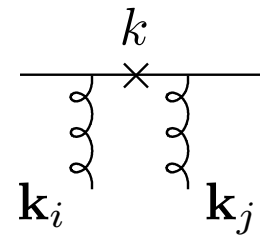

$$\sim \; \mathrm{tr}(\ldots \not{p} \not{k} \not{p} \ldots)\, \delta(k^2)\,. \tag{3.49}$$

The $\delta(k^2)$ comes in since the quark has to be set on–shell. Using a Sudakov decomposition $k = \alpha q' + \beta p + k_t$ with $q' = q + xp$, $q'^2 = p^2 = 0$, $2p \cdot q = s$, and $k_t^2 = -\mathbf{k}^2$ one finds for this expression

$$\mathrm{tr}(\ldots \not{p} \ldots)\, \delta(\beta - \mathbf{k}^2/(\alpha s)) \;\simeq \tag{3.50}$$

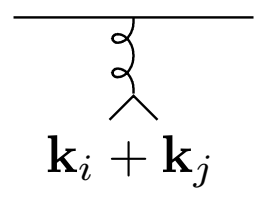

This means that due to energy–momentum conservation only the sum $\mathbf{k}_i + \mathbf{k}_j$ of the two momenta enters. We can apply this to all gluons along the quark loop and thereby reduce the momentum part of each diagram to one corresponding to a diagram in which only two gluons are coupled to the quark loop. The color structure is not affected by this reduction. A remark is in order concerning the contribution in which all gluons are coupled to the quark line or the antiquark line. This term acts as a regularization term. As we will see below it can be added and subtracted in such a way that the full $D_{(n;0)}^{a_1 \ldots a_n}$ can be expressed in terms of $D_{(2;0)}$.

Let us now see how the color structure of the quark loop amplitudes builds up. Each diagram contributing to $D_{(n;0)}$ contains a trace over $n$ su$(N_c)$ generators. The $2^n$ diagrams come in pairs in the following sense. Consider a diagram with $k$ gluons coupled to the quark and $n-k$ gluons coupled to the antiquark. Then there is also a diagram with the $k$ gluons coupled to the antiquark instead of the quark and the other $n-k$ gluons now coupled to the quark. The momentum structure of the two diagrams is the same up to a factor $(-1)^n$. (This is because the coupling of a gluon to a quark effectively differs from that to an antiquark by a sign.) The color part of the second diagram is again a trace over generators $t^{a_i}$, but in the trace they now appear in reversed order compared to the first diagram. Adding the two diagrams one thus finds a color tensor of the kind $d^{a_1 \ldots d_n}$ for an even number $n$ of gluons and a tensor of the kind $f^{a_1 \ldots d_n}$ for an odd number $n$ of gluons (cf. (3.30), (3.31) for the definition of the $d$- and $f$-tensors). There are $2^{n-1}$ pairs of such diagrams. Having in mind that due to the photons at the two ends of the loop the color tensor is not altered if the first or $n$th gluon is coupled to the quark instead of the antiquark and vice versa, we conclude that the number of different color tensors contributing to the coupling of $n$ gluons to the quark loop is in general $2^{n-3}$ if $n \geq 3$. The color structures for two and three gluons attached to the quark loop are more or less trivial: in both cases there is only one color tensor ($\delta_{a_1 a_2}$ and $f_{a_1 a_2 a_3}$, respectively).

We have seen that the diagrams contributing to the quark loop come in $2^{n-1}$ pairs. Among them is a special pair, namely the one consisting of the two regularization terms mentioned earlier, in which all gluons couple to the quark line or the antiquark line. This pair can be added and subtracted with different color structures in such a way that the remaining $2^{n-1}-1$ pairs are regularized to give a $D_{(2;0)}$ each. Therefore the quark loop $D_{(n;0)}^{a_1\ldots a_n}$ can be expressed as a sum of $2^{n-1}-1$ amplitudes $D_{(2;0)}$ .

In this paper, we will need the expressions for the quark loop with up to six gluons attached. For three gluons coupled to the quark loop we find

$$D_{(3;0)}^{a_1a_2a_3}(\mathbf{k}_1,\mathbf{k}_2,\mathbf{k}_3) = \frac{1}{2} g f_{a_1a_2a_3} \left[D_{(2;0)}(12,3) - D_{(2;0)}(13,2) + D_{(2;0)}(1,23)\right], \quad (3.51)$$

where we use the notation introduced in (3.9). In the case of four gluons the amplitude contains two different color structures,

$$\begin{aligned} D_{(4;0)}^{a_1a_2a_3a_4}(\mathbf{k}_1,\mathbf{k}_2,\mathbf{k}_3,\mathbf{k}_4) = & -g^2 d^{a_1a_2a_3a_4}\left[D_{(2;0)}(123,4) + D_{(2;0)}(1,234) - D_{(2;0)}(14,23)\right] \\ & -g^2 d^{a_2a_1a_3a_4}\left[D_{(2;0)}(134,2) + D_{(2;0)}(124,3) - D_{(2;0)}(12,34)\right. \\ & \left. -D_{(2;0)}(13,24)\right]. \end{aligned} \quad (3.52)$$

When five gluons are coupled to the quark loop there appear four different color structures in the corresponding amplitude,

$$\begin{aligned} & D_{(5;0)}^{a_1a_2a_3a_4a_5}(\mathbf{k}_1,\mathbf{k}_2,\mathbf{k}_3,\mathbf{k}_4,\mathbf{k}_5) = \quad (3.53) \\ & = -g^3\{f^{a_1a_2a_3a_4a_5}\left[D_{(2;0)}(1234,5) + D_{(2;0)}(1,2345) - D_{(2;0)}(15,234)\right] \\ & \quad + f^{a_2a_1a_3a_4a_5}\left[D_{(2;0)}(1345,2) - D_{(2;0)}(12,345) + D_{(2;0)}(125,34) - D_{(2;0)}(134,25)\right] \\ & \quad + f^{a_1a_2a_3a_5a_4}\left[D_{(2;0)}(1235,4) - D_{(2;0)}(14,235) + D_{(2;0)}(145,23) - D_{(2;0)}(123,45)\right] \\ & \quad + f^{a_1a_2a_4a_5a_3}\left[D_{(2;0)}(1245,3) - D_{(2;0)}(13,245) + D_{(2;0)}(135,24) - D_{(2;0)}(124,35)\right]\}. \end{aligned}$$

For six gluons attached to the quark loop we find the following result. Now eight different color structures contribute,

$$\begin{aligned} & D_{(6;0)}^{a_1a_2a_3a_4a_5a_6}(\mathbf{k}_1,\mathbf{k}_2,\mathbf{k}_3,\mathbf{k}_4,\mathbf{k}_5,\mathbf{k}_6) = \\ & = g^4\{d^{a_1a_2a_3a_4a_5a_6}\left[D_{(2;0)}(12345,6) + D_{(2;0)}(1,23456) - D_{(2;0)}(16,2345)\right] \\ & \quad + d^{a_2a_1a_3a_4a_5a_6}\left[D_{(2;0)}(13456,2) - D_{(2;0)}(1345,26) + D_{(2;0)}(126,345)\right. \\ & \qquad \left. -D_{(2;0)}(12,3456)\right] \\ & \quad + d^{a_1a_2a_3a_4a_6a_5}\left[D_{(2;0)}(12346,5) - D_{(2;0)}(1234,56) + D_{(2;0)}(156,234)\right. \\ & \qquad \left. -D_{(2;0)}(15,2346)\right] \\ & \quad + d^{a_2a_1a_3a_4a_6a_5}\left[-D_{(2;0)}(1256,34) - D_{(2;0)}(1346,25) + D_{(2;0)}(125,346)\right. \\ & \qquad \left. +D_{(2;0)}(134,256)\right] \\ & \quad + d^{a_3a_1a_2a_4a_5a_6}\left[D_{(2;0)}(12456,3) - D_{(2;0)}(1245,36) + D_{(2;0)}(136,245) + \right. \end{aligned}$$

$$
\begin{aligned}
&\qquad\qquad\qquad -D_{(2;0)}(13,2456)] \\
&+ d^{a_1a_2a_3a_5a_6a_4}\,[D_{(2;0)}(12356,4) - D_{(2;0)}(1235,46) + D_{(2;0)}(146,235) \\
&\qquad\qquad\qquad -D_{(2;0)}(14,2356)] \\
&+ d^{a_2a_1a_3a_5a_6a_4}\,[-D_{(2;0)}(1246,35) - D_{(2;0)}(1356,24) + D_{(2;0)}(124,356) \\
&\qquad\qquad\qquad +D_{(2;0)}(135,246)] \\
&+ d^{a_1a_2a_3a_6a_5a_4}\,[-D_{(2;0)}(1236,45) - D_{(2;0)}(1456,23) + D_{(2;0)}(123,456) \\
&\qquad\qquad\qquad +D_{(2;0)}(145,236)]\}\,. \qquad (3.54)
\end{aligned}
$$

### 3.5 Reggeon momentum diagrams

We now introduce a further diagrammatic notation for the momentum space integrals occurring in our analysis of the integral equations. It will be applied in the next section where we will present the integral kernels $K_{2\to m}^{\{b\}\to\{a\}}$. With the help of so–called reggeon momentum diagrams we hope to make our results more transparent and easier to read. A reggeon momentum integral looks like the following example:

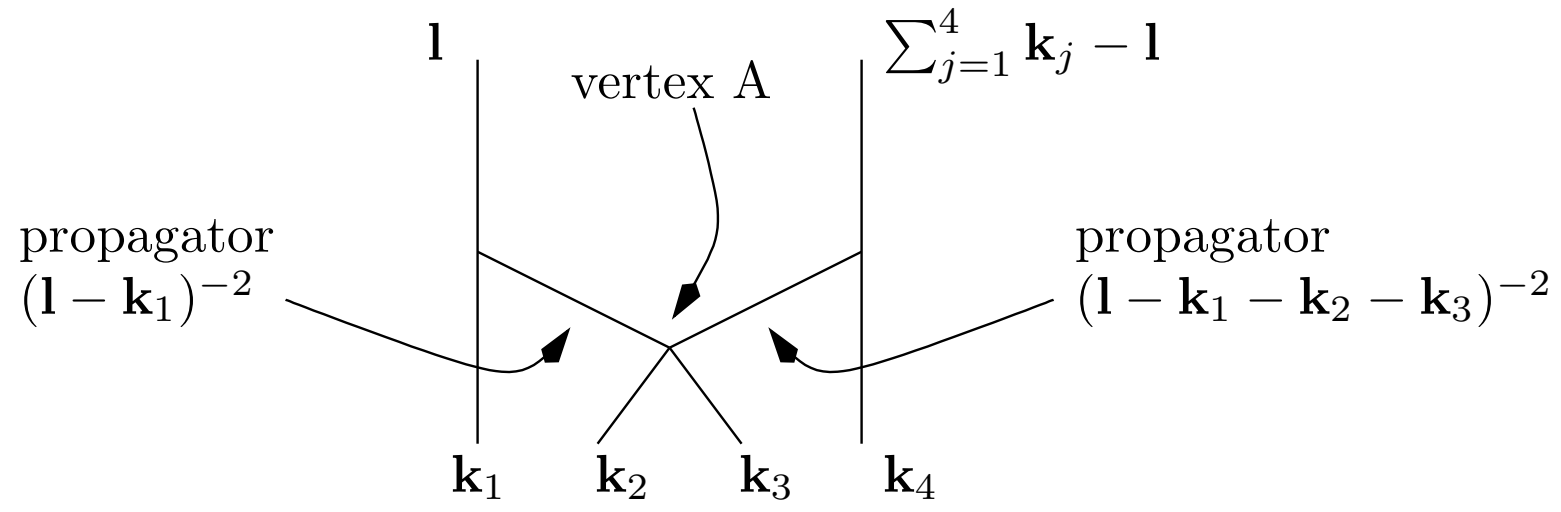

(3.55)

We now state the rules for translating these diagrams back to explicit integrals. At the same time we apply the rules step by step to the above example.

(i) Assign momentum variables to all lines according to momentum conservation (see diagram).

(ii) Write down an integral $\int \frac{d^2\mathbf{l}}{(2\pi)^3}$ over the loop momentum. (There is always a loop since some amplitude, for instance $D_2$, has to be attached to the upper two lines and thus be written under the integral.)

(iii) Find in the diagram the vertex which has two lines attached from above (vertex A in (3.55)).

(iv) Write down the square of the sum of the momenta attached to this vertex from below. (In our example $(\mathbf{k}_2+\mathbf{k}_3)^2$.)

(v) Write down propagators for the two lines attached to this vertex from above. (In our example $(\mathbf{l}-\mathbf{k}_1)^{-2}(\mathbf{l}-\mathbf{k}_1-\mathbf{k}_2-\mathbf{k}_3)^{-2}$.)

For the diagram (3.55) this results in

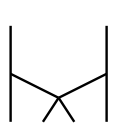

$$
= \int \frac{d^2\mathbf{l}}{(2\pi)^3} \frac{(\mathbf{k}_2+\mathbf{k}_3)^2}{(\mathbf{l}-\mathbf{k}_1)^2(\mathbf{l}-\mathbf{k}_1-\mathbf{k}_2-\mathbf{k}_3)^2}\,. \qquad (3.56)
$$

These rules can easily be inverted in order to construct the reggeon momentum diagram from a given momentum space integral. The reggeon momentum diagrams have to be understood as integral operators. The integration has to be carried out with a function of the two upper momenta. We emphasize that our notation implies only two propagators for a given reggeon momentum diagram.

A few more examples of the diagrammatic notation for momentum space integrals are contained in section 3.7.

### 3.6 Integral kernels

The integral kernels $K_{2\to m}^{\{b\}\to\{a\}}$ were calculated in [15] by means of $s$-channel dispersion relations. As explained in section 3.2 the kernels are convoluted with different amplitudes in the integral equations. Only two of the gluons in the respective amplitude actually interact with each other. The kernel acts trivially on the momenta and color labels of the other gluons. We will therefore discuss only the non–trivial action of the kernel here.

The kernel for the transition from two gluons with transverse momenta $\mathbf{q}_1, \mathbf{q}_2$ and color labels $\{b\} = \{b_1, b_2\}$ to $m$ gluons with transverse momenta $\mathbf{k}_1, \ldots, \mathbf{k}_m$ and color labels $\{a\} = \{a_1, \ldots, a_m\}$ is given by

$$K_{2\to m}^{\{b\}\to\{a\}}(\mathbf{q}_1, \mathbf{q}_2; \mathbf{k}_1, \ldots, \mathbf{k}_m) =$$

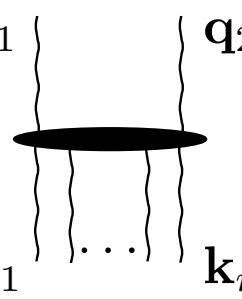

$$= f_{b_1a_1c_1} f_{c_1a_2c_2} \ldots f_{c_{m-1}a_mb_2}\, g^m \left[ (\mathbf{k}_1 + \ldots + \mathbf{k}_m)^2 - \frac{\mathbf{q}_2^2(\mathbf{k}_1 + \ldots + \mathbf{k}_{m-1})^2}{(\mathbf{k}_m - \mathbf{q}_2)^2} - \frac{\mathbf{q}_1^2(\mathbf{k}_2 + \ldots + \mathbf{k}_m)^2}{(\mathbf{k}_1 - \mathbf{q}_1)^2} + \frac{\mathbf{q}_1^2\mathbf{q}_2^2(\mathbf{k}_2 + \ldots + \mathbf{k}_{m-1})^2}{(\mathbf{k}_1 - \mathbf{q}_1)^2(\mathbf{k}_m - \mathbf{q}_2)^2} \right] . \quad (3.57)$$

For the kernels that are needed for up to six gluons in the $t$-channel this means in our diagrammatic notation for color tensors and momentum integral kernels

$K_{2\to 2}^{\{b\}\to\{a\}}(\mathbf{q}_1, \mathbf{q}_2; \{\mathbf{k}_i\}) =$ 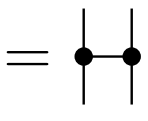 $g^2\mathbf{q}_1^2\mathbf{q}_2^2$ 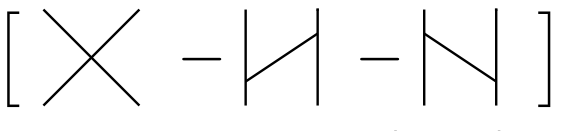 (3.58)

$K_{2\to 3}^{\{b\}\to\{a\}}(\mathbf{q}_1, \mathbf{q}_2; \{\mathbf{k}_i\}) =$ 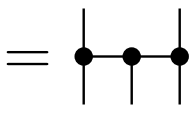 $g^3\mathbf{q}_1^2\mathbf{q}_2^2$ 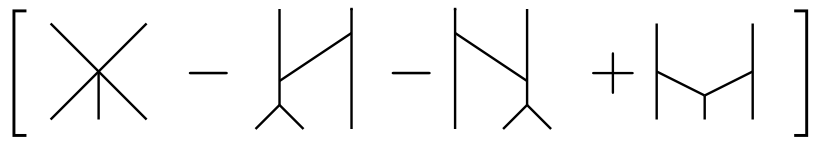 (3.59)

$K_{2\to 4}^{\{b\}\to\{a\}}(\mathbf{q}_1, \mathbf{q}_2; \{\mathbf{k}_i\}) =$ 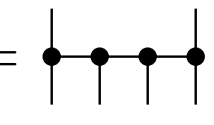 $g^4\mathbf{q}_1^2\mathbf{q}_2^2$ 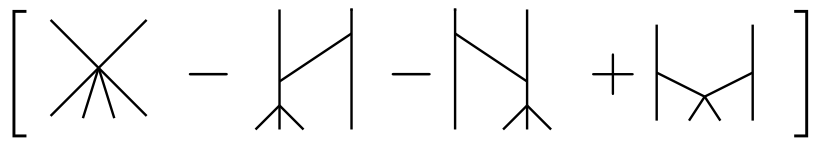 (3.60)

$K_{2\to 5}^{\{b\}\to\{a\}}(\mathbf{q}_1, \mathbf{q}_2; \{\mathbf{k}_i\}) =$ 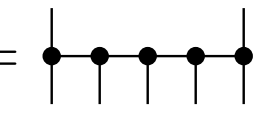 $g^5\mathbf{q}_1^2\mathbf{q}_2^2$ 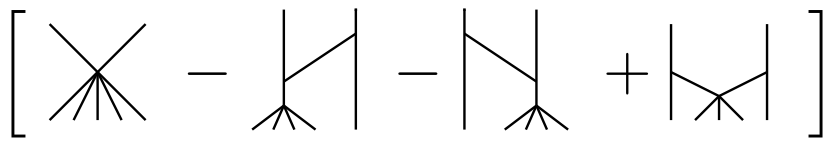 (3.61)

$K_{2\to 6}^{\{b\}\to\{a\}}(\mathbf{q}_1, \mathbf{q}_2; \{\mathbf{k}_i\}) =$ 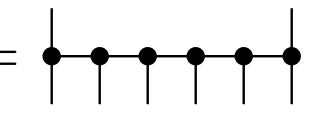 $g^6\mathbf{q}_1^2\mathbf{q}_2^2$ 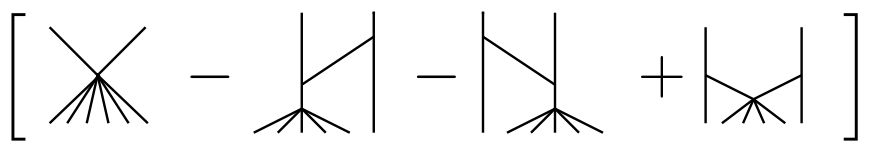 (3.62)

Here the inverse propagators $\mathbf{q}_1^2\mathbf{q}_2^2$ are required in order to cancel the propagators implied by our graphical notation. This is necessary because the kernels $K_{2\to m}^{\{b\}\to\{a\}}$ in the integral equations are defined without propagators; there the propagators come in via the convolution denoted by the symbol $\otimes$.

We emphasize that the kernel $K_{2\to 2}^{\{b\}\to\{a\}}$ is not identical to the full BFKL kernel (2.6) as it does not contain the trajectory functions $\beta$. In equation (3.10) they have been moved to the left–hand side to make the generalization of this equation to $n > 2$ more transparent. Also the kernels $K_{2\to n}^{\{b\}\to\{a\}}$ for $n > 2$ do not contain any trajectories and are not infrared finite. It is only in the full integral equations that the infrared divergences cancel.

### 3.7 Standard integrals

When convoluting the amplitudes $D_n$ with the kernels $K_{2\to m}$ according to the integral equations, we have to deal with a large number of momentum space integrals. It turns out that all these integrals can be classified by a small number of standard integrals. These five standard types of integrals are even sufficient for addressing the case of $n$-gluon amplitudes for any value of $n$. (Strictly speaking this applies to the so–called reggeizing pieces of the amplitudes as will be explained in the next section.) We will therefore give a complete list of the five different standard integrals (or diagrams) which at the same time serve as a few more examples of our notation.

All five standard integrals occur already when dealing with three outgoing gluons. We therefore give the list for $n = 3$ here, the generalization to larger $n$ being obvious. First we have

$a(2,1) =$ 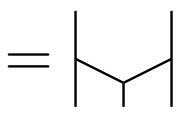 $$= \int \frac{d^2\mathbf{l}}{(2\pi)^3} \frac{\mathbf{k}_2^2}{(\mathbf{l}-\mathbf{k}_1)^2[\mathbf{l}-(\mathbf{k}_1+\mathbf{k}_2)]^2} D_2\left(\mathbf{l}, \sum_{j=1}^{3}\mathbf{k}_j - \mathbf{l}\right) . \qquad (3.63)$$

The first argument of the function $a$ is the index of the momentum attached to the vertex A from below. (As in section 3.5 vertex A is the vertex in the diagram with two lines attached from above.) The momentum structure of the diagram is then fixed by giving one of the other two outgoing momenta since the diagram has to be folded with a symmetric function of the upper two momenta, namely with the BFKL amplitude $D_2$. We choose the outgoing momentum with the lowest index as the second argument of the function $a$ thereby completely fixing the corresponding momentum integral since the momentum carried by the third outgoing gluon line can easily be inferred from the total number of gluons. Therefore our notation for $a$ has to be supplied with the total number of $t$-channel gluons it describes. (In the case of more than three gluons we here choose the group of momenta containing the lowest index as the second argument of $a$.) Applying this notation, our earlier example given in (3.55) would be assigned the expression $a(23,1)$ when applied to $D_2$. In section 3.5 the reggeon momentum diagrams stood for integral operators. In this and in the following sections we use the same diagrams also for the convolution with BFKL amplitudes. The meaning should be clear from the context. The function $a$ and the functions $b$, $c$, $s$, $t$ to be defined below always mean the integrals with $D_2$ included. The second type of diagram is

$b(12) =$ 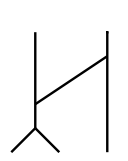 $= \int \frac{d^2\mathbf{l}}{(2\pi)^3} \frac{(\mathbf{k}_1+\mathbf{k}_2)^2}{\mathbf{l}^2[\mathbf{l}-(\mathbf{k}_1+\mathbf{k}_2)]^2} D_2\left(\mathbf{l}, \sum_{j=1}^{3} \mathbf{k}_j - \mathbf{l}\right) .$ (3.64)

The third one is the contact term

$c(123) =$ 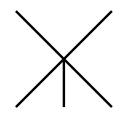 $= \int \frac{d^2\mathbf{l}}{(2\pi)^3} \frac{(\mathbf{k}_1+\mathbf{k}_2+\mathbf{k}_3)^2}{\mathbf{l}^2\left(\mathbf{l}-\sum_{j=1}^{3}\mathbf{k}_j\right)^2} D_2\left(\mathbf{l}, \sum_{j=1}^{3} \mathbf{k}_j - \mathbf{l}\right) ,$ (3.65)

which is local in impact parameter space. The integrals $a$, $b$, and $c$ correspond to real corrections, that is to $s$-channel gluon production.

Further we have two integral types corresponding to virtual corrections. These are factorizing and are connected with what we already know as the trajectory function $\beta$ of the reggeized gluon:

$t(12) =$ 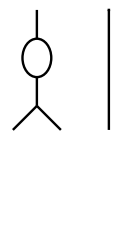 $= \int \frac{d^2\mathbf{l}}{(2\pi)^3} \frac{(\mathbf{k}_1+\mathbf{k}_2)^2}{\mathbf{l}^2[\mathbf{l}-(\mathbf{k}_1+\mathbf{k}_2)]^2} D_2(\mathbf{k}_1+\mathbf{k}_2, \mathbf{k}_3)$

$$= \frac{2}{N_c g^2} \beta(\mathbf{k}_1+\mathbf{k}_2) D_2(\mathbf{k}_1+\mathbf{k}_2, \mathbf{k}_3) \tag{3.66}$$

and

$s(1,2) =$ 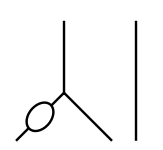 $= \int \frac{d^2\mathbf{l}}{(2\pi)^3} \frac{\mathbf{k}_1^2}{\mathbf{l}^2(\mathbf{l}-\mathbf{k}_1)^2} D_2(\mathbf{k}_1+\mathbf{k}_2, \mathbf{k}_3)$

$$= \frac{2}{N_c g^2} \beta(\mathbf{k}_1) D_2(\mathbf{k}_1+\mathbf{k}_2, \mathbf{k}_3) . \tag{3.67}$$

This means that the bubble diagram corresponds to a gluon trajectory function $\beta$ for the line it is drawn on.

It suggests itself here to write the BFKL kernel using the standard integrals introduced above. When the kernel is applied to the BFKL amplitude $D_2$ we find

$$K_{\mathrm{BFKL}} \otimes D_2 = N_c g^2 \left[ c(12) - b(1) - b(2) + \frac{1}{2} t(1) + \frac{1}{2} t(2) \right] \tag{3.68}$$

$= N_c g^2 \Big[$ 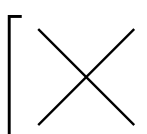 $-$ 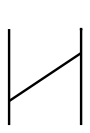 $-$ 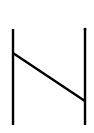 $+ \frac{1}{2}$ 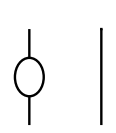 $+ \frac{1}{2}$ 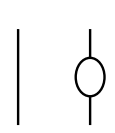 $\Big] .$ (3.69)

The notation introduced here makes implicit use of the fact that the BFKL amplitude $D_2$ is symmetric in its two momentum arguments. Without this assumption it would, for example, be necessary to give explicitly a third argument to fully specify the integral $a$ in (3.63). We have chosen to restrict the short–hand notation to the case of a symmetric function since we will only apply it to the case of $D_2$.

As an example of how the convolutions of $n$-gluon amplitudes with kernels can be reduced to the standard integrals may serve

$$D_2(\mathbf{l}_1+\mathbf{l}_2, \mathbf{l}_3) \otimes K_{2\to 3}(\mathbf{l}_2, \mathbf{l}_3; \mathbf{k}_2, \mathbf{k}_3, \mathbf{k}_4) = g^3 [b(234) - a(23,1) - b(34) + a(3,12)] . \tag{3.70}$$

This can be easily checked using the explicit definition (3.59) or (3.57) of the kernel. When studying the $n$-gluon amplitudes $D_n$ we encounter a problem connected with these convolutions in momentum space. For each convolution it is relatively easy to find a representation in terms of the standard integrals, as the above example shows. The actual problem is the rapidly increasing number of convolutions we have to deal with when coming to larger $n$. We will explain this problem in more detail in the following sections. For $n = 5$ gluons the problem is at the edge of being tractable by hand. For $n = 6$ gluons the problem has to be attacked with the help of a computer. It is exactly for this reason that we introduce the classification of momentum space integrals in this section. In appendix B we give an algorithm suited for implementation on a computer, for instance in the PERL script language. The example above is intended to illustrate that the notation indeed allows us to handle the rather complicated integrals in compact form.

Closing this section, we remark that even the five standard types of integrals mentioned above are not completely independent if we take our definitions (3.63)–(3.67) literally. Relations between them occur in the case that one of the outgoing legs of the diagrams has zero momentum. One finds, for example, that the function $b$ emerges from $a$ when the two arguments of $a$ exhaust the outgoing gluons, that is for three gluons we would get

$$a(1, 23) = b(1)\,. \tag{3.71}$$

Similarly, for three outgoing gluons

$$b(123) = c(123)\,. \tag{3.72}$$

In addition, we find the function $t$ from $s$ when the second argument of $s$ vanishes, for example

$$s(1, -) = t(1)\,. \tag{3.73}$$

In spite of these relations we prefer to treat the five integral types as fundamentally different because they correspond to very different locality properties in impact parameter space. The function $c$, for instance, corresponds to a contact interaction whereas the function $a$ contains a non–locality.

This concludes our discussion of the elements of the integral equations and we can now proceed to solving them (at least partially).

# 4. Three and four gluons, the transition vertex $V_{2\to4}$

## 4.1 The three-gluon amplitude

The amplitudes with three and four gluons, $D_3$ and $D_4$, were first investigated in [18, 19]. It was found that the integral equation (3.11) for the three gluon amplitude can be solved, the solution being

$$D_3^{a_1a_2a_3}(\mathbf{k}_1,\mathbf{k}_2,\mathbf{k}_3) = \frac{1}{2} g f_{a_1a_2a_3} \left[ D_2(12,3) - D_2(13,2) + D_2(1,23) \right] , \tag{4.1}$$

which can be shown by direct computation. In addition to performing the color algebra contractions and the convolutions in momentum space one has to make use of the integral equation (3.10) for the two–gluon amplitude $D_2$.

The result (4.1) means that an actual three–gluon state does not appear.[5] In the contrary, the amplitude turns out to be a superposition of two–gluon states. We call this phenomenon the reggeization of the amplitude. It generalizes the notion of reggeization previously attributed to the fact that the BFKL equation in the color octet channel can be solved by a pole–ansatz and thus describes a one–reggeon state. In the case of the three–gluon amplitude reggeization again occurs in a channel corresponding to the adjoint representation, i. e. in the octet channel. We should emphasize that the reggeization of $D_3$ is a property of the momentum space part of the amplitude. The analytic properties correspond to those of a superposition of two–gluon compound states. Nevertheless, $D_3$ remains a three–gluon amplitude carrying three color labels, i. e. the color part of the amplitude is not affected.

It is worth noting that the solution (4.1) is obtained from the lowest order term (3.51) by replacing the quark loop amplitudes $D_{(2;0)}$ by the full BFKL amplitudes $D_2$ while keeping the color and momentum structure.

We do not give a proof of (4.1) here. In section 5 we will discover that the reggeization of the three–gluon amplitude $D_3$ is actually a special case of a more general identity that we will discuss in detail.

### 4.2 The four-gluon amplitude and the two-to-four transition vertex $V_{2\to4}$

Our method for analyzing the structure of the $n$-gluon amplitudes is the following. One starts with an educated guess for the solution or at least a part of it, assuming the full solution to be a sum of the part we have guessed and a remaining term. That ansatz is inserted into the integral equation and a new integral equation for the unknown part is derived. If the guess was good the new integral equation is in a certain sense simple, and allows one to gain further information about the unknown part. If the guess was not optimal, on the other hand, the new integral equation will be complicated and not allow us to extract further information. Of course, this procedure is not uniquely determined. As we will see, the quark loop contains very useful information that can be used for choosing a promising ansatz. Let us now see how this method works in practice.

In [18, 19] the four–gluon amplitude was split into two parts,

$$D_4 = D_4^R + D_4^I \, , \tag{4.2}$$

[5]This does not affect the existence of the Odderon, since in our case the three–gluon system has even $C$-parity.

a reggeizing part $D_4^R$ and a part $D_4^I$ that for reasons to be explained below is called the irreducible part of the four–gluon amplitude. The reggeizing part is — in analogy with the three–gluon case — chosen as the superposition of two–gluon amplitudes,

$$\begin{aligned} D_4^{R\,a_1a_2a_3a_4}(\mathbf{k}_1,\mathbf{k}_2,\mathbf{k}_3,\mathbf{k}_4) = \\ = -g^2 d^{a_1a_2a_3a_4}\left[D_2(123,4)+D_2(1,234)-D_2(14,23)\right] \\ -g^2 d^{a_2a_1a_3a_4}\left[D_2(134,2)+D_2(124,3)-D_2(12,34)-D_2(13,24)\right]. \end{aligned} \tag{4.3}$$

Again, this is obtained from the lowest order term $D_{(4;0)}$, see (3.52), by the replacement $D_{(2;0)} \to D_2$. The ansatz for the reggeizing part and thus the decomposition (4.2) is to some extend arbitrary. Recently, a different ansatz for the reggeizing part was investigated in [31, 32]. That ansatz, as also discussed in [19], is more convenient for the analysis of high mass diffractive processes. We will not further discuss other possible choices for the splitting (4.2) of the amplitude here. All choices will, of course, lead to equivalent results for the complete amplitude. The choice given above is singled out because it leads to a simple picture for the remaining part $D_4^I$, especially when interpreted in view of a field theory structure of unitarity corrections.

The next step is to derive a new integral equation for the unknown irreducible part $D_4^I$. To this end we insert the ansatz (4.2),(4.3) for the full amplitude into the original integral equation (3.12). The known result (4.1) for the three gluon amplitude is inserted as well. Due to the choice of $D_4^R$ we can then apply the equation (3.10) for the two–gluon amplitude to the expression $\omega D_4^R$ on the left hand side. That exactly eliminates the lowest order term $D_{(4;0)}$ on the right hand side and produces additional terms containing only the convolution of $D_2$ amplitudes with the kernel $K_{2\to2}^{\{b\}\to\{a\}}$ or products of $D_2$ with trajectory functions $\beta$. Besides the terms containing $D_4^I$ all other contributions to the right hand side consist of convolutions of two–gluon amplitudes $D_2$ with the integral kernels. Our new equation thus takes the form

$$\left(\omega-\sum_{i=1}^{4}\beta(\mathbf{k}_i)\right) D_4^{I\,a_1a_2a_3a_4}(\mathbf{k}_1,\mathbf{k}_2,\mathbf{k}_3,\mathbf{k}_4) = V_{2\to4}^{a_1a_2a_3a_4} D_2 + \sum K_{2\to2}^{\{b\}\to\{a\}} \otimes D_4^{I\,b_1b_2b_3b_4}. \tag{4.4}$$

The sum on the right–hand side of this new equation extends over all pairwise interactions of the four gluons. In the inhomogeneous term $V_{2\to4}D_2$ we collect all the terms containing $D_2$, hence the notation. $V_{2\to4}$ should be understood as an integral operator acting on the two–gluon amplitude. As we will explain in more detail below, it describes the transition from the two–gluon state to a fully interacting four–gluon system in the $t$-channel. The explicit expression for the two–to–four transition vertex was computed in [18, 19]. To arrive at this explicit result the following steps have to be done. First we have to contract the color tensors of the amplitudes with those of the kernels. This is done along the lines described in appendix A. The second step is to bring the momentum space integrals to their standard form, that is to classify them according to the scheme explained in section 3.7. Both steps result in lengthy calculations because of the large number of contractions involved. The results of the su($N_c$) tensor contractions are then multiplied with the corresponding momentum space integrals. Finally, all terms can be collected to give the vertex $V_{2\to4}$. Due to cancellations the resulting expression is comparatively compact. Remarkably, all terms belonging to the color tensors $d^{a_1a_2a_3a_4}$ and $d^{a_2a_1a_3a_4}$ drop out. One finds the following color and momentum structure for the vertex:

$$\begin{aligned} V_{2\to4}^{a_1a_2a_3a_4}(\{\mathbf{q}_j\},\mathbf{k}_1,\mathbf{k}_2,\mathbf{k}_3,\mathbf{k}_4) = {} & \delta_{a_1a_2}\delta_{a_3a_4}V(\{\mathbf{q}_j\},\mathbf{k}_1,\mathbf{k}_2;\mathbf{k}_3,\mathbf{k}_4) \\ & + \delta_{a_1a_3}\delta_{a_2a_4}V(\{\mathbf{q}_j\},\mathbf{k}_1,\mathbf{k}_3;\mathbf{k}_2,\mathbf{k}_4) \\ & + \delta_{a_1a_4}\delta_{a_2a_3}V(\{\mathbf{q}_j\},\mathbf{k}_1,\mathbf{k}_4;\mathbf{k}_2,\mathbf{k}_3)\,. \end{aligned} \tag{4.5}$$

The function $V$ is the same in all three terms on the right hand side. The $\mathbf{q}_j$ are the two momenta entering from above. Since throughout this paper $V_{2\to4}$ is almost always contracted with a BFKL amplitude $D_2$ from above, we will omit these arguments in the following and consider the quantity $V_{2\to4}D_2$. The vertex function $V(\mathbf{k}_1,\mathbf{k}_2;\mathbf{k}_3,\mathbf{k}_4)$ has the explicit form

$$\begin{aligned} (VD_2)(\mathbf{k}_1,\mathbf{k}_2;\mathbf{k}_3,\mathbf{k}_4) = \frac{g^4}{4} \times \{ & 2\,[\,c(1234) \\ & - b(123) - b(124) - b(134) - b(234) + b(12) + b(34) \\ & + a(13,2) + a(14,2) + a(23,1) + a(24,1) \\ & - a(1,2) - a(2,1) - a(3,12) - a(4,12)] \\ & +[\,t(123) + t(124) + t(134) + t(234) - t(12) - t(34) \\ & - s(13,2) - s(13,4) - s(14,2) - s(14,3) \\ & - s(23,1) - s(23,4) - s(24,1) - s(24,3) \\ & + s(1,2) + s(1,34) + s(2,1) + s(2,34) \\ & + s(3,12) + s(3,4) + s(4,12) + s(4,3)]\} \end{aligned} \tag{4.6}$$

where we have made use of the notation introduced in section 3.7.

Let us now describe the known properties of the vertex function $V$ and of the full transition vertex $V_{2\to4}$. The first observation is that $V(\mathbf{k}_1,\mathbf{k}_2;\mathbf{k}_3,\mathbf{k}_4)$ is symmetric in its first two and in its last two arguments

$$\begin{aligned} V(\mathbf{k}_1,\mathbf{k}_2;\mathbf{k}_3,\mathbf{k}_4) &= V(\mathbf{k}_2,\mathbf{k}_1;\mathbf{k}_3,\mathbf{k}_4) \\ &= V(\mathbf{k}_1,\mathbf{k}_2;\mathbf{k}_4,\mathbf{k}_3)\,, \end{aligned} \tag{4.7}$$

and symmetric under the exchange of the first pair of arguments and the second pair of arguments (that is why our notation separates these pairs by a semicolon)

$$V(\mathbf{k}_1,\mathbf{k}_2;\mathbf{k}_3,\mathbf{k}_4) = V(\mathbf{k}_3,\mathbf{k}_4;\mathbf{k}_1,\mathbf{k}_2)\,. \tag{4.8}$$

Therefore, according to (4.5) the full vertex $V_{2\to4}$ is completely symmetric under the simultaneous exchange of color and momentum of the gluons.

The combination of integrals in $VD_2$ vanishes when one of the outgoing momenta vanishes,

$$(VD_2)(\mathbf{k}_1,\mathbf{k}_2;\mathbf{k}_3,\mathbf{k}_4)|_{\mathbf{k}_i=0} = 0 \qquad (i \in \{1,\dots,4\})\,. \tag{4.9}$$

This result can be proven easily using identities of the kind mentioned at the end of section 3.7 and the fact that the BFKL amplitude $D_2$ vanishes at zero–momentum argument. This property of $VD_2$ is carried over to the full vertex,

$$(V_{2\to4}D_2)^{a_1a_2a_3a_4}(\mathbf{k}_1,\mathbf{k}_2,\mathbf{k}_3,\mathbf{k}_4)|_{\mathbf{k}_i=0} = 0 \qquad (i \in \{1,\dots,4\})\,. \tag{4.10}$$

Further, the function $VD_2$ is infrared finite, i. e. the infrared divergences in the different integrals contributing to (4.6) cancel in the sum. This can be easily seen after noticing that already certain combinations of very few standard integrals are infrared finite, nicely showing the cancellation of divergences between real and virtual corrections. The combination

$$b(\mathbf{l}) - \frac{1}{2}\,t(\mathbf{l}) \tag{4.11}$$

is infrared finite for any sum of momenta that is substituted for $\mathbf{l}$, as is clear from the corresponding integrals (see section 3.7). The factor 1/2 comes about because the integrand of the trajectory function $\beta$ (see (3.66) and (2.7)) exhibits two divergences of the same form. Similarly, one can show that the combination

$$a(\mathbf{l}_2,\mathbf{l}_1) - \frac{1}{2}\,s(\mathbf{l}_2,\mathbf{l}_1) - \frac{1}{2}\,s(\mathbf{l}_2,\mathbf{l}_3) \tag{4.12}$$

is infrared finite separately for any partition of the momenta $\{\mathbf{k}_i\}$ into three sums $\mathbf{l}_1,\mathbf{l}_2,\mathbf{l}_3$. Finally, the term $c(\mathbf{l})$ is infrared finite separately since in this term (see (3.65)) the poles of the propagators are cancelled by the zeros of the BFKL amplitude $D_2$ in the integral. The finiteness of these three groups is independent of the total number of momenta $\mathbf{k}_i$ that are split into the groups denoted by $\mathbf{l}_j$. It should be obvious from equation (4.6) that all integrals in the vertex function come in exactly these infrared finite combinations.

We now come back to the main problem of understanding the full four–gluon amplitude $D_4$. When the ansatz (4.2) was made the goal was to arrive at a simple equation for the yet unknown part $D_4^I$. The new integral equation (4.4) is in fact simple: It contains only the vertex $V_{2\to4}$ and a homogeneous part, and can therefore now be iterated. The structure arising from this is

$$D_4^I = G_4 \cdot V_{2\to4} \cdot D_2\,, \tag{4.13}$$

$G_4$ being the Green function of the four–gluon state. The Green function obeys the BKP equation with four $t$-channel gluons, which is a four–particle Schrödinger equation. Its Hamiltonian is given by the homogeneous part of the integral equation (4.4),

i. e. by the sum of all pairwise interactions $K_{2\to 2}$ of the four gluons. Unfortunately, the eigenvalues and eigenfunctions of the Hamiltonian are not known and (4.13) remains a formal solution only. Though, some properties of the four–gluon state have been worked out in [33]. We will return to the interpretation of the structure inherent in (4.13) momentarily.

Even without knowing an analytic formula for $D_4^I$ we can deduce two important properties. Like the two–gluon amplitude, the irreducible part of the four–gluon amplitude vanishes (modulo logarithms) when one of the outgoing gluon momenta is set to zero,

$$D_4^{I\,a_1a_2a_3a_4}(\mathbf{k}_1,\mathbf{k}_2,\mathbf{k}_3,\mathbf{k}_4)\Big|_{\mathbf{k}_i=0} = 0 \quad (i \in \{1,\dots,4\})\,. \tag{4.14}$$

To prove this we proceed order by order in the iteration of the Hamiltonian $\sum K_{2\to 2}$. The identity holds in lowest order since the vertex itself has this property (see (4.10)). In the next order, (4.14) holds because $K_{2\to 2}$ also vanishes if one of the outgoing momenta becomes zero, etc. Similarly, we can show that the irreducible part $D_4^I$ is completely symmetric in the four gluons, that is under the simultaneous exchange of color and momentum,

$$\begin{aligned} D_4^{I\,a_1a_2a_3a_4}(\mathbf{k}_1,\mathbf{k}_2,\mathbf{k}_3,\mathbf{k}_4) &= D_4^{I\,a_2a_1a_3a_4}(\mathbf{k}_2,\mathbf{k}_1,\mathbf{k}_3,\mathbf{k}_4) \\ &= D_4^{I\,a_3a_2a_1a_4}(\mathbf{k}_3,\mathbf{k}_2,\mathbf{k}_1,\mathbf{k}_4) \\ &= D_4^{I\,a_4a_2a_3a_1}(\mathbf{k}_4,\mathbf{k}_2,\mathbf{k}_3,\mathbf{k}_1)\,. \end{aligned} \tag{4.15}$$

### 4.3 Field theory structure

Although we do not have an analytic expression for the irreducible part, we have gathered by now quite some knowledge about the structure of the four–gluon amplitude $D_4$. Neglecting for a moment color and normalization factors, this structure is illustrated in the following diagram

$$D_4(\mathbf{k}_1,\mathbf{k}_2,\mathbf{k}_3,\mathbf{k}_4) = \sum$$

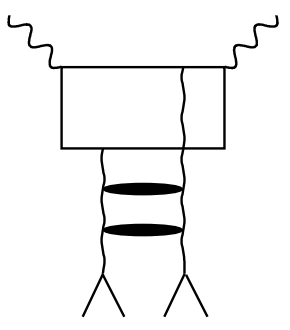

$+$

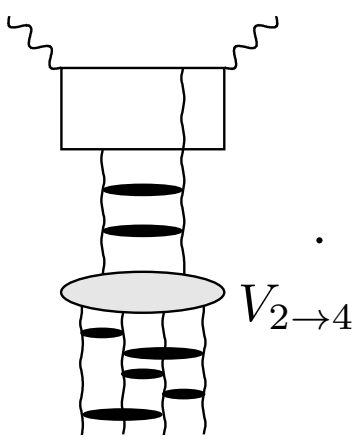

. (4.16)

The first part is the superposition of two–gluon states coupled to the quark loop (the reggeizing part $D_4^R$). In the second (irreducible) part $D_4^I$ a two–gluon system couples to the quark loop and then undergoes a transition to a four–gluon compound state via the vertex $V_{2\to 4}$. From this we learn that the complete amplitude consists of only a few basic building blocks: a quark loop allowing the coupling of the $t$-channel gluons to external particles, the two–gluon Green function (BFKL amplitude), the four–gluon Green function, and the two–to–four transition vertex. The Green functions describe the quantum mechanical propagation of bound states of $t$-channel gluons. With the transition vertex we have in addition a number–changing element connecting different $n$-reggeon states. This turns the quantum mechanical problem of $n$-gluon states into that of a quantum field theory of reggeized gluons. All this takes place in the 2-dimensional space of transverse momenta. The complex angular momentum $\omega$ plays the role of an energy variable. Its conjugate variable, i. e. rapidity, plays the role of the time variable.

We would like to emphasize that phenomenon of reggeization is crucial for the emergence of a field theory structure. Without reggeization in the three–gluon amplitude and in the part $D_4^R$ of the four–gluon amplitude one would not have been able to arrive at the simple structure (4.16).

So far only the simplest elements of a potential effective field theory have been identified: the two–gluon compound state, the two–to–four transition vertex $V_{2\to 4}$ and the four–gluon compound state. An analytic formula for the latter is still missing. The concept of an effective field theory has not been derived from first principles. It has to be tested and further elements should be derived. To achieve this is seems natural to proceed to higher $n$-gluon amplitudes. This constitutes the main goal of this paper, namely to deepen our understanding of the field theory structure of unitarity corrections by studying the five– and six–gluon amplitudes.

## 5. Five gluons

### 5.1 A reggeizing part and the integral equation for the remaining part

In the first step, our analysis of the five–gluon amplitude $D_5$ follows the same lines as the study of the three– and four–gluon amplitudes. To get started we identify a reggeizing part $D_5^R$ of the amplitude and split the amplitude accordingly,

$$D_5 = D_5^R + D_5^I \,. \tag{5.1}$$

With a well–chosen $D_5^R$ we will come to a new integral equation for the yet unknown quantity $D_5^I$. Again, this decomposition is not unique. Our ansatz will lead to an equation for $D_5^I$ that can even be solved. This situation is the best we can hope for and further justification for the ansatz is certainly not needed. The natural choice for $D_5^R$ is once more suggested by the inhomogeneous term $D_{(5;0)}$. This means that our ansatz has exactly the same color and momentum structure as $D_{(5;0)}$ in (3.54), but we replace $D_{(2;0)}$ by $D_2$ resulting in

$$\begin{aligned} D_5^{R\,a_1a_2a_3a_4a_5}(\mathbf{k}_1,\mathbf{k}_2,\mathbf{k}_3,\mathbf{k}_4,\mathbf{k}_5) = \qquad\qquad\qquad\qquad\qquad\qquad\qquad\qquad\qquad\qquad (5.2)\\ = -g^3\{f^{a_1a_2a_3a_4a_5}\,[D_2(1234,5)+D_2(1,2345)-D_2(15,234)] \\ + f^{a_2a_1a_3a_4a_5}\,[D_2(1345,2)-D_2(12,345)+D_2(125,34)-D_2(134,25)] \\ + f^{a_1a_2a_3a_5a_4}\,[D_2(1235,4)-D_2(14,235)+D_2(145,23)-D_2(123,45)] \\ + f^{a_1a_2a_4a_5a_3}\,[D_2(1245,3)-D_2(13,245)+D_2(135,24)-D_2(124,35)]\}\,. \end{aligned}$$

This is inserted into the integral equation (3.13). We insert into that equation the expressions (4.1), (4.2), (4.3) for $D_3$ and $D_4$ as well. In order to find the new integral equation for $D_5^I$ we have to simplify and collect all terms not involving $D_5^I$ and $D_4^I$. These terms will contribute to the inhomogeneous term of the new equation for $D_5^I$ and we will now discuss them. From the left–hand side of (3.13) we get $\omega D_5^R$, which can be treated using the BFKL equation (3.10). Due to this, the inhomogeneous term $D_{(5;0)}$ in (3.13) is exactly cancelled and we get further terms involving only convolutions of $D_2$ functions with kernels $K_{2\to 2}$ or trajectories $\beta$. From the right–hand side of (3.13) we get contributions of the type $K_{2\to 5} \otimes D_2$, $\sum K_{2\to 4} \otimes D_3$, $\sum K_{2\to 3} \otimes D_4^R$, and $\sum K_{2\to 2} \otimes D_5^R$. All of these can be written as sums of convolutions of $D_2$ with the kernels $K_{2\to m}$. The corresponding contractions of color tensors are performed using the diagrammatic method described in appendix A. In that appendix we also give the explicit formulae for some of the contractions required. The total number of contractions needed here is close to one hundred, and they can be easily obtained from those in the appendix. The momentum integrals are brought to their standard forms as classified in section 3.7. The respective momentum integrals and color contractions are then multiplied and can be collected. This last step amounts to collecting several thousand terms and sorting them according to the different color tensors, and we do this with the help of a computer algebra program.

In the derivation of the new integral equation the terms involving $D_4^I$ remain unchanged. They will be treated at a later stage of the analysis. The same is true for the homogeneous term containing $D_5^I$. The combinations of the $D_4^I$ and $D_5^I$ amplitudes with the kernels are therefore the same as in the original equation (3.13). We thus find the following equation for the unknown part $D_5^I$ of the five–gluon amplitude:

$$\left(\omega - \sum_{i=1}^{5} \beta(\mathbf{k}_i)\right) D_5^I = \sum f_{a_1 a_2 a_3} \delta_{a_4 a_5} H(1,2,3;4,5) + \sum K_{2\to 3}^{\{b\}\to\{a\}} \otimes D_4^{I\, b_1 b_2 b_3 b_4} + \sum K_{2\to 2}^{\{b\}\to\{a\}} \otimes D_5^{I\, b_1 b_2 b_3 b_4 b_5} \, . \tag{5.3}$$

The first term on the right hand side is the result of the computation described above. We will now treat it in more detail.

The first interesting observation concerns its color structure. All terms proportional to $f^{a_1 a_2 a_3 a_4 a_5}$ (and the other three permutations of this occurring in (5.3)) are cancelled between the different contributions to this inhomogeneous term and drop out. Something similar happened in the case of $D_4$ where there is no term proportional to $d^{a_1 a_2 a_3 a_4}$ in the vertex $V_{2\to 4}$ and only lower tensors (i. e. products of $\delta$-tensors) contribute.

Secondly, we observe the following symmetry of the new inhomogeneous term that we have calculated. The sum extends over all (ten) possibilities to have a pair of gluons in a color singlet. For each of these permutations of the gluons color and momentum labels are exchanged simultaneously, i. e. the sum in (5.3) stands for

$$\sum f_{a_1a_2a_3}\delta_{a_4a_5}H(1,2,3;4,5) = f_{a_1a_2a_3}\delta_{a_4a_5}H(1,2,3;4,5) + f_{a_1a_2a_4}\delta_{a_3a_5}H(1,2,4;3,5) + \ldots + \delta_{a_1a_2}f_{a_3a_4a_5}H(3,4,5;1,2)\,. \quad (5.4)$$

The function $H$ is the same in all ten permutations. This symmetry is an outcome of our computation, and it has not been used to derive (5.3). On the other hand, it is not an unexpected property of the inhomogeneous term. Already in the corresponding equation (4.4) in the four–gluon case the inhomogeneous term, i. e. the vertex $V_{2\to4}$, had a similar symmetry.

A closer inspection of the function $H$ reveals that it is actually a superposition of vertex functions $V$ which we encountered in the discussion of the two–to–four vertex $V_{2\to4}$. Namely,

$$H(1,2,3;4,5) = \frac{g}{2}[(VD_2)(12,3;4,5) - (VD_2)(13,2;4,5) + (VD_2)(1,23;4,5)]\,. \quad (5.5)$$

To obtain this striking result is was necessary to go through the full calculation of all convolutions of amplitudes with kernels as described. It is only afterwards that we are able to discover the simple structure in terms of $V$. Unfortunately, we do not know a way leading to (5.3), (5.5) that avoids this tedious calculation.

### 5.2 Solving the equation for the remaining part

Up to this point, our analysis of the five–gluon amplitude followed essentially the same lines as in the case of four gluons. Whereas there the new integral equation could simply be iterated, this is not possible here. To find the solution for $D_5^I$ we now have to go beyond the procedure applied for $n = 3$ and 4 gluons.

Taking a close look at the integral equation (5.3) for $D_5^I$ we discover that its structure bears a strong resemblance to the equation (3.11) for the three–gluon amplitude $D_3$. In the second term on the right hand side of (5.3) a pair of gluons of the amplitude $D_4^I$ is convoluted with a two–to–three kernel. In the corresponding term in (3.11) it was the two–gluon (BFKL) amplitude $D_2$ that was convoluted with the same kernel. There the first term on the right hand side, i. e. $D_{(3;0)}$, was the superposition of quark loop amplitudes $D_{(2;0)}$ that are the lowest order terms in the ladder expansion of $D_2$. In (5.3) the corresponding term (5.4) is, according to (5.5), the superposition of functions $VD_2$. (In fact it is even the superposition of full two–to–four reggeon vertices $V_{2\to4}D_2$ as we will see below.) These vertex functions, in turn, constitute the lowest order terms in the ladder expansion[6] of the irreducible amplitude $D_4^I$, cf. (4.13). More specifically, it is for each of the ten terms in (5.4) that we find in a three–gluon subsystem exactly the momentum structure that also determines $D_{(3;0)}$. The similarity is also evident for the color structure, namely the three–gluon subsystem comes with a tensor $f_{abc}$.

[6]This statement has to be taken with some care, since the terms $VD_2$ are of course not of lowest order in the coupling constant $g$. In the contrary, $D_2$ already contains an infinite series of ladder diagrams. What is meant here is that each diagram in the ladder expansion of $D_4^I$ starts with a full two–gluon ladder and a vertex attached to this.

Clearly, this suggests to construct a solution $D_5^I$ in analogy with the three–gluon amplitude. While $D_3$ is a superposition of BFKL amplitudes $D_2$ we now should choose $D_5^I$ as a similar superposition of irreducible four–gluon amplitudes $D_4^I$. The following combination of $D_4^I$ amplitudes is of this kind and in fact is a solution to equation (5.3). We will outline the proof of this fact momentarily.

$$\begin{aligned} D_5^{I\,a_1a_2a_3a_4a_5}(\mathbf{k}_1,\mathbf{k}_2,\mathbf{k}_3,\mathbf{k}_4,\mathbf{k}_5) &= \frac{g}{2}\times \\ \times \Big\{ & f_{a_1a_2c} D_4^{I\,ca_3a_4a_5}(12,3,4,5) + f_{a_1a_3c} D_4^{I\,ca_2a_4a_5}(13,2,4,5) \\ & + f_{a_1a_4c} D_4^{I\,ca_2a_3a_5}(14,2,3,5) + f_{a_1a_5c} D_4^{I\,ca_2a_3a_4}(15,2,3,4) \\ & + f_{a_2a_3c} D_4^{I\,a_1ca_4a_5}(1,23,4,5) + f_{a_2a_4c} D_4^{I\,a_1ca_3a_5}(1,24,3,5) \\ & + f_{a_2a_5c} D_4^{I\,a_1ca_3a_4}(1,25,3,4) + f_{a_3a_4c} D_4^{I\,a_1a_2ca_5}(1,2,34,5) \\ & + f_{a_3a_5c} D_4^{I\,a_1a_2ca_4}(1,2,35,4) + f_{a_4a_5c} D_4^{I\,a_1a_2a_3c}(1,2,3,45) \Big\} \end{aligned} \tag{5.6}$$

In each of the terms one pair $(i,j)$ of gluons is merged[7] into one gluon which then enters the irreducible four–gluon amplitude from below. This gluon in $D_4^I$ has momentum $(\mathbf{k}_i+\mathbf{k}_j)$ and color label $c$. The merging of the two gluons in color space happens via a $f_{a_ia_jc}$ tensor $(i<j)$. The position in the amplitude $D_4^I$ at which the 'composite' gluon with color $c$ and momentum $(\mathbf{k}_i+\mathbf{k}_j)$ enters does not matter since $D_4^I$ is completely symmetric in the four gluons, cf. (4.15). All possible pairs of gluons are treated in the same way. The way pairs of gluons are merged (or arise from splittings) becomes more transparent when (5.6) is written using birdtrack notation,

$D_5^{I\,a_1a_2a_3a_4a_5}(\mathbf{k}_1,\mathbf{k}_2,\mathbf{k}_3,\mathbf{k}_4,\mathbf{k}_5) = \frac{g}{2}\times$

$\times\Big\{$ 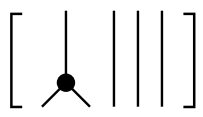 $\star D_4^{I\,b_1b_2b_3b_4}(12,3,4,5) +$ 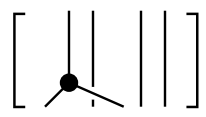 $\star D_4^{I\,b_1b_2b_3b_4}(13,2,4,5)$

$+$ 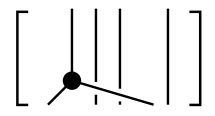 $\star D_4^{I\,b_1b_2b_3b_4}(14,2,3,5) +$ 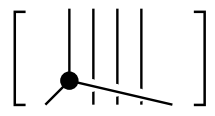 $\star D_4^{I\,b_1b_2b_3b_4}(15,2,3,4)$

$+$ 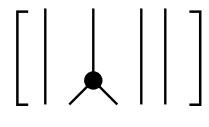 $\star D_4^{I\,b_1b_2b_3b_4}(1,23,4,5) +$ 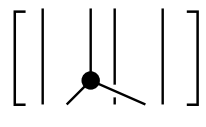 $\star D_4^{I\,b_1b_2b_3b_4}(1,24,3,5)$

$+$ 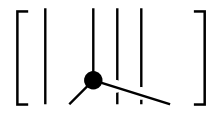 $\star D_4^{I\,b_1b_2b_3b_4}(1,25,3,4) +$ 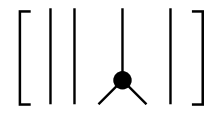 $\star D_4^{I\,b_1b_2b_3b_4}(1,2,34,5)$

$+$ 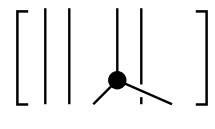 $\star D_4^{I\,b_1b_2b_3b_4}(1,2,35,4) +$ 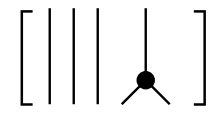 $\star D_4^{I\,b_1b_2b_3b_4}(1,2,3,45)\Big\}.$ (5.7)

We will come to the interpretation of this structure in section 5.3.

[7] Depending on the context one would like to use different words for the formula (5.6). From the point of view of constructing the solution it is clearly a 'merging' of two gluons, with the concept of a $t$-channel evolution in mind one would prefer to speak of a 'splitting' of gluons.

Now we explain how (5.6) can be shown to solve the integral equation (5.3). The only pieces of information about the irreducible four–gluon amplitude $D_4^I$ we need for the purpose of this proof are its complete symmetry in the four outgoing gluons, cf. (4.15), and the integral equation (4.4) it fulfills. Fortunately, an analytic solution of the latter is not required.

We start from the integral equation (5.3) derived previously and insert the conjectured solution (5.6). On the left hand side we then find ten terms of the kind $\omega D_4^I$. To these we apply (4.4). Thereby we produce different terms and we first concentrate on the terms involving the vertex function $V$. For example, applying (4.4) to

$$\omega f_{a_1 a_2 c} D_4^{I\,c a_3 a_4 a_5}(12, 3, 4, 5) \tag{5.8}$$

produces, due to (4.5), the expression

$$\begin{aligned} f_{a_1 a_2 c}(V_{2\to 4} D_2)^{c a_2 a_3 a_4}(12, 3, 4, 5) = \; & f_{a_1 a_2 a_3}\delta_{a_4 a_5}(VD_2)(12, 3; 4, 5) \\ & + f_{a_1 a_2 a_4}\delta_{a_3 a_5}(VD_2)(12, 4; 3, 5) \\ & + f_{a_1 a_2 a_5}\delta_{a_3 a_4}(VD_2)(12, 5; 3, 4) \end{aligned} \tag{5.9}$$

containing three different vertex functions $V$. Similar expressions are obtained from the other $\omega D_4^I$ terms on the left hand side. In some cases a minus sign arises due to the antisymmetry of the structure constant $f_{abc}$. For instance, from the second term in (5.6),

$$f_{a_1 a_3 c} D_4^{I\,c a_2 a_4 a_5}(13, 2, 4, 5)\,, \tag{5.10}$$

in which the pair (1,3) of gluons is merged we get

$$\begin{aligned} & -f_{a_1 a_2 a_3}\delta_{a_4 a_5}(VD_2)(13, 2; 4, 5) + f_{a_1 a_3 a_4}\delta_{a_2 a_5}(VD_2)(13, 4; 2, 5) \\ & + f_{a_1 a_3 a_5}\delta_{a_2 a_4}(VD_2)(13, 5; 2, 4)\,. \end{aligned} \tag{5.11}$$

Therefore we find exactly the same thirty vertex functions that occur also on the right hand side of (5.3) according to (5.4),(5.5). We have thus confirmed that the conjectured solution (5.6) indeed reproduces the correct lowest order term in the integral equation, namely the combination of vertex functions $VD_2$ given above. Moreover, we see that the first term on the right hand side of the integral equation (5.3) is not only a superposition of vertex functions $VD_2$ but of full transition vertices $V_{2\to 4}$ (applied to $D_2$ as usual).

Let us now consider further terms in the integral equation (5.3) that we have not treated yet, namely those involving the irreducible four–gluon amplitude $D_4^I$. Having applied (4.4) to the $\omega D_4^I$ terms on the left hand side the homogeneous term of that equation produces convolutions of $D_4^I$ amplitudes with kernels $K_{2\to 2}$. In these, a kernel acts on $D_4^I$ first and then the splitting of one gluon into a pair happens according to the combinations in (5.6). On the right hand side of the integral equation (5.3) the last term also gives us convolutions of $D_4^I$ amplitudes with kernels $K_{2\to2}$, but here the order of the convolution and the splitting of gluons is interchanged: first one gluons splits into two and then two of the now five gluons interact via a kernel $K_{2\to2}$. Among the terms just mentioned a subclass cancels immediately. Consider the case that the two–to–two kernel acts between two gluons none of which undergoes a splitting (LHS) or has emerged from a splitting (RHS). Then the order of interaction and splitting along the $t$-channel evolution is irrelevant and these terms are in fact the same on both sides.

The next terms in the integral equation that we look at are the products of $D_4^I$'s with trajectories $\beta$. These arise on the left hand side either from $\omega D_4^I$ via (4.4) or from the original $\left[\sum_{i=i}^{5}\beta(\mathbf{k}_i)\right] D_5^I$ after (5.6) is inserted. Those in which the argument of the trajectory function does not correspond to a gluon undergoing or arising from a splitting cancel directly between these contributions. It can be easily checked that the others are exactly cancelled by the terms from the right hand side in which the two gluons emerging from a splitting interact with each other via a kernel $K_{2\to2}$.

It is a bit more complicated to study the expressions still left in the integral equation after the cancellations discussed so far. These are $D_4^I$'s undergoing a two–to–three transition via the kernel $K_{2\to3}$ on the RHS and convolutions of $D_4^I$ functions with kernels $K_{2\to2}$ in which one of the gluons undergoing or emerging from a splitting is involved in the interaction (both sides). The latter do no longer include such convolutions in which the interaction is between the two gluons emerging from the splitting. In the terms under consideration three of the five outgoing gluons participate in the splitting or in the interaction. The other two gluons do not interact and can be in an arbitrary color state. Among the five gluons there can be a total of ten different three–gluon subgroups, and we will argue that the cancellation takes place in each of these subgroups separately. To this end let us concentrate on one of these subgroups, say the one with the first three of the outgoing gluons affected.

The mechanism that makes these contributions cancel between the two sides of the integral equation is the same that already caused reggeization in the three–gluon amplitude $D_3$. This does not come as a surprise since it was just the similarity of the corresponding integral equations that lead us to the ansatz (5.6). The identity actually bringing about the reggeization of the three–gluon subsystem is in pictorial language

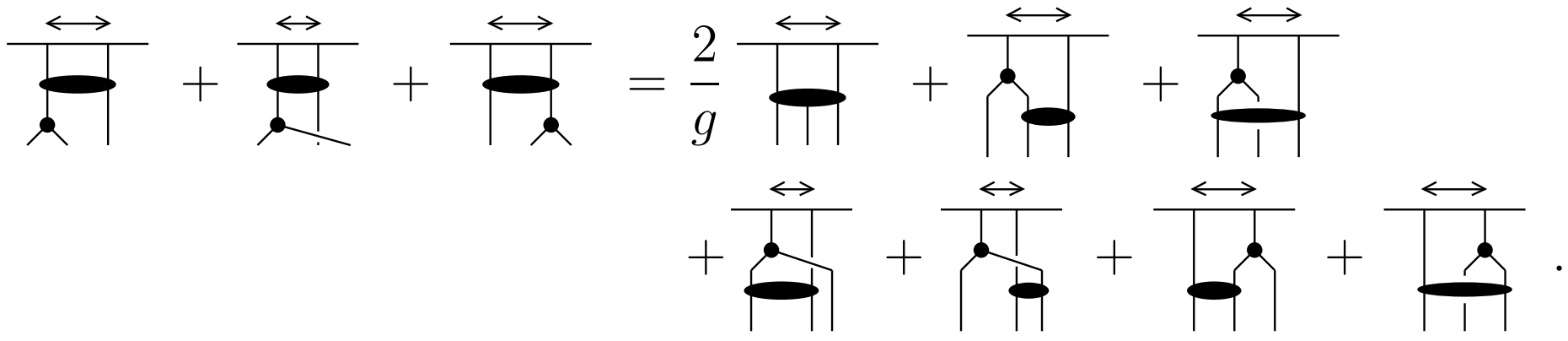
(5.12)

Only the three–gluon subsystem is shown, and the horizontal lines at the top are meant to suggest the irreducible amplitude $D_4^I$ that the gluons enter. The arrows indicate the symmetry of this amplitude under the simultaneous exchange of color and momentum of the two gluons. The kernels are the ones defined in section 3.6.

The splitting of a gluon is depicted here by the corresponding color diagram and is meant to indicate the behavior in momentum space[8] as well. The terms on the left (right) hand side of (5.12) are exactly the ones that occur on the left (right) hand side of the integral equation (5.3). To prove (5.12), the convolutions are evaluated as described in the preceding sections. However, here the situation is slightly complicated by the fact that the two gluons entering from above can be in an arbitrary color state. In the case of $D_3$ these two gluons were in a color singlet state, effectively reducing all color tensors to an overall $f_{a_1 a_2 a_3}$. Here we have to be more careful and treat three independent color classes separately. (Of course, this could have been done already for $D_3$ but there it was not necessary.) The three classes are

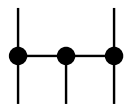 , 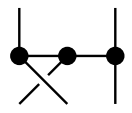 , 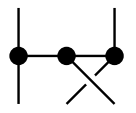 . (5.13)

For some of the terms in (5.12) it is necessary to use the symmetry in the upper two gluons which is a property of $D_4^I$. Each of the three terms on the left hand side of (5.12) contributes to two of the three color classes in (5.13) via the Jacobi identity (3.43). Having dissected the integral equation this far, it is finally a comparatively short calculation to check that (5.12) holds. Thereby we have finished the proof that $D_5^I$, as given in (5.6), in fact is the solution of the integral equation.

## 5.3 Interpretation of the result

In the preceding section we have been able to solve the equation for the five–gluon amplitude. Now we want to interpret our findings in view of a possible field theory of unitarity corrections. Let us first summarize the essential results we have obtained. We have split the five–gluon amplitude into two parts. The first part was the reggeizing part $D_5^R$ that is the superposition of two–gluon amplitudes $D_2$. We have found an integral equation for the remaining part $D_5^I$ and have solved it. It turned out that the remaining part is the superposition of irreducible four–gluon amplitudes $D_4^I$. Neglecting all normalization factors and color tensors, this situation can be sketched in the following way:

$D_5(\mathbf{k}_1, \mathbf{k}_2, \mathbf{k}_3, \mathbf{k}_4, \mathbf{k}_5) = \sum$ 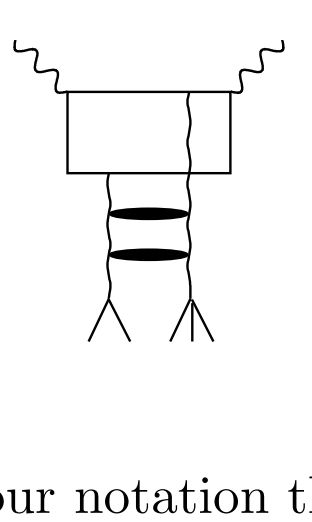 $+ \sum$ 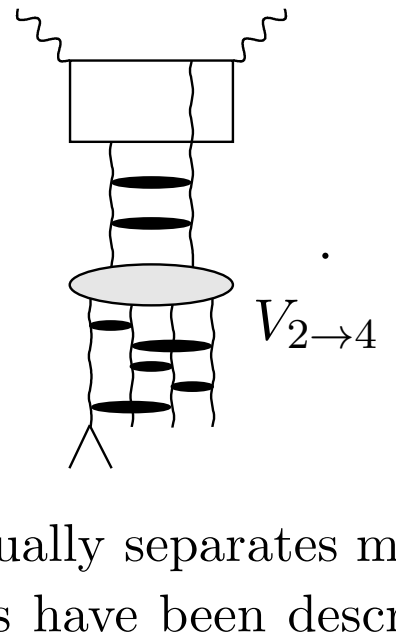

. (5.14)

[8] Strictly speaking, this is done in abuse of our notation that usually separates momentum and color space. Confusion should hardly be possible here as all terms have been described in detail before.

The first term on the right hand side is the reggeizing part $D_5^R$ of the amplitude (see (5.3)). The sum extends over all partitions of the five gluons into two groups. The second term is the one found in the preceding section, see (5.6). Here the sum includes all possible pairs of gluons that then merge into one.

In the five–gluon amplitude, we only find elements that are already known: the amplitudes $D_2$ and $D_4^I$, and the two–to–four transition vertex $V_{2\to4}$. There is no new transition vertex and no new irreducible amplitude that would include a five–gluon compound state. The absence of new elements is an intriguing result. It shows that the five–gluon amplitude reggeizes completely. This result (5.14) clearly constitutes a generalization of the concept of reggeization and proves that reggeization also takes place in more complicated amplitudes. Especially interesting is the reggeization in the second part that turns out to be a superposition of irreducible four–gluon amplitudes. The mechanism at work here is exactly the same as in the three–gluon amplitude.

Reggeization was an important prerequisite for the emergence of the two–to–four vertex and thus of the field theory structure in the four–gluon amplitude. That the phenomenon of reggeization also occurs in the five–gluon amplitude gives us confidence that the idea of a field theory structure will be a good guiding line also for the investigation of the six–gluon amplitude.

Given that the three– and five–gluon amplitudes exhibit complete reggeization caused by the same mechanism, one is naturally lead to the question whether the same is true for each odd number $n$ of gluons. Indeed, as we have seen, the mechanism leading to reggeization in a three–gluon subsystem is very general. It is completely independent of the structure of the quark loop that we started with in the analysis of the integral equations. In deriving (5.12) we only made use of the fact that the amplitude to which the upper two gluons are attached — in this case $D_4^I$ — is symmetric in the two gluons. Therefore we can conclude that for a given odd $n$ one important condition for the reggeization of the $n$-gluon amplitude is fulfilled as soon as the irreducible part of the $(n-1)$-gluon amplitude is symmetric. However, we have to keep in mind that this is only one of the two conditions leading to complete reggeization. The second condition necessary for the reggeization of $D_5^I$ was that the inhomogeneous term in the integral equation (5.3) had the specific form (5.5), i. e. could be written as a special superposition of transition vertices $V_{2\to4}$. For a part of an arbitrary $n$-gluon amplitude with odd $n$ to reggeize it is obviously necessary that the respective inhomogeneous term has a very specific form. With our present knowledge, we are not able to derive this specific form of the inhomogeneous term for general $n$. The complete reggeization of amplitudes with an odd number of gluons is therefore at present only a (plausible) conjecture.

We would like to add a remark that concerns our choice of notation, but actually goes beyond a pure issue of notation. In the splitting of the four–gluon amplitude into two parts in (4.2) the superscript $I$ in $D_4^I$ was meant to indicate that this part of the amplitude is irreducible — in contrast to the other part. For four gluons this was a good choice of notation since that part in fact contains a new irreducible compound state of four reggeized gluons. In the case of five gluons, we again split the amplitude into two parts, cf. (5.1). The first part $D_5^R$ is a reggeizing one as the superscript $R$ indicates. But now we have discovered that the remaining part $D_5^I$ reggeizes as well. (For this reason we have avoided to call this part 'irreducible'.) Nevertheless our notation makes perfect sense when extended in an appropriate way. The first superscript $I$ or $R$ should be understood as specifying the (non–)reggeization of the respective part with respect to the two–gluon state. We can then introduce a second superscript to indicate the (non–)reggeization with respect to the four–gluon state. (We will actually be forced to do so when considering the six–gluon amplitude in the next section.) For a proper notation we should thus identify $D_5^I = D_5^{I,R}$, the second superscript now indicating that this part reggeizes with respect to the four–gluon state. The notation can easily be extended to accommodate reggeization with respect to a potential six–gluon state or even higher compound states.

## 6. Six gluons

### 6.1 A reggeizing part

Encouraged by the success we have had so far with that procedure we again use the quark loop amplitude $D_{(6;0)}$ to construct from it a reggeizing part $D_6^R$ as a superposition of two–gluon (BFKL) amplitudes. The full six–gluon amplitude is then split into two parts,

$$D_6 = D_6^R + D_6^I\,, \tag{6.1}$$

and it will be our first task to find a new integral equation for the remaining part $D_6^I$. In detail, the reggeizing part $D_6^R$ is

$$\begin{aligned}
&D_6^{R\,a_1a_2a_3a_4a_5a_6}(\mathbf{k}_1,\mathbf{k}_2,\mathbf{k}_3,\mathbf{k}_4,\mathbf{k}_5,\mathbf{k}_6) = \qquad\qquad (6.2)\\
&= g^4\{d^{a_1a_2a_3a_4a_5a_6}\,[D_2(12345,6)+D_2(1,23456)-D_2(16,2345)]\\
&\quad + d^{a_2a_1a_3a_4a_5a_6}\,[D_2(13456,2)-D_2(1345,26)+D_2(126,345)-D_2(12,3456)]\\
&\quad + d^{a_1a_2a_3a_4a_6a_5}\,[D_2(12346,5)-D_2(1234,56)+D_2(156,234)-D_2(15,2346)]\\
&\quad + d^{a_2a_1a_3a_4a_6a_5}\,[-D_2(1256,34)-D_2(1346,25)+D_2(125,346)+D_2(134,256)]\\
&\quad + d^{a_3a_1a_2a_4a_5a_6}\,[D_2(12456,3)-D_2(1245,36)+D_2(136,245)-D_2(13,2456)]\\
&\quad + d^{a_1a_2a_3a_5a_6a_4}\,[D_2(12356,4)-D_2(1235,46)+D_2(146,235)-D_2(14,2356)]\\
&\quad + d^{a_2a_1a_3a_5a_6a_4}\,[-D_2(1246,35)-D_2(1356,24)+D_2(124,356)+D_2(135,246)]\\
&\quad + d^{a_1a_2a_3a_6a_5a_4}\,[-D_2(1236,45)-D_2(1456,23)+D_2(123,456)+D_2(145,236)]\}
\end{aligned}$$

as obtained from (3.54) by the replacement $D_{(2;0)} \to D_2$ while keeping the color and momentum structure. This expression already indicates one of the major difficulties we have to overcome during the treatment of the six–gluon amplitude: the large number of terms we have to take care of.

## 6.2 The integral equation for the remaining part

The original integral equation (3.14) for the six–gluon amplitude is now used to derive a new integral equation for the unknown part $D_6^I$. The method in this step is exactly the same as for the four– and five–gluon amplitudes. We insert into the integral equation our complete knowledge about the reggeizing parts $D_n^R$ of the amplitudes $D_n$ with up to $n = 6$ gluons, including the ansatz (6.1), (6.3) for the six–gluon amplitude. The corresponding formulae for $n \leq 5$ can be found in the preceding sections. Then we apply the BFKL equation (3.10) to the expression $\omega D_6^R$ on the left hand side. This is possible because $D_6^R$ was chosen as a superposition of BFKL amplitudes, cf. (6.3). We thereby produce convolutions of $D_2$ amplitudes with two–to–two kernels and products of $D_2$ amplitudes with trajectory functions $\beta$. The insertion of the reggeizing parts $D_n^R$ of the amplitudes on the right hand side leads to convolutions of $D_2$ amplitudes with the integral kernels. We have to perform the corresponding contractions of color tensors and have to bring the integrals to their standard forms as classified in section 3.7. The main problem consists in the huge number of combinations of amplitudes with kernels. We have to perform close to 250 contractions of color tensors, and we have to find the standard form of more than 3500 integrals. Whereas the color tensors can still be calculated by hand this is no longer possible for the huge number of momentum space integrals. We have therefore developed an algorithm for this purpose that is suited for the implementation on a computer. The algorithm is explained in detail in appendix B. We have written a PERL script based on this algorithm that produces an output which can directly be used as an input for a computer algebra program like MAPLE. The tensor contractions are calculated with the help of the method described in appendix A. Some of the contractions are given explicitly in that appendix. Many other contractions are obtained from these by permutations of the gluon color labels. The computer algebra program is then used to multiply the resulting sums of elementary tensors with the corresponding integrals, and to finally collect all terms. In the final step more than $2 \cdot 10^4$ integrals have to be sorted according to their color tensor coefficients. (This shows that our method of dealing with the integral equations will in its practical applicability be limited to relatively small numbers $n$ of gluons.) Having collected all terms in the equation which contain the amplitude $D_2$ we have found the inhomogeneous term of the new integral equation for $D_6^I$.

In the derivation of the new integral equation the terms containing the irreducible four–gluon amplitude $D_4^I$ and the second part $D_5^I$ of the five–gluon amplitude remain unchanged. Their combinations with the kernels are the same as in the original equation (3.14). The resulting integral equation for $D_6^I$ is then found to have the

form

$$
\begin{aligned}
&\left(\omega - \sum_{i=1}^{6} \beta(\mathbf{k}_i)\right) D_6^{I\,a_1a_2a_3a_4a_5a_6}(\mathbf{k}_1, \mathbf{k}_2, \mathbf{k}_3, \mathbf{k}_4, \mathbf{k}_5\mathbf{k}_6) = \\
&= (W^{a_1a_2a_3a_4a_5a_6} D_2)(\mathbf{k}_1, \mathbf{k}_2, \mathbf{k}_3, \mathbf{k}_4, \mathbf{k}_5\mathbf{k}_6) \\
&\quad + \sum f_{a_1a_2a_3} f_{a_4a_5a_6} L(1,2,3;4,5,6) \\
&\quad + \sum d^{a_1a_2a_3a_4} \delta_{a_5a_6} I(1,2,3,4;5,6) \\
&\quad + \sum d^{a_2a_1a_3a_4} \delta_{a_5a_6} J(1,2,3,4;5,6) \\
&\quad + \sum K_{2\to 4}^{\{b\}\to\{a\}} \otimes D_4^{I\,b_1b_2b_3b_4} + \sum K_{2\to 3}^{\{b\}\to\{a\}} \otimes D_5^{I\,b_1b_2b_3b_4b_5} \\
&\quad + \sum K_{2\to 2}^{\{b\}\to\{a\}} \otimes D_6^{I\,b_1b_2b_3b_4b_5b_6} \,. 
\end{aligned} \tag{6.3}
$$

The first four terms on the right hand side are the result of the computation outlined above. We will now describe them in detail.

The first observation we make is again that certain color structures are completely cancelled in the equation. All terms proportional to $d^{a_1a_2a_3a_4a_5a_6}$ (and the other seven permutations of this occurring in (6.3)) are cancelled between the different contributions to the inhomogeneous term and drop out. The same was observed for the tensors $d^{a_1a_2a_3a_4}$ and $f^{a_1a_2a_3a_4a_5}$ in the equations for the parts $D_4^I$ and $D_5^I$ of the four– and five–gluon amplitudes, respectively.

As in the integral equations for $D_4^I$ and $D_5^I$ the inhomogeneous term has a high degree of symmetry which we will explain for each of the terms below. This symmetry is not only nice by itself, but it is also a possibility to check our calculation.

The first term on the right hand side of (6.3) differs in its structure from the other terms and will be treated separately in section 6.3. Here we mention already that it is symmetric in the sense that it is the sum of terms that are obtained from each other by permutations of the gluons.

The same is true for the second term on the right hand side of the new integral equation. The sum extends over all partitions of the six gluons into two groups each of which contains three gluons,

$$
\begin{aligned}
\sum f_{a_1a_2a_3} f_{a_4a_5a_6} L(1,2,3;4,5,6) = &\; f_{a_1a_2a_3} f_{a_4a_5a_6} L(1,2,3;4,5,6) \\
&+ f_{a_1a_2a_4} f_{a_3a_5a_6} L(1,2,4;3,5,6) \\
&+ \ldots + f_{a_1a_5a_6} f_{a_2a_3a_4} L(1,5,6;2,3,4) \,. 
\end{aligned} \tag{6.4}
$$

The function $L$ is the same in all terms in the sum and only its arguments are exchanged in the different terms. A closer inspection reveals that the function $L$ permits a decomposition into vertex functions $V$ known from the two–to–four transition vertex (see section 4),

$$
L(1,2,3;4,5,6) = \frac{g^2}{4} \times
$$

$$\times[(VD_2)(12,3;45,6) - (VD_2)(12,3;46,5) + (VD_2)(12,3;4,56)$$
$$-(VD_2)(13,2;45,6) + (VD_2)(13,2;46,5) - (VD_2)(13,2;4,56)$$
$$+(VD_2)(1,23;45,6) - (VD_2)(1,23;46,5) + (VD_2)(1,23;4,56)] \,. \qquad (6.5)$$

The sum in the third term on the right hand side of the integral equation extends over all partitions of the six gluons into one group containing four and one group containing two gluons,

$$\sum d^{a_1a_2a_3a_4}\delta_{a_5a_6}I(1,2,3,4;5,6) = d^{a_1a_2a_3a_4}\delta_{a_5a_6}I(1,2,3,4;5,6) + d^{a_1a_2a_3a_5}\delta_{a_4a_6}I(1,2,3,5;4,6) + \ldots + \delta_{a_1a_2}d^{a_3a_4a_5a_6}I(3,4,5,6;1,2) \,. \qquad (6.6)$$

Also in this case we find that the function $I$ is the same in all terms in the sum. Remarkably, also this function can be written in terms of the vertex function $V$,

$$I(1,2,3,4;5,6) = -g^2[(VD_2)(1,234;5,6) + (VD_2)(123,4;5,6) - (VD_2)(14,23;5,6)] \,. \qquad (6.7)$$

The sum in the fourth term on the right hand side of the new integral equation (6.3),

$$\sum d^{a_2a_1a_3a_4}\delta_{a_5a_6}J(1,2,3,4;5,6) \,, \qquad (6.8)$$

extends over the same permutations of gluons as the term discussed before (see (6.6)). Again the function $J$ is the same in all terms in the sum, and it can be written as a superposition of vertex functions $V$ as

$$J(1,2,3,4;5,6) = -g^2[(VD_2)(134,2;5,6) + (VD_2)(124,3;5,6) - (VD_2)(12,34;5,6) - (VD_2)(13,24;5,6)] \,. \qquad (6.9)$$

We would like to emphasize that the symmetry of the sums contributing to the inhomogeneous term of the new integral equation is an outcome of our calculation. We have not used it to derive the new equation. That we find the symmetry in the resulting equation gives us confidence that we did not make any errors in the long and tedious calculation leading to (6.3). We also would like to stress that the representation of a part of the inhomogeneous term as a superposition of well–known vertex function $V$ is an outcome of our calculation and was not used to derive the new equation. Unfortunately, we do not know a way that directly leads to the comparatively simple structure arising in the terms discussed above.

### 6.3 A new piece in the field theory

The first term on the right hand side of the integral equation (6.3) differs in its structure from the other terms. We will therefore discuss it separately in this section. The study of the other terms will be resumed in section 6.4. We start by giving an explicit representation of the term under consideration and proceed by listing its properties. After that we will elaborate on the question how the new term has to be interpreted in the context of the effective field theory structure of unitarity corrections.

### 6.3.1 Explicit representation and properties

The first term on the right hand side in (6.3) has the following color and momentum structure:

$$(W^{a_1a_2a_3a_4a_5a_6}D_2)(\mathbf{k}_1,\mathbf{k}_2,\mathbf{k}_3,\mathbf{k}_4,\mathbf{k}_5,\mathbf{k}_6) = \sum d_{a_1a_2a_3}d_{a_4a_5a_6}(WD_2)(1,2,3;4,5,6)\,. \tag{6.10}$$

The sum extends over all (ten) partitions of the six gluons into two groups containing three gluons each,

$$\begin{aligned}\sum d_{a_1a_2a_3}d_{a_4a_5a_6}(WD_2)(1,2,3;4,5,6) = {} & d_{a_1a_2a_3}d_{a_4a_5a_6}(WD_2)(1,2,3;4,5,6)\\ & + d_{a_1a_2a_4}d_{a_3a_5a_6}(WD_2)(1,2,4;3,5,6) + \dots\\ & + d_{a_1a_5a_6}d_{a_2a_3a_4}(WD_2)(1,5,6;2,3,4)\,.\end{aligned} \tag{6.11}$$

The function $WD_2$ is the same in all permutations. (Again this is an outcome of our calculation and was not assumed at any stage when the equation (6.3) was derived.) The sum thus contains the same permutations of the six gluons as the second term (6.4) on the right hand side of (6.3). For the notation to be consistent the function $W^{a_1a_2a_3a_4a_5a_6}$ should be understood as an integral operator acting on a BFKL amplitude $D_2$. It thus carries two more momentum arguments $\mathbf{q}_j$ for the momenta entering from above. We will again suppress these two momenta in the following.

In contrast to the other terms in (6.3) discussed so far the function $W$ does not permit a decomposition into the vertex functions $V$ known from the two–to–four transition vertex. We therefore give its full momentum space representation as we have obtained it as a result of our calculation. We again use the standard integrals defined in section 3.7. Then $WD_2$ has the explicit representation

$$\begin{aligned}&(WD_2)(\mathbf{k}_1,\mathbf{k}_2,\mathbf{k}_3;\mathbf{k}_4,\mathbf{k}_5,\mathbf{k}_6) = \frac{g^6}{16}\times\\ &\quad \{2\,[\,c(123456)\\ &\qquad - b(12345) - b(12346) - b(12356) - b(12456) - b(13456) - b(23456)\\ &\qquad + b(1234) + b(1235) + b(1236) + b(1456) + b(2456) + b(3456)\\ &\qquad + a(1245,3) + a(1246,3) + a(1256,3) + a(1345,2) + a(1346,2) + a(1356,2)\\ &\qquad + a(2345,1) + a(2346,1) + a(2356,1)\\ &\qquad - b(123) - b(456)\\ &\qquad - a(124,3) - a(125,3) - a(126,3) - a(134,2) - a(135,2) - a(136,2) +\end{aligned}$$

$$
\begin{aligned}
&- a(234,1) - a(235,1) - a(236,1) \\
&- a(145,23) - a(146,23) - a(156,23) - a(245,13) - a(246,13) - a(256,13) \\
&- a(345,12) - a(346,12) - a(356,12) \\
&+ a(12,3) + a(13,2) + a(23,1) + a(45,123) + a(46,123) + a(56,123) \\
&+ a(14,23) + a(15,23) + a(16,23) + a(24,13) + a(25,13) + a(26,13) \\
&+ a(34,12) + a(35,12) + a(36,12) \\
&- a(1,23) - a(2,13) - a(3,12) - a(4,123) - a(5,123) - a(6,123)] \\
+[\, & t(12345) + t(12346) + t(12356) + t(12456) + t(13456) + t(23456) \\
&- t(1234) - t(1235) - t(1236) - t(1456) - t(2456) - t(3456) \\
&- s(1245,3) - s(1245,6) - s(1246,3) - s(1246,5) - s(1256,3) - s(1256,4) \\
&- s(1345,2) - s(1345,6) - s(1346,2) - s(1346,5) - s(1356,2) - s(1356,4) \\
&- s(2345,1) - s(2345,6) - s(2346,1) - s(2346,5) - s(2356,1) - s(2356,4) \\
&+ t(123) + t(456) \\
&+ s(124,3) + s(124,56) + s(125,3) + s(125,46) + s(126,3) + s(126,45) \\
&+ s(134,2) + s(134,56) + s(135,2) + s(135,46) + s(136,2) + s(136,45) \\
&+ s(234,1) + s(234,56) + s(235,1) + s(235,46) + s(236,1) + s(236,45) \\
&+ s(145,23) + s(145,6) + s(146,23) + s(146,5) + s(156,23) + s(156,4) \\
&+ s(245,13) + s(245,6) + s(246,13) + s(246,5) + s(256,13) + s(256,4) \\
&+ s(345,12) + s(345,6) + s(346,12) + s(346,5) + s(356,12) + s(356,4) \\
&- s(12,3) - s(12,456) - s(13,2) - s(13,456) - s(23,1) - s(23,456) \\
&- s(45,123) - s(45,6) - s(46,123) - s(46,5) - s(56,123) - s(56,4) \\
&- s(14,23) - s(14,56) - s(15,23) - s(15,46) - s(16,23) - s(16,45) \\
&- s(24,13) - s(24,56) - s(25,13) - s(25,46) - s(26,13) - s(26,45) \\
&- s(34,12) - s(34,56) - s(35,12) - s(35,46) - s(36,12) - s(36,45) \\
&+ s(1,23) + s(1,456) + s(2,13) + s(2,456) + s(3,12) + s(3,456) \\
&+ s(4,123) + s(4,56) + s(5,123) + s(5,46) + s(6,123) + s(6,45)]\} \,. \qquad (6.12)
\end{aligned}
$$

On first sight this expression appears to be very complicated. Closer inspection reveals that it has a series of very interesting properties. Some of them very much resemble those of the function $V$ we have described in section 4.2.

Let us first have a look at the symmetry properties of the function $W$. We find that $W$ is fully symmetric in its first three arguments

$$
\begin{aligned}
W(\mathbf{k}_1,\mathbf{k}_2,\mathbf{k}_3;\mathbf{k}_4,\mathbf{k}_5,\mathbf{k}_6) &= W(\mathbf{k}_2,\mathbf{k}_1,\mathbf{k}_3;\mathbf{k}_4,\mathbf{k}_5,\mathbf{k}_6) \\
&= W(\mathbf{k}_3,\mathbf{k}_2,\mathbf{k}_1;\mathbf{k}_4,\mathbf{k}_5,\mathbf{k}_6) \qquad (6.13)
\end{aligned}
$$

as well as in its last three arguments,

$$\begin{aligned} W(\mathbf{k}_1,\mathbf{k}_2,\mathbf{k}_3;\mathbf{k}_4,\mathbf{k}_5,\mathbf{k}_6) &= W(\mathbf{k}_1,\mathbf{k}_2,\mathbf{k}_3;\mathbf{k}_5,\mathbf{k}_4,\mathbf{k}_6) \\ &= W(\mathbf{k}_1,\mathbf{k}_2,\mathbf{k}_3;\mathbf{k}_6,\mathbf{k}_5,\mathbf{k}_4)\,. \end{aligned} \tag{6.14}$$

Notably, the color structure corresponding to that permutation of momenta, i. e. the tensor $d_{a_1a_2a_3}d_{a_4a_5a_6}$, has exactly the same symmetry properties. Further, $W$ is symmetric under the exchange of the first three and last three arguments,

$$W(\mathbf{k}_1,\mathbf{k}_2,\mathbf{k}_3;\mathbf{k}_4,\mathbf{k}_5,\mathbf{k}_6) = W(\mathbf{k}_4,\mathbf{k}_5,\mathbf{k}_6;\mathbf{k}_1,\mathbf{k}_2,\mathbf{k}_3)\,. \tag{6.15}$$

From these symmetries of the function $W$ and from the permutations that enter the sum in (6.10) we can conclude that the full expression $W^{a_1a_2a_3a_4a_5a_6}(1,2,3,4,5,6)$ is completely symmetric in the six outgoing gluons, i. e. under the simultaneous exchange of color labels and momentum arguments.

Next we look at the behavior of $W$ when one of its momentum arguments vanishes. Not unexpectedly, we find that $W$ vanishes whenever one of the six gluons carries zero transverse momentum,

$$W(\mathbf{k}_1,\mathbf{k}_2,\mathbf{k}_3;\mathbf{k}_4,\mathbf{k}_5,\mathbf{k}_6)|_{\mathbf{k}_i=0} = 0 \qquad (i \in \{1,\ldots,6\})\,. \tag{6.16}$$

Starting from the explicit representation (6.12) the proof is straightforward. Of course, we again have to use the relations between the different standard integrals mentioned at the end of section 3.7 and the fact that the gluon trajectory function $\beta(\mathbf{k})$ vanishes for $\mathbf{k}=0$.

Further we find that the function $W$ is infrared finite. The proof requires just a close inspection of the standard integrals occurring in (6.12). It is easily seen that the integrals come in the infrared finite combinations (4.11) and (4.12) discussed already in section 4.2. (Again, the integrals have been arranged in (6.12) in a way that hopefully makes this transparent.)

We expect the function $W$ to be conformally invariant in impact parameter space, but we will not discuss this issue further here.

### 6.3.2 Interpretation in view of an effective field theory

It is now natural to ask where the new piece $(WD_2)^{a_1a_2a_3a_4a_5a_6}$ finds its place in the effective field theory of unitarity corrections. At present, we are not yet able to give a conclusive answer to this question. However, the following two possible answers naturally arise.

As we have seen in the preceding section the new piece has properties that very much resemble those of the two–to–four transition vertex $V_{2\to4}$. It is fully symmetric in the six gluons, it is infrared finite, and it vanishes when one of the gluon momenta vanishes. It is well possible that the new piece $W^{a_1a_2a_3a_4a_5a_6}$ is a new two–to–six gluon transition vertex $V_{2\to6}$, and that it constitutes a new element of the effective field theory.

The second possibility is the following. A coupling scheme can be chosen in the first three gluons and in the remaining three gluons. Then the function $W$ is split into several parts according to the symmetry or antisymmetry under the exchange of the gluons 1 and 2, say, and under the exchange of the pair (12) of gluons with gluon 3, and analogously for the other three gluons. Based on the result of this procedure it is possible to define new two–to–four vertices with symmetry properties different from those of the well–known two–to–four vertex $V_{2\to4}$. Based on these one can in turn define new four–gluon amplitudes that then become basic elements of the effective field theory. Due to this a direct transition from two to six reggeized gluons is avoided, and the new piece becomes a superposition of different two–to–four vertices. However, this second possibility has not yet been fully investigated.

The two possibilities have considerably different implications for the emerging field theory of unitarity corrections, especially in view of the necessary extension to amplitudes with more gluons in the $t$-channel. The first possibility corresponds to a picture in which an infinite number of new transition vertices $V_{2\to2n}$ occurs in the field theory, one for each even number of gluons. The second possibility, in contrast, leads to a picture in which there are only four two–to–four vertices with different symmetry properties, and no other vertices of the type $V_{2\to2n}$. Both possibilities should be investigated in more detail. We expect that especially a better understanding of the conformal invariance of the expected effective field theory will help to clarify the status of the new piece.

### 6.4 Further reggeization

Now we come to discuss the other terms that are present in the new integral equation (6.3) for the part $D_6^I$ of the six–gluon amplitude. In section 6.2 we have already shown that these terms can be written as superpositions of well–known vertex functions $V$. We will in this section disregard the new piece discussed in the previous section.

Already in section 5 we encountered a situation similar to the one which we find here in the integral equation (6.3). Also there the inhomogeneous term of the integral equation for $D_5^I$ could be written as a superposition of vertex functions $V$. It was a characteristic indication for the occurrence of a further reggeization of the amplitude $D_5^I$ with respect to the irreducible four–gluon amplitude $D_4^I$. This idea even allowed us to find the exact solution of the equation. We cannot expect that the remaining part $D_6^I$ of the six–gluon amplitude reggeizes completely. But the occurrence of the vertex functions in the inhomogeneous term of its equation strongly suggests that a part of $D_6^I$ will reggeize. To gain further insight we should therefore construct an ansatz for the remaining part in order to simplify the integral equation (6.3). The remaining part $D_6^I$ should thus be split into a reggeizing part and an irreducible part,

$$D_6^I = D_6^{I,R} + D_6^{I,I} , \tag{6.17}$$

where this time the term 'reggeizing' refers to the reggeization with respect to the four–gluon compound state, cf. the discussion at the end of section 5.3. The reggeizing part should be a superposition of irreducible four–gluon amplitudes, symbolically

$$D_6^{I,R} = \sum D_4^I . \tag{6.18}$$

The problem is now to find the correct color and momentum structure for the right hand side of this symbolic equation.

We should have in mind that the inhomogeneous term in the integral equations for $D_n$, i. e. the quark loop, always suggests the best choice of a reggeizing part $D_n^R$. To make a good guess for the reggeizing part $D_6^{I,R}$ we should therefore have a close look at the inhomogeneous term of the new integral equation (6.3).

Let us first look at the terms (6.7) and (6.9) containing the functions $I$ and $J$. We will pick one permutation in the sums (6.6) and (6.8) only, the other permutations can then be treated in analogy. We see immediately that the color and momentum structure in the first four gluons in the terms

$$d^{a_1a_2a_3a_4}\delta_{a_5a_6}I(1,2,3,4;5,6) \tag{6.19}$$

and

$$d^{a_2a_1a_3a_4}\delta_{a_5a_6}J(1,2,3,4;5,6) \tag{6.20}$$

is exactly the same as in $D_4^R$.

A second observation is a certain mismatch between $f_{abc}$ and $d_{abc}$ tensors. While there are terms of the kind

$$\sum f_{a_1a_2a_3}f_{a_4a_5a_6}L(1,2,3;4,5,6) \tag{6.21}$$

present in the equation which can be written in terms of $V$, the corresponding terms with $d_{abc}$ tensors (the new piece, see section 6.3) cannot be written in a similar way. This already indicates that the $f$- and $d$-tensors have to be treated differently.

In order to come from an ansatz of the form (6.18) to the cancellation of the inhomogeneous term in the integral equation we have to use the integral equation (4.4) for the irreducible part $D_4^I$ of the four–gluon amplitude. We want to write the arguments of the $D_4^I$'s in the ansatz in such a way that they have exactly the same momentum structure which we find in the vertex functions $V$ in (6.3). This is completely analogous to the parts $D_n^R$ in which the momentum structure was taken from the quark loop. In the case of $D_n^R$ we could also keep the color structure. This was possible because the color structure of the two–gluon amplitude $D_2$ was trivial, i. e. the two gluons were always in a color singlet state. That allowed us to factorize the two–gluon amplitude into a color part ($\delta_{a_1a_2}$) and a momentum part. Now the situation is more complicated since such a factorization is not possible for the irreducible four–gluon amplitude. We have to use the full amplitude $D_4^{I\,b_1b_2b_3b_4}$. Since we certainly need the six color labels $a_1, \dots, a_6$ in the ansatz (6.18), the technical procedure we have to use is the contraction with a tensor $\Theta$,

$$\Theta^{a_1a_2a_3a_4a_5a_6;b_1b_2b_3b_4} D_4^{I\,b_1b_2b_3b_4} \,. \tag{6.22}$$

The tensor $\Theta$ is an invariant tensor in the ten–fold tensor product $\bigotimes_{i=1}^{10}[\mathrm{su}(N_c)]$ of the Lie algebra. It will obviously be very difficult to find the correct tensors for the contractions in this huge tensor space without having additional information. We have to hope that the situation is in a certain sense more simple. It will be necessary to find restrictions on the tensors from the inhomogeneous term in the new integral equation. Now a problem arises. In the term

$$d^{a_1a_2a_3a_4}\delta_{a_5a_6}V(123,4;5,6) \tag{6.23}$$

for example the first four gluons are in an overall color singlet state, as are the last two gluons. Obviously, the color tensor necessary for the contraction with $D_4^{I\,b_1b_2b_3b_4}$ is fixed by the inhomogeneous term only in the case in which the gluons with labels $b_1$ and $b_2$ are in a color singlet state. For the other irreducible representations we have no hint from the inhomogeneous term which would restrict the tensor $\Theta$. This problem seems to be a conceptual one in our approach. In the first step, that is for identifying a reggeizing part $D_n^R$ in the $n$-gluon amplitude, the quark loop was sufficient to fix the reggeizing part. In the present situation it is not completely excluded that the correct color tensor for the non–singlet states cannot be fixed unambiguously. Possibly the solution of this problem requires a better knowledge of the irreducible four–gluon amplitude $D_4^I$. In spite of this conceptual problem we expect that one can find a simple ansatz that leads to further insight.

Unfortunately, we have not yet been able to find a satisfying and unique solution. It appears that a better understanding of the phenomenon of reggeization, especially in the irreducible parts of the amplitudes, will help to resolve this problem. First steps in this direction have been done in [61].

In view of the discussion above and in section 6.3.2 it is certainly to early to draw final conclusions concerning the field theory structure of the six–gluon amplitude. The most important result of this section is contained in the integral equation (6.3) for the remaining part $D_6^I$ of the six–gluon amplitude. The occurrence of the vertex function $V$ in this equation is an extremely strong indication for the fact that further reggeization with respect to the four–gluon compound state takes place in the six–gluon amplitude. Exactly this is the necessary condition for the emergence of the field theory structure in the unitarity corrections. We regard this as strong evidence for the existence of an effective field theory of unitarity corrections.

### 6.5 The Pomeron-Odderon-Odderon vertex

We now turn to the question which place the Odderon finds in the effective field theory. The Odderon is the $C = -1$ partner of the Pomeron, i. e. it carries negative charge parity. In perturbative QCD it consists of a compound state of three reggeized gluons described by the three–particle BKP equation [15, 20]. The three gluons are in a completely symmetric state, and the color part of the wavefunction is a $d_{abc}$ tensor. Given the quantum numbers of these states, it is a natural question whether a BFKL Pomeron can be coupled to two Odderons. The six–gluon amplitude is the obvious place to look for such a Pomeron–Odderon–Odderon vertex. The triple Pomeron vertex is obtained from the two–to–four gluon transition vertex $V_{2\to4}$ by projecting it onto three BFKL eigenfunctions [33, 41]. We can therefore in analogy try to project the inhomogeneous term of the new integral equation (6.3) onto two Odderon wavefunctions from below. The inhomogeneous term in that equation consists of several contributions, and we will concentrate on the piece discussed in section 6.3. We therefore ask whether the integral

$$V_{POO} = \int \left( \prod_{i=1}^{6} d^2\mathbf{k}_i \right) (W^{a_1a_2a_3a_4a_5a_6} D_2)(\mathbf{k}_1, \mathbf{k}_2, \mathbf{k}_3, \mathbf{k}_4, \mathbf{k}_5, \mathbf{k}_6) \\ \times d_{a_1a_2a_3} d_{a_4a_5a_6} \Psi_1(\mathbf{k}_1, \mathbf{k}_2, \mathbf{k}_3) \Psi_2(\mathbf{k}_4, \mathbf{k}_5, \mathbf{k}_6) \tag{6.24}$$

is different from zero. Unfortunately, the wavefunction $\Psi$ of the Odderon is not known explicitly, but conformal invariance places strong constraints on it. In [26] the wavefunction of the Odderon in impact parameter space was found to have the general form

$$\Psi(\rho_1, \rho_2, \rho_3) = \left( \frac{\rho_{12}\rho_{13}\rho_{23}}{\rho_{10}^2\rho_{20}^2\rho_{30}^2} \right)^{h/3} \psi(x)\,. \tag{6.25}$$

$h$ is the conformal weight of the Odderon state and $x$ is the anharmonic ratio

$$x = \frac{\rho_{12}\rho_{30}}{\rho_{13}\rho_{20}} \tag{6.26}$$

with $\rho_{ij} = \rho_i - \rho_j$, the $\rho_i$ being the two–dimensional coordinates in impact parameter space. The wavefunction $\Psi$ vanishes when two of the coordinates of the three gluons in the Odderon coincide. This property of the Odderon drastically reduces the number of terms in $(W^{a_1a_2a_3a_4a_5a_6} D_2)$ that can give a non–vanishing contribution to the integral (6.24). If a term in $(WD_2)$ depends only on the sum of two momenta, say $\mathbf{k}_1$ and $\mathbf{k}_2$, then this term is after Fourier transformation to impact parameter space proportional to a delta–function of the two corresponding coordinates, i. e. proportional to $\delta(\rho_1 - \rho_2)$. This implies a zero in the Odderon wavefunction and the corresponding term does not contribute to (6.24). Only very few standard integrals in $(WD_2)$ can actually give non–vanishing contributions. Some of them give identical results in the integral (6.24) due to the symmetry of the Odderon wavefunction.

(They are in this sense equivalent to each other as far as their contribution to (6.24) is concerned.) The possible contributions to the above integral can in this way be reduced to the following infrared finite combination of standard integrals:

$$2a(14,25) - s(14,25) - s(14,36)\,. \tag{6.27}$$

We now have to look for these terms in the new piece $(W^{a_1a_2a_3a_4a_5a_6}D_2)$. Interestingly, the first of the permutations in (6.11) — the color tensor of which exactly matches the color structure of the two Odderons — does not contain terms equivalent to the above combination. But the other 9 permutations do contain such terms. The term

$$d_{a_1a_2a_5}d_{a_3a_4a_6}(WD_2)(1,2,5;3,4,6) \tag{6.28}$$

for example, contains exactly the terms (6.27). As can be easily shown, each of those other permutations contains exactly four combinations equivalent to (6.27). The corresponding color contraction gives a factor

$$d_{a_1a_2a_5}d_{a_3a_4a_6}d_{a_1a_2a_3}d_{a_4a_5a_6} = \left(\frac{N_c^2-4}{N_c}\right)^2 (N_c^2-1)\,. \tag{6.29}$$

We can therefore conclude that a perturbative Pomeron–Odderon–Odderon vertex exists, and collecting all terms it becomes

$$\begin{aligned} V_{POO} = 9\,\cdot\,4\cdot\frac{g^6}{16}\left(\frac{N_c^2-4}{N_c}\right)^2(N_c^2-1)\int\left(\prod_{i=1}^{6}d^2\mathbf{k}_i\right)\times \\ \times[2a(14,25)-s(14,25)-s(14,36)]\Psi_1(\mathbf{k}_1,\mathbf{k}_2,\mathbf{k}_3)\Psi_2(\mathbf{k}_4,\mathbf{k}_5,\mathbf{k}_6)\,. \end{aligned} \tag{6.30}$$

This vertex certainly deserves further study. For example, it should be possible to write it in the form of a conformal three–point function as it was possible for the triple–Pomeron vertex. A more detailed knowledge of the Odderon wavefunction will eventually allow one to determine the numerical value of the integral (6.31). Another interesting question is whether the Pomeron–Odderon–Odderon vertex can also be calculated in the dipole picture of high energy QCD, as it was possible for the coupling of three Pomerons. This immediately raises the more fundamental question of how the Odderon arises in the dipole picture at all.

## 7. Summary and outlook

We have studied unitarity corrections in high energy QCD in the generalized leading logarithmic approximation. The objects of interest in this framework are amplitudes describing the production of $n$ gluons in the $t$-channel. These $n$-gluon amplitudes obey a tower of coupled integral equations. The equation for the two–gluon amplitudes coincides with the BFKL equation, and each integral equation involves all amplitudes with a lower number of gluons. The equation for a given $n$-gluon amplitude can therefore only be approached after the equations with less gluons have been solved.

A systematic approach to solving the equations successively has been presented. The first step is the identification of a reggeizing part $D_n^R$ of the amplitude under consideration. It is a superposition of two–gluon (BFKL) amplitudes and can be obtained from the lowest order term in the corresponding integral equation, i. e. the quark loop with $n$ gluons attached to it. We have given the corresponding expressions explicitly for up to six gluons. In the case of the three–gluon amplitude this part already solves the integral equation, whereas for $n \geq 4$ a new integral equation for the remaining part has to be derived. That derivation is technically involved since the number of convolutions of amplitudes with integral kernels increases rapidly with the number of gluons. We have identified a small set of standard integrals that can be used to classify all momentum space integrals occurring in the derivation. A combinatorial method suitable for implementation on a computer has been developed to bring all integrals to their standard form. Birdtrack notation is used for performing the contractions in color space. The resulting equation can then be studied in a second step.

We have reviewed the known results about the three– and four–gluon amplitudes, and discussed the field theory structure discovered in these amplitudes. The three–gluon amplitude reggeizes completely and is a superposition of BFKL amplitudes. The four–gluon amplitude consists of a reggeizing part (a superposition of BFKL amplitudes) and an irreducible part. The $t$-channel evolution of the latter starts with a two–gluon state that couples to a four–gluon state via the two–to–four vertex $V_{2\to 4}$. The emerging picture is that of an effective field theory with $n$-gluon states coupled to each other via number–changing vertices.

Using the methods described above we have then investigated the amplitudes with five and six gluons in the $t$-channel. After extracting a reggeizing part, i. e. a superposition of BFKL amplitudes, from the five–gluon amplitude we have derived the equation for the remaining part. The solution of this equation can be found as a superposition of irreducible four–gluon amplitudes. Thus the five–gluon amplitude again reggeizes completely. The mechanism causing this is the same as in the case of the three–gluon amplitude. It appears therefore very likely that every amplitude with an odd number of gluons exhibits complete reggeization. This would imply that only $t$-channel states with even numbers of gluons occur in the effective field theory.

A part of the six–gluon amplitude reggeizes again and is a superposition of BFKL amplitudes. As one of our main results, the integral equation for the remaining part has been derived, see eq. (6.3). It contains a new piece with very interesting symmetry properties. However, we have not yet been able to clarify whether it should be interpreted as a new two–to–six vertex or as a superposition of new two–to–four vertices. The other terms in the equation have been shown to be superpositions of

two–to–four vertices $V_{2\to4}$. We have calculated the perturbative Pomeron–Odderon–Odderon vertex from the new piece.

The emergence of the vertex $V_{2\to4}$ in some of the terms in the equation strongly suggests that a further part of the six–gluon amplitude is a superposition of irreducible four–gluon states, i. e. reggeizes with respect to the four–gluon state. We have not yet been able to find the correct color structure in this superposition. Here we encounter for the first time a situation in which reggeization occurs in a subsystem of two gluons which do not form a color singlet. In contrary to the first stage of reggeization (with respect to the two–gluon state) our equation does not fix that color structure uniquely. Knowing it would immediately allow one to generalize also the two–to–four vertex to the non–singlet case. The solution of this problem will probably require a better and more general understanding of the phenomenon of reggeization. Also information gained in other approaches could be very helpful here, especially the the dipole picture seems to be promising in this respect.

The most important property of the five– and six–gluon amplitudes found in the present paper is that they exhibit reggeization. On the other hand, we have seen that exactly reggeization is the necessary condition for the emergence of the field theory structure in the unitarity corrections. Therefore the five– and six–gluon amplitudes fit nicely into the picture of a potential effective field theory. We regard this as strong evidence for the conjecture that the whole set of unitarity corrections can be formulated as an effective field theory in $2 + 1$ dimensions, with rapidity acting as the time–like variable.

An important step will be to prove the conformal invariance of that effective field theory in impact parameter space also for the amplitudes with more than four gluons. The proof should obviously start with the five– and six–gluon amplitudes, but can hopefully be extended to all possible elements of the theory. We hope that the conformal symmetry can also help to answer the open questions about reggeization, about the two–to–four vertex in the color non–singlet, and about the meaning of the new piece found in the six–gluon amplitude. In summary, we expect that the unitarity corrections can be cast into the form of a conformal field theory. This opens the fascinating possibility of applying the powerful methods of conformal field theory and — once the effective conformal theory is identified — to derive the general properties of high energy QCD, now bypassing the laborious explicit calculation of higher $n$-gluon amplitudes. These general features should certainly not depend on the process under consideration. An effective field theory found in the amplitudes describing virtual photon–photon scattering is expected to be relevant also to other scattering processes in the high energy limit.

The NLO corrections to the BFKL equation have recently become available [62, 63], and the understanding of these corrections has been rapidly improved, see for example [64]. On a long–term basis it would be interesting to compute the NLO order corrections to all elements of the effective field theory of unitarity corrections, although this appears very difficult.

In the present paper we have concentrated on more theoretical aspects of the unitarity corrections. We would like to point out that unitarity corrections are also interesting from a phenomenological point of view. Unitarity is violated only at asymptotically high energies in any measurable quantity calculated in the BFKL formalism. However, this does by not means imply that the unitarity corrections, for example the four–gluon state, give a negligible contribution to observable quantities at presently accessible energies. The presence of the four–gluon state might well have a sizable effect on many observables, among them the total cross–section in virtual photon–photon scattering. A study of these effects would be very valuable.

## Acknowledgments

We would like to thank Gregory Korchemsky, Lev Lipatov, Alan White, and especially Hans Lotter and Mark Wüsthoff for most helpful discussions. We have also benefitted from very enjoyable and encouraging discussions with the late Vladimir Gribov.

## A. Colors

In this appendix we focus on more technical details of color algebra. In section A.1 we explain how the notation introduced in section 3.3 can be used to contract su($N_c$) tensors of arbitrary rank. We have applied the method for tensor contractions with up to six external gluon lines. Some of the basic results needed for the investigation of the $n$-gluon amplitudes are collected in section A.2.

### A.1 A method for contractions in su($N_c$) algebra

When doing the calculations sketched in sections 4 through 6 a standard task would be to calculate contractions of the type

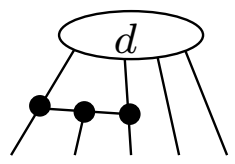

$$= d^{knde} f_{kal} f_{lbm} f_{mcn} \, . \tag{A.1}$$

We will now outline an algorithm to solve problems of this kind. We restrict ourselves to contractions in which the outgoing lines correspond to gluon color representations. The method is, however, readily extended to arbitrary tensors involving quark representations as well. The following prescription can be carried out diagrammatically.

Let us call 'standard tensors' such tensors that are (up to overall factors) the sum or the difference of traces of generators of the form

$$\mathrm{tr}(t^a \dots t^z) \pm \mathrm{tr}(t^z \dots t^a) \tag{A.2}$$

like the ones defined in (3.30), (3.31). The typical examples that occur in the analysis of the integral equations are $f_{abc}$, $d_{abc}$, $d^{abcd}$, $f^{abcde}$, and $d^{abcdef}$. The first step is to express all standard tensors occurring in the diagram by their representation in terms of generators, that is – diagrammatically speaking – by quark loops according to their respective definitions. (Here the terms 'quark line' and 'gluon line' refer to their respective color representation only.) For each of the standard tensors we then get two quark loops. The whole diagram is thus transferred to a sum of $2^m$ diagrams, $m$ being the number of standard tensors involved. Each of these diagrams contains only gluon lines and closed quark lines. It is natural to call all gluon lines starting on some closed quark line and ending on some closed quark line 'inner' gluon lines.

The key ingredient for our method is the decomposition of a quark–antiquark state into a singlet and an adjoint representation, also known as the Fierz identity,

$$\delta^\alpha_\gamma \delta^\delta_\beta = 2(t^a)^\alpha_\beta (t^a)^\delta_\gamma + \frac{1}{N_c}\delta^\alpha_\beta \delta^\delta_\gamma \,, \tag{A.3}$$

$\alpha, \ldots, \delta$ being color labels in the fundamental representation. After rearranging terms, it is in birdtracks

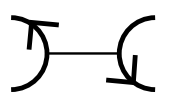 $= \frac{1}{2}$ 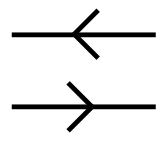 $- \frac{1}{2N_c}$ 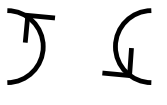 , (A.4)

which is applied to all inner gluon lines. To do this properly one has to draw all quark loops in the diagrams counterclockwise before, which is not a mathematical operation although it might be quite some exercise in drawing. Applying (A.4) again considerably increases the number of diagrams, but we now can read off the result. The reason is the following. The use of (A.4) replaces each inner gluon line by two diagrams. In one of them the two quark loops[9] joined by the gluon line are disconnected, in the other one they are joined to one closed quark loop. Having applied (A.4) to all inner gluon lines, we are left with diagrams that only contain closed quark loops on which the outer gluon lines end. We can now join the diagrams back into rather compact expressions using the identity

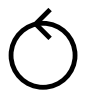 $= \mathrm{tr}\,\mathbf{1} = N_c \,,$ (A.5)

the vanishing of the trace of su($N_c$) generators,

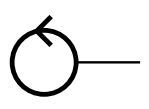 $= \mathrm{tr}\, t^a = 0 \,,$ (A.6)

and the definitions of the standard tensors given in section 3.3. The latter are supplied by

$$d_{abc}d_{def} + f_{abc}f_{def} = 8\,[\mathrm{tr}(t^a t^b t^c)\mathrm{tr}(t^f t^e t^d) + \mathrm{tr}(t^c t^b t^a)\mathrm{tr}(t^d t^e t^f)] \tag{A.7}$$

$$d_{abc}d_{def} - f_{abc}f_{def} = 8\,[\mathrm{tr}(t^a t^b t^c)\mathrm{tr}(t^d t^e t^f) + \mathrm{tr}(t^c t^b t^a)\mathrm{tr}(t^f t^e t^d)] \tag{A.8}$$

[9]They might be one and the same quark line, but that does not change our argument.

which is readily proved using the definition of the structure constants. In general, it is also necessary to use standard tensors of the type

$$f^{abcd} = -i\,[\mathrm{tr}(t^a t^b t^c t^d) - \mathrm{tr}(t^d t^c t^b t^a)] \tag{A.9}$$

$$d^{abcde} = \mathrm{tr}(t^a t^b t^c t^d t^e) + \mathrm{tr}(t^e t^d t^c t^b t^a) \tag{A.10}$$

— that is $f$-type tensors with an even number of color labels or $d$-type tensors with an odd number of color labels — to the ones mentioned below equation (A.2). For the identities needed in this paper (see next section), however, this is not necessary.

In the case of the above example (A.1) the result of this procedure is

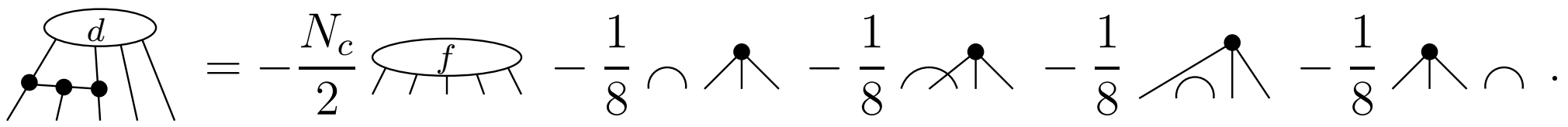
(A.11)

In many cases the above prescription can be shortened: if a subdiagram can be reduced or vanishes, if the whole diagram can be obtained from a known diagram by permutation of outgoing gluon lines, or using the invariance of subdiagrams under cyclic permutations.

## A.2 Useful contractions of color tensors

In this section we collect a series of su($N$) identities[10] obtained with the help of the method explained above. The list does not exhaust the contractions needed for the calculations described in this paper. Instead, we provide a list of identities from which many others can be easily derived.

For two external gluons we have

$$f_{lak} f_{kbl} = -N\delta_{ab} \tag{A.12}$$

$$d_{lak} d_{kbl} = \frac{N^2-4}{N}\,\delta_{ab}\,. \tag{A.13}$$

For three external gluons the following identities hold:

$$f_{kal} f_{lbm} f_{mck} = -\frac{N}{2} f_{abc} \tag{A.14}$$

$$d_{kal} f_{lbm} f_{mck} = -\frac{N}{2}\, d_{abc} \tag{A.15}$$

$$d_{kal} d_{lbm} f_{mck} = \frac{N^2-4}{2N} f_{abc} \tag{A.16}$$

$$d_{kal} d_{lbm} d_{mck} = \frac{N^2-12}{2N}\, d_{abc}\,. \tag{A.17}$$

The use of the last two identities can be avoided for the problems under consideration in this paper. They have been added for the sake of completeness here. For

[10] To avoid possible confusion of the subscript $c$ in $N_c$ with a color label $c$ we omit the subscript and give all results for su($N$) in this appendix.

considering the case of four external gluons the following identities are helpful:

$$f_{kal}f_{lbm}f_{mcn}f_{ndk} = Nd^{abcd} + \frac{1}{2}(\delta_{ab}\delta_{cd} + \delta_{ac}\delta_{bd} + \delta_{ad}\delta_{bc}) \tag{A.18}$$

$$d^{klcd}f_{kam}f_{mbl} = -\frac{N}{2}d^{abcd} - \frac{1}{4}\delta_{ab}\delta_{cd} \tag{A.19}$$

$$d^{kbld}f_{kam}f_{mcl} = \frac{1}{4}(\delta_{ab}\delta_{cd} + \delta_{ad}\delta_{bc}) \tag{A.20}$$

The following identities apply to five external gluons:

$$\begin{aligned} f_{kal}f_{lbm}f_{mcn}f_{ndo}f_{oek} = {} & Nf^{abcde} \\ & +\frac{1}{4}(\delta_{ab}f_{cde} + \delta_{ac}f_{bde} + \delta_{ad}f_{bce} + \delta_{ae}f_{bcd} \\ & \quad +f_{ade}\delta_{bc} + f_{ace}\delta_{bd} + f_{acd}\delta_{be} + f_{abe}\delta_{cd} \\ & \quad +f_{abd}\delta_{ce} + f_{abc}\delta_{de}) \end{aligned} \tag{A.21}$$

$$\begin{aligned} d^{klde}f_{kam}f_{mbn}f_{ncl} = {} & -\frac{N}{2}f^{abcde} \\ & -\frac{1}{8}(\delta_{ab}f_{cde} + \delta_{ac}f_{bde} + f_{ade}\delta_{bc} + f_{abc}\delta_{de}) \end{aligned} \tag{A.22}$$

$$\begin{aligned} d^{kcle}f_{kam}f_{mbn}f_{ndl} = {} & \frac{1}{8}(\delta_{ac}f_{bde} - \delta_{ae}f_{bcd} + f_{ade}\delta_{bc} - f_{acd}\delta_{be} \\ & \quad +f_{abe}\delta_{cd} + f_{abc}\delta_{de}) \end{aligned} \tag{A.23}$$

$$f^{klcde}f_{kam}f_{mbl} = -\frac{N}{2}f^{abcde} - \frac{1}{8}\delta_{ab}f_{cde} \tag{A.24}$$

$$f^{kblde}f_{kam}f_{mcl} = \frac{1}{8}(\delta_{ab}f_{cde} + f_{ade}\delta_{bc} - f_{abc}\delta_{de}) \tag{A.25}$$

Six external gluons require these identities:

$$\begin{aligned} f_{kal}f_{lbm}f_{mcn}f_{ndo}f_{oep}f_{pfk} = {} & -Nd^{abcdef} \\ & -\frac{1}{2}(\delta_{ab}d^{cdef} + \delta_{ac}d^{bdef} + \delta_{ad}d^{bcef} + \delta_{ae}d^{bcdf} \\ & \quad +\delta_{af}d^{bcde} + d^{adef}\delta_{bc} + d^{acef}\delta_{bd} + d^{acdf}\delta_{be} \\ & \quad +d^{acde}\delta_{bf} + d^{abef}\delta_{cd} + d^{abdf}\delta_{ce} + d^{abde}\delta_{cf} \\ & \quad +d^{abcf}\delta_{de} + d^{abce}\delta_{df} + d^{abcd}\delta_{ef}) \\ & +\frac{1}{8}[(d_{abc}d_{def} + f_{abc}f_{def}) + (d_{abd}d_{cef} + f_{abd}f_{cef}) \\ & \quad +(d_{abe}d_{cdf} + f_{abe}f_{cdf}) + (d_{abf}d_{cde} + f_{abf}f_{cde}) \\ & \quad +(d_{acd}d_{bef} + f_{acd}f_{bef}) + (d_{ace}d_{bdf} + f_{ace}f_{bdf}) \\ & \quad +(d_{acf}d_{bde} + f_{acf}f_{bde}) + (d_{ade}d_{bcf} + f_{ade}f_{bcf}) \\ & \quad +(d_{adf}d_{bce} + f_{adf}f_{bce}) + (d_{aef}d_{bcd} + f_{aef}f_{bcd})] \end{aligned} \tag{A.26}$$

$$d^{klef}f_{kam}f_{mbn}f_{nco}f_{odl} = \frac{N}{2}d^{abcdef} + \tag{A.27}$$

$$
\begin{aligned}
&+\frac{1}{4}(\delta_{ab}d^{cdef}+\delta_{ac}d^{bdef}+\delta_{ad}d^{bcef}+d^{adef}\delta_{bc} \\
&\quad +d^{acef}\delta_{bd}+d^{abef}\delta_{cd}+d^{abcd}\delta_{ef}) \\
&-\frac{1}{16}[(d_{abc}d_{def}+f_{abc}f_{def})+(d_{aef}d_{bcd}+f_{aef}f_{bcd}) \\
&\quad +(d_{acd}d_{bef}+f_{acd}f_{bef})+(d_{abd}d_{cef}+f_{abd}f_{cef})]
\end{aligned}
$$

$$
\begin{aligned}
d^{kdlf}f_{kam}f_{mbn}f_{nco}f_{oel} = &-\frac{1}{4}(\delta_{ad}d^{bcef}+d^{acef}\delta_{bd}+d^{abef}\delta_{cd}+d^{abcf}\delta_{de} \\
&\quad +d^{abcd}\delta_{ef}+\delta_{af}d^{cbde}+d^{bade}\delta_{cf}+d^{cade}\delta_{bf}) \\
&+\frac{1}{16}[(d_{abd}d_{cef}+f_{abd}f_{cef})+(d_{acd}d_{bef}+f_{acd}f_{bef}) \\
&\quad +(d_{aef}d_{bcd}+f_{aef}f_{bcd})+(d_{ade}d_{bcf}-f_{ade}f_{bcf}) \\
&\quad +(d_{acf}d_{bde}-f_{acf}f_{bde}) \\
&\quad +(d_{abf}d_{cde}-f_{abf}f_{cde})] \qquad \text{(A.28)}
\end{aligned}
$$

$$
\begin{aligned}
f^{kldef}f_{kam}f_{mbn}f_{ncl} = &\frac{N}{2}d^{abcdef}+\frac{1}{4}(\delta_{ab}d^{cdef}+\delta_{ac}d^{bdef}+d^{adef}\delta_{bc}) \\
&-\frac{1}{16}(d_{abc}d_{def}+f_{abc}f_{def}) \qquad \text{(A.29)}
\end{aligned}
$$

$$
\begin{aligned}
f^{kclef}f_{kam}f_{mbn}f_{ndl} = &-\frac{1}{4}(\delta_{ac}d^{bdef}+d^{adef}\delta_{bc}+d^{abef}\delta_{cd}+d^{bacd}\delta_{ef}) \\
&+\frac{1}{16}[(d_{abc}d_{def}+f_{abc}f_{def})+(d_{acd}d_{bef}-f_{acd}f_{bef}) \\
&\quad +(d_{aef}d_{bcd}-f_{aef}f_{bcd})] \qquad \text{(A.30)}
\end{aligned}
$$

$$
d^{klcdef}f_{kam}f_{mbl} = -\frac{N}{2}d^{abcdef}-\frac{1}{4}\delta_{ab}d^{cdef} \qquad \text{(A.31)}
$$

$$
\begin{aligned}
d^{kbldef}f_{kam}f_{mcl} = &\frac{1}{4}(\delta_{ab}d^{cdef}+\delta_{bc}d^{adef}) \\
&-\frac{1}{16}(d_{abc}d_{def}-f_{abc}f_{def}) \qquad \text{(A.32)}
\end{aligned}
$$

$$
\begin{aligned}
d^{kbclef}f_{kam}f_{mdl} = &-\frac{1}{4}(d^{abcd}\delta_{ef}+\delta_{bc}d^{adef}) \qquad \text{(A.33)} \\
&+\frac{1}{16}[(d_{abc}d_{def}-f_{abc}f_{def})+(d_{aef}d_{bcd}-f_{aef}f_{bcd})]
\end{aligned}
$$

# B. A combinatorial method for the momentum space integrals

A main step in our investigation of a given $n$–gluon amplitude $D_n$ is to split it into a reggeizing part $D_n^R$ and a remaining part. The reggeizing part is a superposition of BFKL amplitudes $D_2$. Starting from this ansatz a new integral equation for the remaining part of the amplitude is derived. In order to calculate its inhomogeneous term it is necessary to convolute the reggeizing parts $D_l^R$ of the $l$-gluon amplitudes (with $l \leq n$) with the integral kernels $K_{2\to m}^{\{b\}\to\{a\}}$ $(l+m-2=n)$ according to the original integral equations. In this appendix we present an algorithm for performing the momentum space integrals. It relies on the classification of possible momentum space integrals given in section 3.7, and allows us to bring all occuring integrals to their standard form. For obtaining a part of the results in this paper we have implemented the algorithm in the PERL script language. After that a computer algebra program (like MAPLE for example) can be used to multiply the integrals by the corresponding color tensors and to finally collect all terms.

The main purpose of our method is to reduce the problem of convoluting amplitudes with kernels to a purely combinatorial task. We therefore use notation known from the theory of sets in this appendix. Below we will give a rule for the treatment of the convolution of one specific term in the amplitude $D_l^R$ with the transition kernel $K_{2\to m}^{\{b\}\to\{a\}}$ $(l+m-2=n)$, and only the momentum part of the kernel will be of interest in this appendix. The method can then successively be applied to all possible convolutions of individual terms in the reggeizing parts of the amplitudes with the integral kernels.

Let us now consider one specific term in the reggeizing part $D_l^R$ of the $l$-gluon amplitude. It is given by a BFKL amplitude that has two momentum arguments. Each of them is the sum of a subset of the $l$ momenta $\mathbf{q}_j$. Let us call these two subsets $\mathcal{A}$ and $\mathcal{B}$, respectively. Their union exhausts the $l$ momenta,

$$\mathcal{A}\cup\mathcal{B} = \{\mathbf{q}_1,\dots,\mathbf{q}_l\}\,, \tag{B.1}$$

and each of them contains at least one element,

$$1 \le \#\mathcal{A},\ \#\mathcal{B} \le l-1\,. \tag{B.2}$$

We will in the following identify a set of momenta with the sum of its elements. With this identification the term in the amplitude $D_l^R$ we want to consider is

$$D_2(\mathcal{A},\mathcal{B}) = D_2\left(\sum_{r=1}^{\#\mathcal{A}}\mathbf{q}_{j_r},\sum_{s=1}^{\#\mathcal{B}}\mathbf{q}_{j_s}\right)\,. \tag{B.3}$$

Now we want to convolute this term with an integral kernel. Only two of the $l$ momenta $\mathbf{q}_j$ will actually be affected by the convolution. Let us call these two momenta $\mathbf{v}$ and $\mathbf{w}$. The kernel $K_{2\to m}^{\{b\}\to\{a\}}$ was given explicitly in section 3.6. We neglect the coupling constant $g$ and the color tensor for the purpose of this appendix. The momentum part of the kernel is according to (3.57)

$$\begin{aligned}K_{2\to m}(\mathbf{v},\mathbf{w};\mathbf{k}_{i_1},\dots,\mathbf{k}_{i_m}) &= (\mathbf{k}_{i_1}+\dots+\mathbf{k}_{i_m})^2 - \frac{\mathbf{w}^2(\mathbf{k}_{i_1}+\dots+\mathbf{k}_{i_{m-1}})^2}{(\mathbf{k}_{i_m}-\mathbf{w})^2}\\ &\quad-\frac{\mathbf{v}^2(\mathbf{k}_{i_2}+\dots+\mathbf{k}_{i_m})^2}{(\mathbf{k}_{i_1}-\mathbf{v})^2}+\frac{\mathbf{v}^2\mathbf{w}^2(\mathbf{k}_{i_2}+\dots+\mathbf{k}_{i_{m-1}})^2}{(\mathbf{k}_{i_1}-\mathbf{v})^2(\mathbf{k}_{i_m}-\mathbf{w})^2}\,.\end{aligned} \tag{B.4}$$

The last term is not present if $m=2$. The momenta $\mathbf{k}_{i_t}$ with $(t\in\{1,\dots,m\})$ are $m$ of the $n$ momenta that occur in the integral equation for $D_n$. Due to the condition that $t$-channel gluons do not cross in the integral equations (see section 3.2) they are ordered:

$$1 \le i_1 < \ldots < i_m \le n\,. \tag{B.5}$$

Which $m$ of the $n$ gluons in the integral equation enter the kernel from below depends of course on the term we have chosen in the sums on the right hand side of the integral equations (3.10)–(3.14). The quantity we want to calculate here is the convolution

$$K_{2\to m}(\mathbf{v},\mathbf{w};\mathbf{k}_{i_1},\ldots,\mathbf{k}_{i_m}) \otimes D_2(\mathcal{A},\mathcal{B})\,, \tag{B.6}$$

where the symbol $\otimes$ again includes an integral over the loop momentum and the two propagators $\frac{1}{\mathbf{v}^2}\frac{1}{\mathbf{w}^2}$. The kernel acts trivially on the other $l-2$ momenta in the term $D_2(\mathcal{A},\mathcal{B})$. Our algorithm will leave them unchanged, that is after its application we are still left with some $\mathbf{q}_j$'s in the standard integral. They have to be replaced in the end by the respective $\mathbf{k}_i$'s. Mathematically speaking this is done by the one–to–one map

$$\mathcal{A}\cup\mathcal{B}\setminus\{\mathbf{v},\mathbf{w}\} \longrightarrow \{\mathbf{k}_1,\ldots,\mathbf{k}_n\}\setminus\{\mathbf{k}_{i_1},\ldots,\mathbf{k}_{i_m}\} \tag{B.7}$$

which has to be applied in ascending order according to the occurrence of the momenta on both sides. On the left hand side we have the $l-2$ momenta in $D_2(\mathcal{A},\mathcal{B})$ not affected by the kernel, on the right hand side we find the $n-m=l-2$ momenta that are not attached to the kernel from below.

In addition, one more step has to be performed to finish the result after the rules below have been applied. This is connected with the definition of the second argument of the standard integral $a$. As described in section 3.7 the second argument of the function $a$ is a sum of momenta which has to be chosen out of two sums that occur in the integral. According to our definition the second argument of $a$ is the group of momenta that contains the momentum $\mathbf{k}$ with the lowest index. In our general treatment in this appendix it is not convenient to implement this condition from the beginning. Instead we adjust the resulting standard integrals in the end. This is done very easily. For example, if we have $n=4$ and the algorithm below leads to the result $a(2,34)$ then this should be replaced by $a(2,1)$.

We will treat the four parts of the kernel in (B.4) separately now. The resulting standard integrals have to be added in the end.

### First part of the kernel

The first term in the kernel (B.4) is

$$P(\mathbf{v},\mathbf{w};\mathbf{k}_{i_1},\ldots,\mathbf{k}_{i_m}) = (\mathbf{k}_{i_1}+\ldots+\mathbf{k}_{i_m})^2\,, \tag{B.8}$$

and we want to bring the convolution

$$P(\mathbf{v},\mathbf{w};\mathbf{k}_{i_1},\ldots,\mathbf{k}_{i_m}) \otimes D_2(\mathcal{A},\mathcal{B}) \tag{B.9}$$

to its standard form. Let the set $\mathcal{X}$ be $\mathcal{X}=\{\mathbf{k}_{i_1},\ldots,\mathbf{k}_{i_m}\}$. Then the different possible cases are

1. $\mathbf{v}$ and $\mathbf{w}$ are elements of the same set $\mathcal{A}$ or $\mathcal{B}$. We then denote this set ($\mathcal{A}$ or $\mathcal{B}$) by $\mathcal{C}$.
   (a) $\#\mathcal{C} = 2$: The integral is $t(\mathcal{X})$.
   (b) $\#\mathcal{C} > 2$: The integral is $s(\mathcal{X}, \mathcal{C} \setminus \{\mathbf{v}, \mathbf{w}\})$.
2. $\mathbf{v}$ and $\mathbf{w}$ are *not* elements of the same set $\mathcal{A}$ or $\mathcal{B}$.
   (a) $(\#\mathcal{A} = 1) \wedge (\#\mathcal{B} = 1)$: The integral is $c(\mathcal{X})$.
   (b) $(\#\mathcal{A} = 1) \wedge (\#\mathcal{B} > 1)$: The integral is $b(\mathcal{X})$.
   (c) $(\#\mathcal{A} > 1) \wedge (\#\mathcal{B} = 1)$: The integral is $b(\mathcal{X})$.
   (d) $(\#\mathcal{A} > 1) \wedge (\#\mathcal{B} > 1)$: The integral is $a(\mathcal{X}, \mathcal{A} \setminus \{\mathbf{v}, \mathbf{w}\})$.

**Second part of the kernel**

The second term in the kernel (B.4) is

$$Q(\mathbf{v}, \mathbf{w}; \mathbf{k}_{i_1}, \ldots, \mathbf{k}_{i_m}) = \frac{\mathbf{w}^2(\mathbf{k}_{i_1} + \ldots + \mathbf{k}_{i_{m-1}})^2}{(\mathbf{k}_{i_m} - \mathbf{w})^2}, \tag{B.10}$$

and we want to bring the convolution

$$Q(\mathbf{v}, \mathbf{w}; \mathbf{k}_{i_1}, \ldots, \mathbf{k}_{i_m}) \otimes D_2(\mathcal{A}, \mathcal{B}) \tag{B.11}$$

to its standard form. Let now the set $\mathcal{X}$ denote $\mathcal{X} = \{\mathbf{k}_{i_1}, \ldots, \mathbf{k}_{i_{m-1}}\}$, and let the set $\mathcal{Y}$ be $\mathcal{Y} = \{\mathbf{k}_{i_m}\}$. Then the different possible cases are

1. $\mathbf{v}$ and $\mathbf{w}$ are elements of the same set $\mathcal{A}$ or $\mathcal{B}$. We then denote this set ($\mathcal{A}$ or $\mathcal{B}$) by $\mathcal{C}$. The integral is $s(\mathcal{X}, (\mathcal{C} \setminus \{\mathbf{v}, \mathbf{w}\}) \cup \mathcal{Y})$.
2. $\mathbf{v}$ and $\mathbf{w}$ are *not* elements of the same set $\mathcal{A}$ or $\mathcal{B}$. Let the set ($\mathcal{A}$ or $\mathcal{B}$) containing $\mathbf{v}$ be $\mathcal{C}$.
   (a) $\#\mathcal{C} = 1$: The integral is $b(\mathcal{X})$.
   (b) $\#\mathcal{C} > 1$: The integral is $a(\mathcal{X}, \mathcal{C} \setminus \{\mathbf{v}\})$.

**Third part of the kernel**

The third term in the kernel (B.4) is

$$R(\mathbf{v}, \mathbf{w}; \mathbf{k}_{i_1}, \ldots, \mathbf{k}_{i_m}) = \frac{\mathbf{v}^2(\mathbf{k}_{i_2} + \ldots + \mathbf{k}_{i_m})^2}{(\mathbf{k}_{i_1} - \mathbf{v})^2}, \tag{B.12}$$

and we want to bring the convolution

$$R(\mathbf{v}, \mathbf{w}; \mathbf{k}_{i_1}, \ldots, \mathbf{k}_{i_m}) \otimes D_2(\mathcal{A}, \mathcal{B}) \tag{B.13}$$

to its standard form. Let now the set $\mathcal{X}$ denote $\mathcal{X} = \{\mathbf{k}_{i_2}, \ldots, \mathbf{k}_{i_m}\}$, and let now the set $\mathcal{Y}$ be $\mathcal{Y} = \{\mathbf{k}_{i_1}\}$. Then the different possible cases are

1. $\mathbf{v}$ and $\mathbf{w}$ are elements of the same set $\mathcal{A}$ or $\mathcal{B}$. We then denote this set ($\mathcal{A}$ or $\mathcal{B}$) by $\mathcal{C}$. The integral is $s(\mathcal{X}, (\mathcal{C} \setminus \{\mathbf{v}, \mathbf{w}\}) \cup \mathcal{Y})$.

2. $\mathbf{v}$ and $\mathbf{w}$ are *not* elements of the same set $\mathcal{A}$ or $\mathcal{B}$. Let the set ($\mathcal{A}$ or $\mathcal{B}$) containing $\mathbf{w}$ be $\mathcal{C}$.

   (a) $\#\mathcal{C} = 1$: The integral is $b(\mathcal{X})$.

   (b) $\#\mathcal{C} > 1$: The integral is $a(\mathcal{X}, \mathcal{C} \setminus \{\mathbf{w}\})$.

**Fourth part of the kernel**

The fourth term in the kernel (B.4) is

$$S(\mathbf{v}, \mathbf{w}; \mathbf{k}_{i_1}, \ldots, \mathbf{k}_{i_m}) = \frac{\mathbf{v}^2 \mathbf{w}^2 (\mathbf{k}_{i_2} + \ldots + \mathbf{k}_{i_{m-1}})^2}{(\mathbf{k}_{i_1} - \mathbf{v})^2 (\mathbf{k}_{i_m} - \mathbf{w})^2}\,, \tag{B.14}$$

and we want to bring the convolution

$$S(\mathbf{v}, \mathbf{w}; \mathbf{k}_{i_1}, \ldots, \mathbf{k}_{i_m}) \otimes D_2(\mathcal{A}, \mathcal{B}) \tag{B.15}$$

to its standard form. Let now the set $\mathcal{X}$ denote $\mathcal{X} = \{\mathbf{k}_{i_2}, \ldots, \mathbf{k}_{i_{m-1}}\}$, and let now the set $\mathcal{Y}$ be $\mathcal{Y} = \{\mathbf{k}_{i_1}\}$. Let in addition the set $\mathcal{Z}$ be $\mathcal{Z} = \{\mathbf{k}_{i_m}\}$. Then the different possible cases are

1. $\mathbf{v}$ and $\mathbf{w}$ are elements of the same set $\mathcal{A}$ or $\mathcal{B}$. We then denote this set ($\mathcal{A}$ or $\mathcal{B}$) by $\mathcal{C}$. The integral is $s(\mathcal{X}, (\mathcal{C} \setminus \{\mathbf{v}, \mathbf{w}\}) \cup \mathcal{Y} \cup \mathcal{Z})$.

2. $\mathbf{v}$ and $\mathbf{w}$ are *not* elements of the same set $\mathcal{A}$ or $\mathcal{B}$. Let the set ($\mathcal{A}$ or $\mathcal{B}$) containing $\mathbf{v}$ be $\mathcal{C}$. The integral is $a(\mathcal{X}, (\mathcal{C} \setminus \{\mathbf{v}\}) \cup \mathcal{Y})$.